\documentclass[aps,prl,twocolumn,10pt,amsmath,amssymb,showpacs,superscriptaddress,notitlepage,longbibliography]{revtex4-1}
\usepackage[colorlinks=true,linkcolor=blue,anchorcolor=red,citecolor=blue, urlcolor=blue]{hyperref}
\usepackage{bm}
\usepackage{graphicx}
\usepackage{xcolor}
\usepackage{wasysym}
\usepackage{makecell}
\usepackage{multirow}
\usepackage{comment}

\begin{document}
\title{Landscapes of an out-of-equilibrium anyonic sea}

\author{Gu Zhang}\email{zhanggu@baqis.ac.cn}
\affiliation{Beijing Academy of Quantum Information Sciences, Beijing 100193, China}

\author{Igor Gornyi}
\email{igor.gornyi@kit.edu}
\affiliation{Institute for Quantum Materials and Technologies and Institut f\"ur Theorie der Kondensierten Materie, Karlsruhe Institute of Technology, 76131 Karlsruhe, Germany}

\author{Yuval Gefen}
\email{yuval.gefen@weizmann.ac.il}
\affiliation{Department of Condensed Matter Physics, Weizmann Institute of Science, Rehovot 761001, Israel}

\begin{abstract}

The low-energy dynamics of two-dimensional topological matter hinges on its one-dimensional edge modes. Tunneling between fractional quantum Hall edge modes facilitates the study of anyonic statistics: it induces time-domain braiding that dominates signals from diluted anyon beams. We develop a framework for characterizing one-dimensional out-of-equilibrium anyonic states and define their effective potential and temperature, both arising from anyonic braiding, as well as the landscape of their excitations. Unlike fermions, the effective anyon potential depends on the type of the tunneling quasiparticles;
non-equilibrium anyonic states are underlain by power-law energy distributions. This allows ``hot’’ anyons to tunnel above the chemical potential of the source, which we capture by a measurable universal witness function.
Our analysis raises the prospect of generalizing the kinetic approach to compressible anyonic matter in higher dimensions.

\end{abstract}

\maketitle

The foundational observations associated with non-conventional quantum statistics (see, e.g., Refs.~\cite{LaughlinPRL83, Arovas1984,Kitaev2001UFN, MongPRX14}), and the prospects for harnessing this for the design of quantum information platforms~\cite{HorodeckiRevModPhys09,NielsenChuangBook,WildeBook,GuhneTothReview,StreltsovAdessoPlenioRevModPhys17}, render the physics of anyons a focal point of interest, with recent experiments on fractional quantum Hall (QH) edge states~\cite{NakamuraNatPhys19, NakamuraNatPhys20,BartolomeiScience20, NakamuraNC22, Nakamura2023, HeiblumNP2023, PierrePRX23, PierreNC23, LeeNature23, RuellePRX23, ghosh2024} following a long line of theoretical study~\cite{ThoulessGefenPRL91, SafiDevilardPRL01,  KaneFisherPRB03, VishveshwaraPRL03, KimPRL05, LawPRB06, FeldmanKitaevPRL06, FeldmanPRB07, PonomarenkoAverinPRL07, RosenowHalperinPRL07, CampagnanoPRL12, CampagnanoPRB13, RosenowLevkivskyiHalperinPRL16, CampagnanoPRB16, SimNC16, LeePRL19, RosenowSternPRL20, MartinDeltaT20, MorelPRB22, KyryloPRB22, GuPRB22, LeeSimNC22, schillerPRL23, JonckheerePRL23, JonckheerePRB23, IyerX2023, GuX2023Andreev, Batra2023, KivelsonX2024, ThammBerndPRL24}.
Remarkable breakthrough has been reported in identifying the anyonic braiding phase employing Fabry-Perot interferometry~\cite{NakamuraNatPhys19, NakamuraNatPhys20, NakamuraNC22, Nakamura2023}, Mach-Zehnder~\cite{Deviatov2012, HeiblumNP2023, ghosh2024}, as well as Hong-Ou-Mandel (HOM) collider setups~\cite{BartolomeiScience20, PierrePRX23, LeeNature23, RuellePRX23}.
With the latter, the anyonic braiding phase was produced through ``time-domain braiding''~\cite{schillerPRL23}, a generalization of ``bubble-diagram braiding''~\cite{SimNC16,LeePRL19}, see Supplemental Information (SI) \footnote{see Supplemental Information for technical details} Sec.~S1 for illustration.
This process comprises an exchange between an incoming anyon (part of a source-injected non-equilibrium beam)~\cite{SimNC16,LeePRL19} and a quasiparticle-quasihole pair generated 
at the collider; this is followed by the annihilation of this pair at a later time, hence ``time-domain braiding''.
Admittedly, the results reported in Refs.~\cite{BartolomeiScience20, PierrePRX23, LeeNature23, RuellePRX23} do not provide a direct observation of how time-domain braiding influences individual-anyon tunneling events, in which the anyonic gas is maintained out of equilibrium.

The study reported here introduces the notion of \textit{Laughlin surface}, the anyonic counterpart of electrons' Fermi surface.
Considering such strongly-correlated interacting particles, the very possibility of employing the quantum kinetic approach 
is far from obvious.
The challenging questions here are:
(i) whether the physics of non-equilibrium anyons, with Pauli blockade replaced by fractional statistics, differs from that of a non-equilibrium Fermi landscape,
and (ii) whether
non-equilibrium anyons can be described in terms of an effective surface, dubbed ``Laughlin surface'', characterized by (iii) anyonic distribution functions with (iv) effective chemical potential and temperature.
We show that the answer to these questions is in the affirmative.

Below we investigate the non-equilibrium landscape of excitations
around the Laughlin surface, focusing on the impact of 
anyon-braiding dynamics.
The inclusion of the latter
radically distinguishes our analysis from earlier studies  (Refs.~\cite{HaldanePRL91, WuRPL94}), 
where equilibrium distributions of anyons have been evaluated with standard statistical-mechanics tools.
The braiding-induced kinetics is behind our main findings, as detailed below.

\textbf{\emph{Main results}---}\textbf{(I).}~Anyon braiding induces an effective shift of the potential, $V_\text{eff}$, and an effective temperature, $T_\text{eff}$, 
in an out-of-equilibrium anyonic system (one-dimensional channel A below).
These are diagnosed by 
an equilibrium anyonic
channel B prepared at potential $V_\text{B}$ and temperature $T_\text{B}$, tuned such that no net charge or heat transport 
between the channels implies $V_\text{eff} = V_\text{B}$ and $T_\text{eff} = T_\text{B}$ (cf. Refs.~\cite{EngquistAndersonPRB81,
NosigliaParkRosenowGefenPRB18,
SpanslattParkGefenMirlinPRL19}).
We further show that $V_\text{eff}$ is generated by the non-equilibrium current, 
$I_\text{A}$, 
while $T_\text{eff}$ is influenced by the corresponding non-equilibrium current noise $S_\text{A}$. 
\textbf{(II).}~The non-equilibrium distribution function of excited anyon modes decreases slowly (power law) at high energies, 
in sharp contrast to the exponential decrease with the energy of equilibrium excitations.
This implies finite non-equilibrium particle excitations even above the ``ceiling''
(the energy above which zero-temperature equilibrium particle distribution at the source vanishes).
Hereafter, this is referred to as the generation of hot anyons by braiding.
We construct a universal measurable function, $\mathcal{W}$, 
witnessing and characterizing this
phenomenon. 
\textbf{(III).}~At zero ambient temperature, charge tunneling from the non-equilibrium channel to the equilibrium one is only possible above the Laughlin surface (the surface of the anyonic sea) of the latter;
The tunneling of anyons with energies below this Laughlin surface is blocked, in analogy to Pauli blocking for fermions.
\textbf{(IV).}~The effective potential depends on the type of quasiparticles that tunnel between the anyonic channels.
These remarkable features are unique to out-of-equilibrium anyonic landscapes, which rely on the anyonic statistical phase, and thus expected for compressible anyonic matter beyond quantum Hall edges.

\begin{figure}
  \includegraphics[width=1 \linewidth]{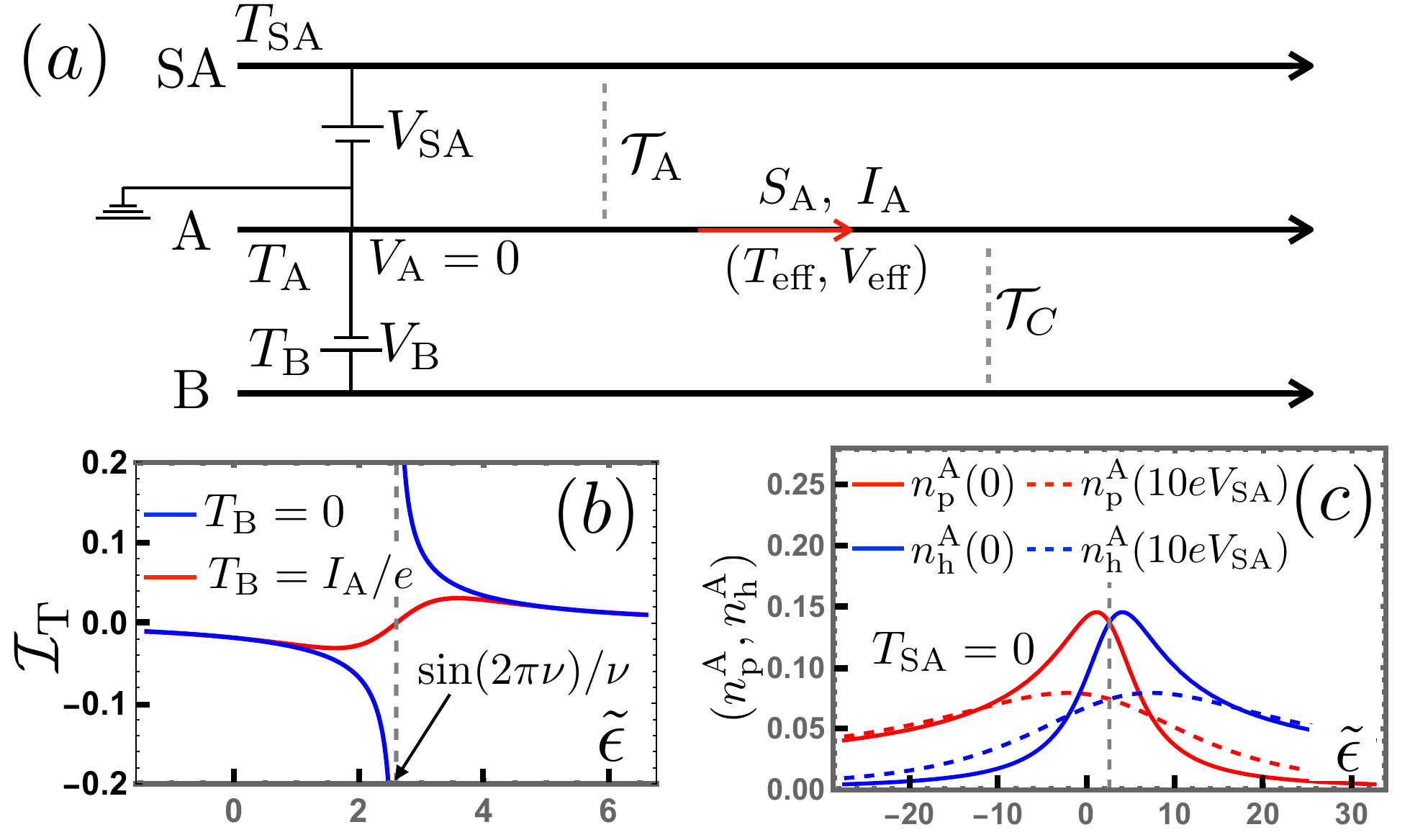} 
  \caption{The model, spectrum, and particle/hole distribution functions, as function of the dimensionless energy $\tilde{\epsilon} \equiv \epsilon/(I_\text{A}/e)$. (a)
Channel A supports non-equilibrium current, $I_\text{A}$, coming
from source SA.
The effective 
potential $V_\text{eff}$ and temperature $T_\text{eff}$
of channel A are diagnosed by the equilibrium channel B.
(b) The tunneling-current spectrum $\mathcal{I}_\text{T} (\epsilon)$ [in units of $(\tau_c I_\text{A}/e)^{2\nu-2}$], when $ V_\text{B} = V_\text{eff} = I_\text{A}\sin (2\pi\nu)/\nu^2e^2$, for zero (blue curve) and finite ($T_\text{B}= I_\text{A}/e$, red curve) temperatures of channel B. In both cases, the charge tunneling current vanishes: $I_\text{T} \equiv \nu e \int d\epsilon \mathcal{I}_\text{T} (\epsilon) = 0$.
(c) Particle and hole distributions in channel A [in units of $(\tau_c I_\text{A}/e)^{\nu-1}$], for ambient $T_\text{A} = 0$ (solid curves) and $T_\text{A} = 10 e V_\text{SA}$ (dashed curves), with $T_\text{SA} = 0$.
  }
  \label{fig:structure}
\end{figure}

\textbf{\emph{The model}---}
Consider the setup in Fig.~\ref{fig:structure}(a) (cf.  Ref.~\cite{KaneFisherPRB03}), representing
three chiral anyonic channels that may support quasiparticles of charge $\nu e$.
Channel A is grounded at its source; non-equilibrium current $I_\text{A}$ is injected from source SA via a tunneling bridge (``diluter'',  $x=0$). Channel B is tunnel-coupled to channel A at the ``collider'' ($x=L$).
Equilibrium channels SA (with temperature $T_\text{SA}$) and B (with temperature $T_\text{B}$) are biased at voltages $V_\text{SA}$ and $V_\text{B}$, respectively.
The three chiral channels are described by the free Hamiltonian $H_\text{edge} = \frac{1}{4\pi} \sum_\alpha\int dx (\partial_x \phi_\alpha)^2$, where $\phi_\alpha$ is the dynamical bosonic field operator, satisfying $[\phi_\alpha (x),\phi_{\alpha'} (x')] =i\pi \delta_{\alpha\alpha'} \text{sgn}(x-x')$, with $\alpha\!=\!\text{SA},\,\text{A},\,\text{B}$ (hereafter, we set $\hbar = k_B = 1$). The diluter Hamiltonian is $H_\text{SA-A} = t_\text{A} \psi^\dagger_\text{SA} (0) \psi_\text{A} (0) + \text{H.c.}$, 
where ${\psi^\dagger_\text{SA}\! (0)\psi_\text{A}\!(0)\!=\!F_\text{SA}^\dagger F_\text{A}\exp[-i\sqrt{\nu} (\phi_\text{SA}\!-\!\phi_\text{A})]/(2\pi \tau_c)}$, with $\tau_c$ the ultraviolet cutoff, $F_\text{SA}^\dagger F_\text{A}$ the product of Klein factors, and $|t_\text{A}|^2 = \mathcal{T}_\text{A}$.
Downstream of the diluter, channel A carries a non-equilibrium current $I_\text{A}$ with noise $S_\text{A} \equiv \int dt [\langle \hat{I}_\text{A} (t) \hat{I}_\text{A} (0)\rangle - I_\text{A}^2]$. The collider Hamiltonian is
$H_\text{T} = t_\text{C} \psi^\dagger_\text{A} (L) \psi_\text{B} (L) + \text{H.c.}$, where $|t_\text{C}|^2 = \mathcal{T}_\text{C}$.

%%%%%%%%%%%%%%%%%%%%%%%%%
%%%%%%%%%%%%%%%%%%%%%%%%%

For strong dilution, $I_A\ll \nu^2 e^2 V_\text{SA}$,
non-equilibrium anyons emanating from the source SA rarely participate in tunneling at the collider.
Instead, they mainly influence tunneling indirectly, via the time-domain braiding~\cite{RosenowLevkivskyiHalperinPRL16,SimNC16,LeeSimNC22,LeeNature23}---braiding between an 
anyon-hole pair excitation generated at the collider and an injected (diluted) anyon beam.
The relevant building-block correlators are derived in Sec.~S2 of SI:
\begin{equation}
 \left.\begin{array}{c}
 \big\langle \!\psi^\dagger_A (L,t) \psi_A (L,0) \!\big\rangle_\text{neq}      \\[-0.05cm]
 \big\langle \!\psi_A (L,t) \psi^\dagger_A (L,0) \!\big\rangle_\text{neq}     
\end{array}
\!\!\!\right\}
\!\!=\! \! \frac{ e^{-\frac{1 - \cos (2\pi\nu) }{\nu^2 e^2}S_\text{A} |t|  \pm i  \frac{\sin (2\pi\nu) }{\nu e}I_\text{A}t } }{2 (\pi \tau_c)^{1\!-\!\nu} \! \sin^\nu [ \pi T_\text{A}(\tau_c \!+\! it) ]/T_\text{A}^\nu}.
\label{eq:correlation_a}
\end{equation}
Here, ``neq'' indicates that correlations are evaluated in the non-equilibrium state at $x>0$
(setting $\mathcal{T}_\text{C} = 0$).
Compared with zero temperature~\cite{RosenowLevkivskyiHalperinPRL16,SimNC16,LeeSimNC22,LeeNature23}, for which $S_\text{A} = \nu e I_\text{A}$, finite temperature enters Eq.~\eqref{eq:correlation_a} through $T_\text{A}$ and $T_\text{SA}$, both rendering
the ratio $S_\text{A}/I_\text{A}$ temperature-dependent.
Note also that Eq.~\eqref{eq:correlation_a} is only valid when $T_\text{A}$ is smaller than $\nu e V_\text{SA}$; otherwise, thermal fluctuations undermine time-domain braiding. Channel B is characterized by the equilibrium correlation functions (labeled by ``0''; again evaluated at $\mathcal{T}_\text{C}=0$):
\begin{equation}
     \left.\begin{array}{c}
 \big\langle \psi^\dagger_{B} (L,t) \psi_{B} (L,0) \big\rangle_0      \\
 \big\langle \psi_{B} (L, t) \psi_{B}^\dagger (L,0)\big\rangle_0     
\end{array}
\!\!\!\right\}
\!=\!  
\frac{ 
e^{\pm i \nu e V_\text{B} t}}{2(\pi \tau_c)^{1\!-\!\nu}\!\sin^\nu [ \pi T_\text{B} (\tau_c + i t) ]/T_\text{B}^\nu}.
\label{eq:correlation_b}
\end{equation}

\textbf{\emph{Anyon distribution functions---}}The tunneling current between channels A and B reads, to leading order in $\mathcal{T}_\text{C}\ll 1$, as
\begin{align}
I_\text{T}&\! \equiv  \nu e \mathcal{T}_\text{C} \!\!\int \!\!dt \Big[\langle \psi^\dagger_A (L,\! t) \psi_A (L,\! 0) \rangle_\text{neq}  \langle \psi_{B} (L,\! t) \psi_{B}^\dagger (L, \!0) \rangle_0
\notag\\[-0.2cm]
   &\ \ -\! \langle \psi_A (L,0) \psi^\dagger_A (L,0) \rangle_\text{neq} \langle \psi^\dagger_{B} (L,0) \psi_{B} (L,0) \rangle_0\Big].
\label{eq:it_definition}
\end{align}
The structure of Eq.~(\ref{eq:it_definition}) suggests introducing the anyonic particle and hole distributions (cf.~Ref.~\cite{GuPRB22}; see Sec.~S3 of SI for explicit expressions):
\begin{equation}
\left.\begin{array}{c}
     n^\alpha_\text{p} (\epsilon)\\
     n^\alpha_\text{h} (\epsilon)
\end{array}    
\right\}
    \equiv \int  dt \, 
    \left\{
\begin{array}{cc}    
   \!e^{-i\epsilon t}& \!\!\langle \psi_\alpha^\dagger (t) \psi_\alpha(0)\rangle_a,\\
   \!e^{i\epsilon t}&  \!\!\langle \psi_\alpha (t) \psi^\dagger_\alpha (0)\rangle_a,
 \end{array}    
 \right.
\label{eq:distribution_definitions}
\end{equation}
where ${a=\text{neq},0,0}$ for ${\alpha=\text{A},\text{SA},\text{B}}$, respectively.
With these distributions, we define the charge tunneling rate $\mathcal{I}_\text{T} (\epsilon)$ at energy $\epsilon$, such that $I_\text{T}\!=\! \nu e\! \int_{-\infty}^\infty\!d\epsilon\,  \mathcal{I}_\text{T} (\epsilon )$:
\begin{equation}\mathcal{I}_\text{T} (\epsilon) \equiv n^\text{A}_\text{p} (\epsilon) n^\text{B}_\text{h}(\epsilon)  - n^\text{A}_\text{h} (\epsilon) n^\text{B}_\text{p} (\epsilon).
\label{IT-rate}
\end{equation}

Although the distributions used in Eq.~\eqref{IT-rate} are defined upstream of the collider, their definition, Eq.~\eqref{eq:distribution_definitions},
applies to any point between the diluter and the collider. In the limit of small $\mathcal{T}_\text{C}$, this also applies to positions downstream of the collider (SI Sec.~S3). 
Crucially, for anyonic systems, $n_\text{h}^\alpha(\epsilon) \neq 1 - n^\alpha_\text{p}(\epsilon)$, unlike the fermionic case. With the anyonic distribution functions ~\eqref{eq:distribution_definitions}, 
we will be able to describe the charge and heat tunneling transport in general kinetic terms, without resorting to from-scratch field-theoretical derivations for every observable.

\textbf{\emph{Effective chemical potential---}}
The effective potential $V_\text{eff}$ of A is defined as the value of $V_\text{B}$, at which the \textit{net} charge current at the collider vanishes~\cite{f0}.
Following this definition, with Eqs.~\eqref{eq:correlation_a} and \eqref{eq:correlation_b}, we obtain   
\begin{equation}
    V_\text{eff} = I_\text{A} \sin (2\pi\nu)/\nu^2 e^2,
    \label{Veff}
\end{equation}
which holds for arbitrary $T_\text{A}$ and/or $T_\text{SA}$.
Setting $V_\text{B}=V_\text{eff}$, we have $\langle \psi^\dagger_{A} (L,t) \psi_{A} (L,0) \rangle_\text{neq} \langle \psi_{B} (L,t) \psi_{B}^\dagger (L,0) \rangle_0 = \langle \psi^\dagger_{B} (L,t) \psi_{B} (L,0) \rangle_0 \langle \psi_{A} (L,t) \psi^\dagger_{A} (L,0) \rangle_\text{neq}$, 
yielding zero for the integrand in Eq.~\eqref{eq:it_definition}. The energy-resolved tunneling rate $\mathcal{I}_\text{T}$ [Eq.~\eqref{IT-rate}], defined through anyonic distributions \eqref{eq:distribution_definitions}, satisfies ${\mathcal{I}_\text{T}(\epsilon-\nu eV_\text{eff}) = - \mathcal{I}_\text{T}(-\epsilon+\nu eV_\text{eff})}$, indicating a vanishing tunneling current $I_\text{T}= 0$  [see Fig.~\ref{fig:structure}(b) for $T_\text{B} = 0$ and $T_\text{B} = I_\text{A}/e$, and more general cases in SI Sec.~S6].
This means that $V_\text{eff}$  depends on the non-equilibrium state 
only through the diluted current $I_\text{A}$. 

The above features follow from the particle-hole symmetry with respect to the induced Laughlin surface:
\begin{equation}
    n_\text{p}^\text{A} (\epsilon - \nu e V_\text{eff}) = n_\text{h}^\text{A} (-\epsilon + \nu e V_\text{eff}),
\end{equation}
see Fig.~\ref{fig:structure}(c).
This particle-hole symmetry is absent in the fermionic case, where braiding is absent.
Even for anyonic systems, this symmetry is undermined when processes of direct tunneling or collisions of
non-equilibrium anyons supplied by SA are taken into account in the distribution functions
(SI Sec.~S8). Their effect on $n^\text{A}_\text{p,h}$ is negligible for tunneling transport in the limiting case of strongly dilute beams \footnote{See, however, Ref.~\cite{GuX2023Andreev}, where similar subleading processes determine entanglement of anyonic channels}: the corresponding correction to the effective potential is $\delta V_\text{eff}\sim \mathcal{T}_\text{A}V_\text{eff}\ll V_\text{eff}$.

%%%%%%%%%%%%%%%%%%%%%%%%%
\begin{figure}
  \includegraphics[width= 1 \linewidth]{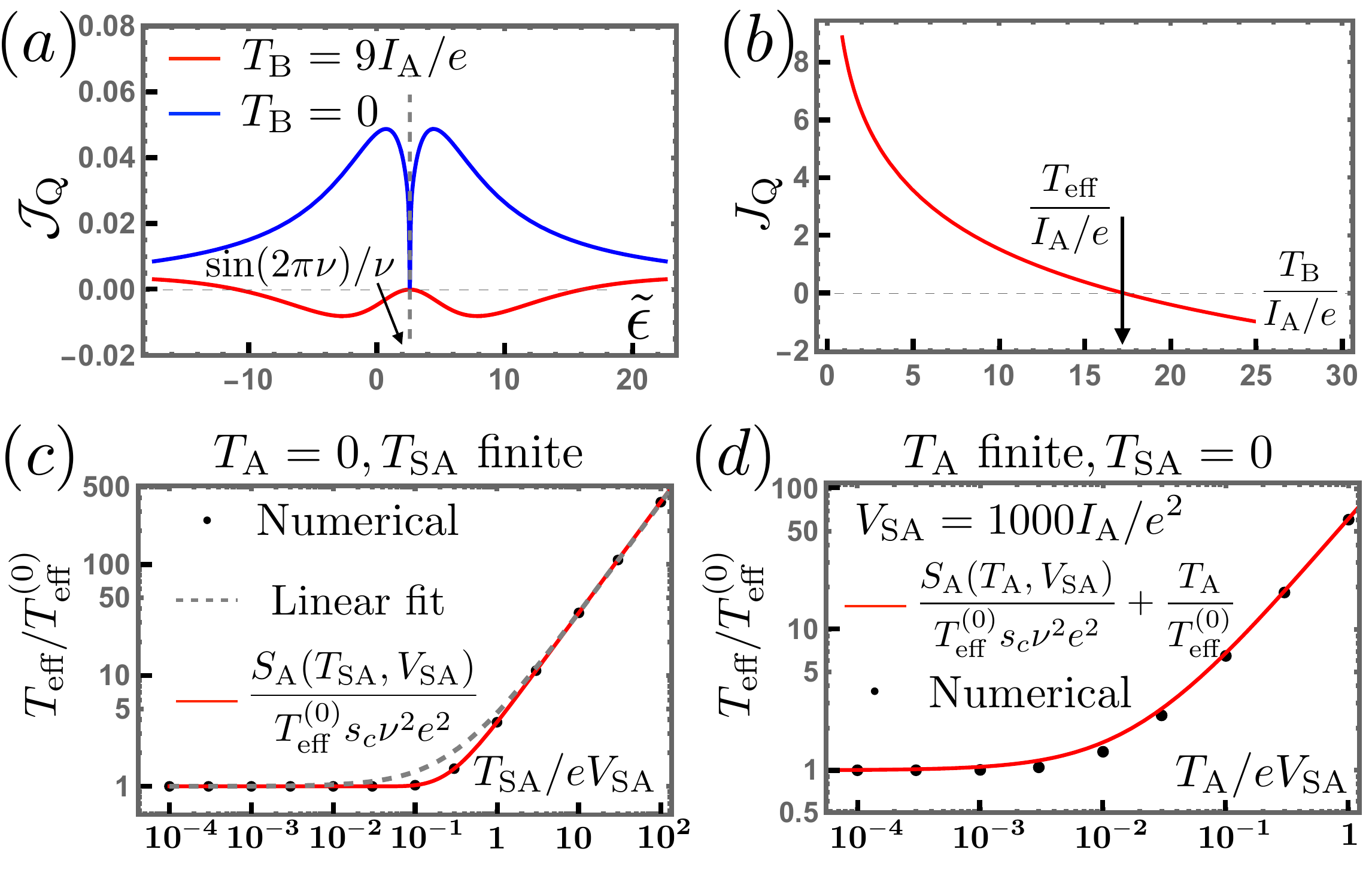}
  \caption{Heat current and effective temperature of channel A, for $\nu=1/3$. (a) The heat-current spectrum $\mathcal{J}_Q(\epsilon)$ [in units of $(\tau_c I_\text{A}/e)^{2\nu-1}/\tau_c$] at energy $\epsilon = (I_\text{A}/e)\tilde{\epsilon}$, when $V_\text{B } = V_\text{eff} = I_\text{A}\sin (2\pi\nu)/\nu^2e^2$, for zero (blue curve) and finite ($T_\text{B}= 9I_\text{A}/e$, red curve) temperatures.
  (b) The total heat current $J_Q = \int d\epsilon \mathcal{J}_Q (\epsilon) $, in units of $(\tau_c I_\text{A}/e)^{2\nu}/\tau_c^2$, as a function of $eT_\text{B}/I_\text{A}$; $J_Q = 0$ when $T_\text{B} = T_\text{eff}$.
 (c) When only $T_\text{SA}$ is finite, the effective temperature $T_\text{eff}$ is proportional to the noise $S_\text{A}$.
(d) When the ambient temperature $T_\text{A}$ is finite, $T_\text{eff}$ is approximately a sum of  $T_\text{A}$ and the noise-induced term.
}
  \label{fig:temperature_features}
\end{figure}
%%%%%%%%%%%%%%%%%%%%%%%%%

\textbf{\emph{Effective temperature---}}To set the
effective temperature $T_\text{eff}$ of channel A, we fix $V_\text{B} = V_\text{eff}$
and calibrate $T_\text{eff}$ against $T_\text{B}$, at which no \textit{net} heat current tunnels between A and B.
The heat current is given by $J_Q = \int_{-\infty}^\infty d\epsilon\ \mathcal{J}_Q (\epsilon)$, where heat refers to the energy counted from the Laughlin surface, $\nu e V_\text{B}$, of channel B:
\begin{equation}
\begin{aligned}
    & \mathcal{J}_Q (\epsilon)\! \equiv\! \left[ n^\text{A}_\text{p} (\epsilon)  n^\text{B}_\text{h} (\epsilon) -  n^\text{A}_\text{h} ( \epsilon)  n^\text{B}_\text{p} (\epsilon)\right] \left(\epsilon - \nu e V_\text{B} \right).
\end{aligned}
\label{eq:heat_current_spectrum}
\end{equation}
We start with Eqs.~(\ref{eq:correlation_a}),(\ref{eq:correlation_b}),(\ref{eq:distribution_definitions}), and first set $T_\text{A} = T_\text{SA} = 0$.
When $T_\text{B} = 0$ [the blue curve of Fig.~\ref{fig:temperature_features}(a)], $\mathcal{J}_Q (\epsilon)$ is positive for all energies, indicating a net heat current from A to B.
With $T_\text{B}$ finite, however, $\mathcal{J}_Q (\epsilon)$ can become negative at low energies [$T_\text{B} = 9 I_\text{A}/e$, the red curve of Fig.~\ref{fig:temperature_features}(a)].
A sign change of $J_Q$ is thus anticipated when $T_\text{B}$ becomes large enough [Fig.~\ref{fig:temperature_features}(b); note that the heat flow vanishes when $T_\text{B} = T_\text{eff} \approx 17.3 I_\text{A} /e$ (SI Sec.~S6E)].

The effective temperature, obtained by requiring $J_Q = 0$, is shown in the lower panels of Fig.~\ref{fig:temperature_features} (for ${\nu=1/3}$), where $T_\text{eff}^{(0)}$ (chosen as the energy unit) is the effective temperature when ${T_\text{A}=T_\text{SA}=0}$.
For ${T_\text{SA}=0}$ [Fig.~\ref{fig:temperature_features}(c)],  $T_\text{eff}$ (doted points) is proportional to $S_\text{A}$ (the red curve), identifying $S_\text{A}$ as the source of $T_\text{eff}$: \begin{equation}T_\text{eff} = S_\text{A}/[s_c(\nu) \nu^2 e^2]\ \ \text{for}\ T_A=T_\text{SA}=0, \label{eq:zeroT-T-eff}
\end{equation}
where ${s_c(\nu)\propto1/[1\!-\!\cos(2\pi \nu)]}$ depends only on the filling fraction $\nu$ of the channels (SI Sec.~S6E).
When $T_\text{A}\neq 0$, the effective temperature in A, downstream of the diluter, has two contributions: the real temperature $T_\text{A}$ and the temperature associated with the noise at the diluter ($S_\text{A}$) through a Johnson-Nyquist relation (SI Sec.~S10).
The effective temperature $T_\text{eff}$ is then approximately the sum of these two temperatures,
as illustrated in Fig.~\ref{fig:temperature_features}(d).
Crucially, for ${\nu= 1}$ (fermions), $s_c$ diverges (SI Sec.~S6E), and the corresponding effective temperature \eqref{eq:zeroT-T-eff} vanishes as $(1-\nu)^2$.
From this perspective, at zero ambient temperature,
$T_\text{eff}$ originates from anyonic statistics: a feature shared by the effective chemical potential \footnote{In fact, $T_\text{eff}$ defined through the vanishing of heat current is finite for fermions, being solely determined by the processes beyond time-domain braiding [neglected in Eq.~(\ref{eq:correlation_a})], which are subleading for strongly diluted anyons}.

\begin{figure}
  \includegraphics[width= 1 \linewidth]{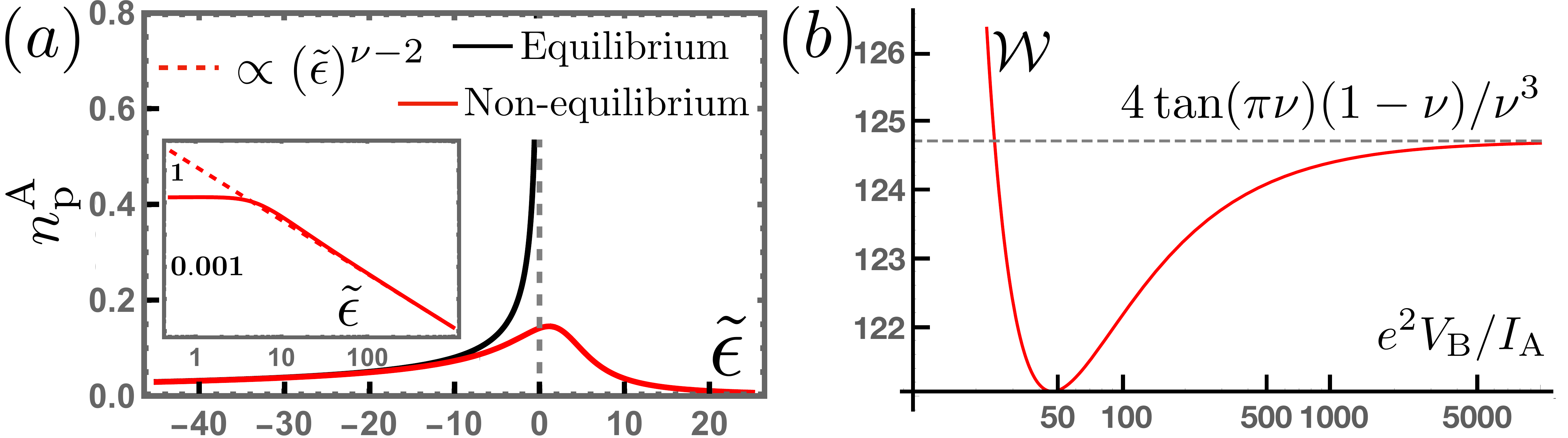}
  \caption{
  Hot anyons for $\nu = 1/3$. (a) Particle distributions [in units of $(\tau_c I_\text{A}/e)^{\nu-1}$, with $\tilde{\epsilon} \equiv \epsilon /(I_\text{A}/e)$], with (red curve) and without (black curve) a non-equilibrium current. Inset: with a non-equilibrium current, particle distributions decay as a power law, $\propto (\tilde{\epsilon})^{\nu - 2}$ for positive energies.
  (b) The value of $\mathcal{W}$ [Eq.~\eqref{eq:f_definition}] as a function of the bias $V_\text{B}$. When $e^2 V_\text{B} \gg I_\text{A}$, $\mathcal{W} \to 4 \tan (\pi\nu) (1-\nu ) /\nu^3$, a universal value that depends only on the filling factor.
  }
  \label{fig:f_factor}
\end{figure}

\textbf{\emph{Hot anyons from time-domain braiding}---}Non-equilibrium noisy anyonic current introduces an effective temperature in an anyonic channel.
One would thus naturally anticipate the presence of ``thermal'' quasiparticle excitations in channel A around the shifted chemical potential.
Indeed, as Fig.~\ref{fig:f_factor}(a) shows, the particle distribution is smeared by $T_\text{eff}$ and has a tail for positive high-energy states (counted from the effective chemical potential $\nu e V_\text{eff}$). This tail however decays slowly in energy $\epsilon$, in a power-law manner: $n_\text{p}^\text{A} \propto (e \epsilon /I_\text{A})^{\nu - 2}$
(see SI Sec.~S9 for the full expression), in contrast to the exponential decay of equilibrium thermal distributions. Most remarkably, the distribution remains finite even for energies larger than $\nu e V_\text{SA}$, the Laughlin surface of source SA.
We refer to these high-energy anyons as hot anyons (similar to hot electrons in semiconductors and superconductors~\cite{SzeJAP1966,AscheBook2006}) generated by time-domain-braiding.
In sharp contrast, for zero-ambient-temperature fermionic systems, particle energies should \emph{never} exceed the Fermi surface of the source.

We define a ``witness function,'' identifying the presence of hot anyons generated by time-domain braiding:
\begin{equation}
\mathcal{W} \equiv  \left(e^2 V_\text{B}/I_\text{A}\right)\,\left(\partial_{V_\text{B}} S_\text{T}\right)^{-1}\, \partial_{V_\text{B}} (e I_\text{T} + S_\text{T}/\nu) .
\label{eq:f_definition}
\end{equation}
Here, $S_\text{T}\equiv\int dt[\langle \hat{I}_\text{T}(t)\hat{I}_\text{T}(0)\rangle -I_\text{T}^2]$ is the tunneling-current noise at the collider, with $\hat{I}_\text{T}$ the tunneling-current operator.
The function $\mathcal{W}$ is accessible by measuring the current auto-correlations in channel B (SI Sec.~S7).
As shown by Fig.~\ref{fig:f_factor}(b), when the bias $V_\text{B}$ in channel B becomes much larger than $I_\text{A}$, $\mathcal{W}$ approaches $4 (1 - \nu) \tan (\pi\nu)/\nu^3$, a ``universal value'' that only depends on the Laughlin filling fraction, with the factor $\tan(\pi \nu)$ stemming from the ``braiding phase'' and $(1-\nu)/\nu^3$ from the fractional charge.

%%%%%%%%%%%%%%%%%%%%%%%%%
%%%%%%%%%%%%%%%%%%%%%%%%%

As stated above, ${\mathcal{W}(\nu=1)=0}$, i.e.,  $\mathcal{W}$ vanishes for fermions, indicating that any finite $\mathcal{W}$ is an anyon-specific feature. In comparison to the ``generalized Fano factor'' of Refs.~\cite{RosenowLevkivskyiHalperinPRL16, LeeSimNC22, LeeNature23}, $\mathcal{W}$ has several advantages. First, when $V_\text{B} > V_\text{SA}$, contributions to $\mathcal{W}$ come entirely from quasiparticles whose energy is above the ``anyonic sea'', representing hot anyons.
Second, to obtain the universal value of $\mathcal{W}$, one requires $e^2 V_\text{B} \gg I_\text{A}$, which is more easily accessible experimentally than the limit of vanishingly small transmission through the diluter.
Focusing on $\mathcal{W}$ at large bias also minimizes the effect of thermal contributions.
Finally, $\mathcal{W}$ is defined in terms of the differential current and noise instead of the total current and noise. With this definition, one overcomes the difficulty of dealing with energy-dependent transmission amplitudes experimentally (see recent discussion in  Refs.~\cite{VeillonNature24, SchillerX2024}).

\textbf{\emph{Role of the type of tunneling quasiparticles}---}We have shown above how the effective chemical potential and effective temperature originate from the non-equilibrium current $I_\text{A}$ and its noise, owing to anyonic braiding. These results are obtained for the setup, in which all three chiral channels are characterized by the same filling factor $\nu$, with the same quasiparticles of charge $\nu e$ tunneling at both the diluter and the collider. However, when different channels have different filling factors and different types of quasiparticles may be involved in inter-channel tunneling, 
the effective chemical potential also depends on the charge of tunneling quasiparticles. This is yet another feature that distinguishes a non-equilibrium anyonic channel from a fermionic one.

%%%%%%%%%%%%%%%%%%%%%%%%%
\begin{figure}
  \includegraphics[width= 0.7 \linewidth]{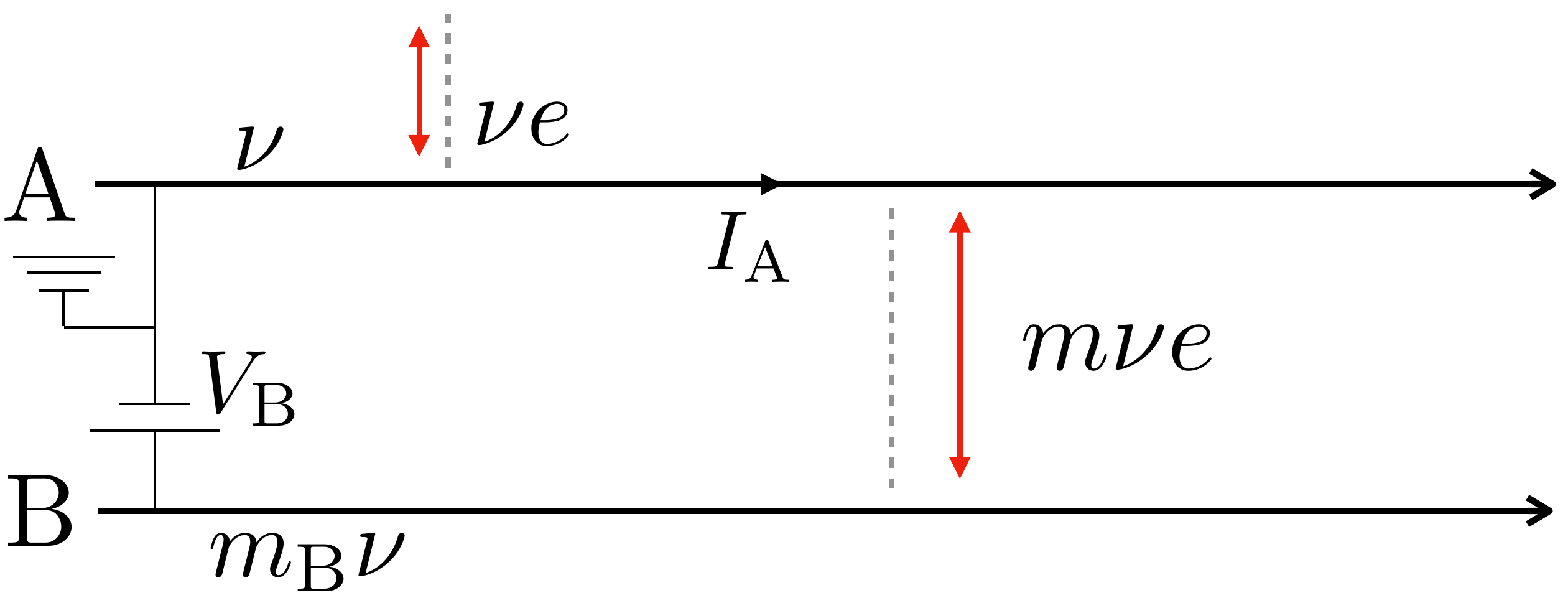}
  \caption{
  An out-of-equilibrium anyonic channel A (with filling fraction $ \nu$ and non-equilibrium current $I_\text{A}$, after receiving charge $\nu e$ quasiparticles from the not-shown source), whose effective chemical potential is calibrated by an equilibrium channel B (with filling fraction $m_\text{B} \nu$ and bias $V_\text{B}$). These two channels communicate via quasiparticles with charge $m\nu e$.
  }
  \label{fig:different_fillings_main}
\end{figure}
%%%%%%%%%%%%%%%%%%%%%%%%%

Let us consider an out-of-equilibrium edge A (filling fraction $\nu$), whose effective chemical potential is calibrated by an equilibrium edge B,
whose filling fraction is $m_\text{B} \nu$, with $m_\text{B}$ an integer (Fig.~\ref{fig:different_fillings_main}).
Channel A carries a diluted current $I_\text{A}$, provided by charge-$\nu e$ quasiparticles from the source SA.
We further assume that these two channels communicate via tunneling of charge-$m\nu e$ quasiparticles, with $m$ an integer; $m_\text{B}$ and $m$ are determined by sample-specific features, see SI Secs.~S11 and S12 for the discussion of specific cases.
Tunneling from A to B (Fig.~\ref{fig:different_fillings_main}) is described by the operator $\propto \exp [ -i( m \sqrt{\nu} \phi_\text{A} - m\sqrt{\nu/m_\text{B}} \phi_\text{B} ) ]$, with $\phi_\text{A}$ and $\phi_\text{B}$ the corresponding bosonic fields. 
The net tunneling current between $\text{A}$ and $\text{B}$ vanishes when $n^\text{A}_\text{p} n^\text{B}_\text{h} = n^\text{B}_\text{p} n^\text{A}_\text{h}$,
which now happens for (SI Sec.~S11)
\begin{equation}
    V_\text{B} = V_\text{eff} = I_\text{A} \sin\left( 2m\pi\nu \right)/(m  \nu^2 e^2).
    \label{eq:veff_different_fillings}
\end{equation}

Equation~\eqref{eq:veff_different_fillings} displays a notable feature: the generalized effective potential induced by time-domain braiding
depends on the statistical phase $m\nu$ of quasiparticles that tunnel at the collider.
This is because the braiding phase of tunneling quasiparticles appears in the exponential factor of the channel-A correlators [cf. Eq.~(\ref{eq:correlation_a})], where $2\pi \nu$ is replaced with $2\pi m \nu$.
In addition, the voltage $V_\text{B}$ (used for calibrating $V_\text{eff}$) appears in the correlation functions [generalizing Eq.~(\ref{eq:correlation_b})] being multiplied with the quasiparticle charge $m\nu e$. In particular, for $\nu=1/3$ and $m=m_B=3$,
the channels A and B communicate by tunneling of charge-$e$ fermions, which results in the absence of time-domain braiding (anyons do not braid with fermions) and, hence, the corresponding shift of the chemical potential vanishes, $V_\text{eff}=0$.
At the same time, $m_\text{B}$ determines only the power in the denominator of the equilibrium correlation function of channel B [cf. Eq.~\eqref{eq:correlation_b}] 
and is thus irrelevant to determining the effective chemical potential.

Remarkably, the above insights indicate that, contrary to common wisdom for fermions,
the effective chemical potential of an anyonic channel only becomes well-defined after specifying the type of quasiparticles with which the anyonic channel is probed. For out-of-equilibrium anyons, the ``transitivity'' of the potentials (for thermodynamic chemical potentials of three systems, if $\mu_1=\mu_2$ and $\mu_2=\mu_3$, then $\mu_1=\mu_3$) only holds when the communication between the systems is established by quasiparticles of the same type. Equivalently, the non-equilibrium distributions \eqref{eq:distribution_definitions} of anyons can only be accessed by tunneling of these same anyons. The above conclusions also apply to setups where channel A has the filling fraction $m_\text{A} \nu$, with $m_\text{A}$ an integer, as well as to the effective temperature in such setups (SI Sec.~S11).

This unique anyonic feature could be relevant to tunneling between out-of-equilibrium anyonic edges with $\nu=1/3$ and $2/3$. In particular, at the Kane-Fisher-Polchinski~\cite{KaneFisherPolchinskiPRL94} fixed point, the $2/3$ channel consists of a neutral node and a charge-$2e/3$ mode, where inter-channel tunneling may rely on either charge-$e/3$ or $2e/3$ quasiparticles,  with some experimental tuning in place~\cite{GoldsteinGefenPRL16}, see details in SI Sec.~S12.

\textbf{\emph{Discussion and Perspective---}}We have generalized the concept of non-equilibrium Fermi sea to anyonic sea, involving non-equilibrium landscapes of Laughlin surface.
The definition and characterization of these new entities establish the basis of new and rich prospects: (i) Generalization to QH states beyond the Laughlin fractions,
e.g., for hole-like QH states, reconstructed edges~\cite{YigalPRL94, WanYangRezayiPRL06, WanHuRezayiYangPRB08, WangMeirGefenPRL13, ZhangWuHutasoitJainPRB14}, and counter-propagating modes;
(ii) The dependence of the above picture on the degree of decoherence and inter-mode equilibration (see, e.g., Refs.~\cite{ProtopopovGefenMirlinAoP17, NosigliaParkRosenowGefenPRB18, SpanslattParkGefenMirlinPRL19, ParkSpanslattGefenMirlinPRL20, AsasiMulliganPRB20, AnindyalPRL21, SpanslattGefenGornyiPolyakovPRB21, DuttaScience22,SrivastavNC22, KumarNC22, MelcerNC22, KhannaGoldsteinGefen22, DasRaoGefenMurthyPRB22, LeBretonPRL22, MannaDasX23,MannaDasGoldsteinX23,YutushuiMrossPRB23, HeinSpanslattPRB23, ParkSpanslattMirlinPRL24, KumarNC24, MannaDasGoldsteinGefenPRL24,MannaX24});
(iii) An extension to the boundaries of non-Abelian QH phases; (iv) Developing an effective linear-response theory on top of the non-equilibrium Laughlin surface, including effective fluctuation-dissipation relations.
Finally, our quantum kinetic approach, based on anyonic distribution functions that reveal fractional statistics through braiding with probe quasiparticles, is potentially relevant to higher-dimensional strongly-correlated systems hosting anyons, such as  fractons~\cite{VijayHaahFuPRB15,NandkishoreHermeleFracton19} or  semions~\cite{LaughlinSemion88,CanrightGirvinBrassPRL89}, in particular, in the context of quantum spin liquids~\cite{Savary2016}.

\textbf{\emph{Acknowledgments---}}We are grateful to Gabriele Campagnano, Domenico Giuliano, Fr\'ed\'eric Pierre, and Bernd Rosenow for discussions.  G.Z. acknowledges the support from National Natural Science
Foundation of China, (Grant No.~12374158), and Innovation Program for Quantum Science and Technology
(Grant No.~2021ZD0302400). I.G. and Y.G. acknowledge the support by the Deutsche Forschungsgemeinschaft (DFG) through grant No. MI\,$658/10$-$2$.  Y.G. acknowledges support from the
DFG Grant RO\,$2247/11$-$1$, the US-Israel Binational Science Foundation, and the Minerva Foundation. Y.G. is the incumbent of the InfoSys chair.

\newpage

\clearpage

\renewcommand{\bibnumfmt}[1]{[S#1]}
\renewcommand{\citenumfont}[1]{S#1}
\global\long\def\theequation{S\arabic{equation}}
\global\long\def\thefigure{S\arabic{figure}}
\setcounter{equation}{0}
\setcounter{figure}{0}

\begin{widetext}

\begin{center}
\textbf{\large Supplementary Information for ``Landscapes of an out-of-equilibrium anyonic sea''
	\vspace{5pt}}\\
\vspace{15pt}
Gu Zhang, Igor Gornyi, and Yuval Gefen\\
(Dated: July 19, 2024)
\end{center}
\vspace{10pt}

In this Supplementary Information, we provide details on: (i) Illustration of time-domain braiding; (ii) Finite-temperature correlation functions of an out-of-equilibrium anyonic channel; (iii) Distribution functions, defined through Fourier transformation of correlation functions, of equilibrium and non-equilibrium anyonic channels; (iv) Interpreting tunneling features in terms of zero-temperature distribution functions; (v) Criteria to define the effective chemical potential of an out-of-equilibrium anyonic channel; 
(vi) Evaluation of the heat current in various cases; (vii) Relation between the tunneling-current noise and the current auto-correlation (the latter is more easily accessible in experiments); (viii) Distribution functions after the inclusion of direct tunneling of non-equilibrium anyons from diluted beams; (ix) Comparison of distribution functions characterized by a real temperature and an effective temperature; (x) Nyquist-Johnson noise in the considered non-equilibrium anyonic system; (xi)  General expression of the effective potential, for Tunneling between channels with different filling fractions; and (xii) Tunneling of different-charge quasiparticles between $\nu = 1/3$ and $2/3$ edges.

\section*{S1. Illustrations of time-domain braiding}

As explained in the main text, the so-called time-domain braiding is a unique process that occurs when ``braiding'' anyonic tunneling operators at the collider with those of the ``passing-by'' anyons from the non-equilibrium (diluted) beam~\cite{SSimNC16,SLeePRL19,SschillerPRL23}.
In this section, we further illustrate time-domain braiding.

More accurately, time-domain braiding can be viewed as the ``interference'' between two tunneling processes.
Here, without loss of generality, we consider the overlap between the ket ($|\Psi_i\rangle$) and the bra ($|\Psi_f\rangle$) states defined in Fig.~\ref{fig:tdb_phase}.
This overlap appears in the expansion for the correlation function of the tunneling operators at the collider to leading-order expansion in diluter's transmission $\mathcal{T}_A$: 
$$
    \langle \Psi_f | \Psi_i\rangle = \langle T_K \psi^\dagger_\text{A} (L,t^-) \psi_\text{B} (L,t^-) \psi^\dagger_\text{B} (L,0^+) \psi_\text{A} (L,0^+) \psi^\dagger_\text{A} (0,s_1^{\eta_1}) \psi_\text{SA} (0,s_1^{\eta_1}) \psi^\dagger_\text{SA} (0,s_2^{\eta_2}) \psi_\text{A} (0,s_2^{\eta_2}) \rangle e^{i\theta (\eta_1,\eta_2)}.
$$
Here, $s_1$ and $s_2$ refer to the time arguments of the anyon operators at the position of the diluter, with $\eta_1$ and $\eta_2$ their corresponding Keldysh indexes.
The phase factor $e^{i\theta (\eta_1,\eta_2)}$ with
$$\theta (\eta_1,\eta_2) = \pi \nu [1 - (\eta_1 - \eta_2)],$$ 
being $\eta_1$ and $\eta_2$-dependent, refers to the difference in phase [see Figs.~\ref{fig:tdb_phase}(c) to \ref{fig:tdb_phase}(f)] between $\langle \Psi_f | \Psi_i\rangle$ and the Keldysh-ordered correlation function.
This phase difference is generated by exchanging the anyonic operators [see Figs.~\ref{fig:tdb_phase}(c) to \ref{fig:tdb_phase}(f)]:
$$
\psi^\dagger_\alpha(x)\psi_\beta(x')= e^{-i\pi\nu}\psi_\beta(x')\psi^\dagger_\alpha(x), \quad \psi^\dagger_\alpha(x)\psi^\dagger_\beta(x')= e^{i\pi\nu} \psi^\dagger_\beta(x')\psi^\dagger_\alpha(x),
$$
where the position $x'$ is downstream of $x$.

%%%%%%%%%%%%%%%%%%%%%%%%%
\begin{figure}
  \includegraphics[width= 0.7\linewidth]{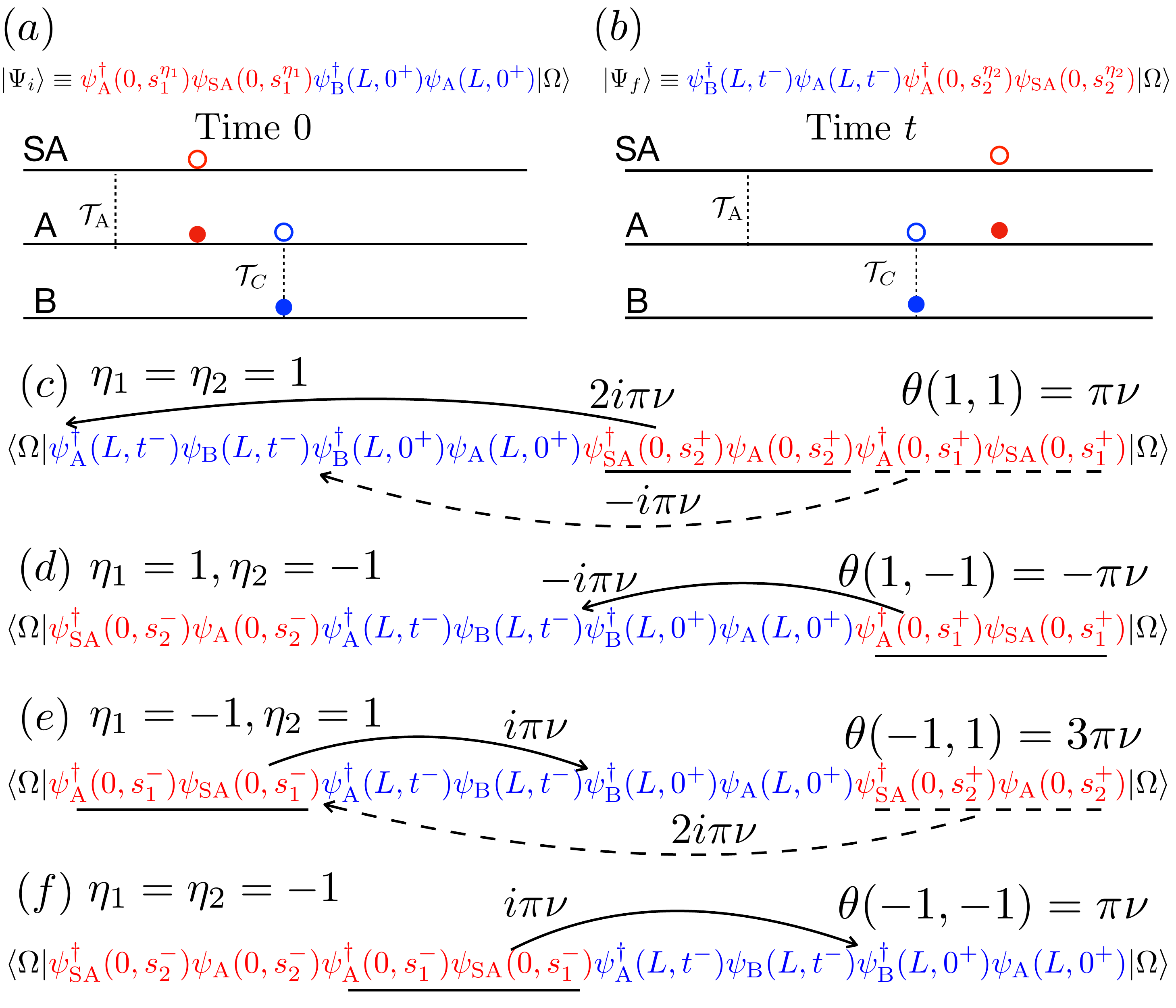}
  \caption{
  \textbf{Illustration of time-domain braiding for the leading-order expansion in the tunneling transmission of the diluter.}
  \textbf{Panels (a) and (b):} Positions of anyonic operators,
  for states $|\Psi_i\rangle$ and $|\Psi_f\rangle$, at times 0 and $t > 0$, respectively.
  Here, red dots refer to operators of the anyons that are generated by the tunneling at the diluter (filled dot: quasiparticle, empty dot: quasihole) and blue dots refer to  the quasiparticle (fillied dot) and quasihole (empty dot) operators involved in the tunneling process at the central QPC. The state $|\Omega\rangle$ is the vacuum state (no quasiparticles in all three channels).
  \textbf{Panels (c) to (f)}: Operators after Keldysh ordering, for different arrangements of the Keldysh indexes $\eta_1$ and $\eta_2$. The exchange of the operators (as indicated by arrows) is needed to reconstruct states $|\Psi_i\rangle$ and $|\Psi_f\rangle$, thus leading to extra phases that are marked out next to corresponding arrows.
  The exchange-induced total phase $\theta(\eta_1,\eta_2)$
  is shown for each case.
  In panels (c) and (e), the exchange operations marked by solid arrows are taken before those indicated by dashed ones.
  }
  \label{fig:tdb_phase}
\end{figure}
%%%%%%%%%%%%%%%%%%%%%%%%%

\color{black}
In Fig.~\ref{fig:tdb_phase}, anyons of the non-equilibrium beam created by tunneling from SA to A at times $s_1$ [panel (a)] and $s_2$ [panel (b)], when moving with constant velocity $v$ downstream of the diluter, would arrive at the collider at times $s_1 + L/v$ and $s_2 + L/v$, satisfying $0< s_1 + L /v, s_2 + L /v <t$, where $L$ is the distance between the diluter and the collider.
As a consequence, in the state $|\Psi_i\rangle$, in which the tunneling event at the collider (generating a quasiparticle-quasihole pair shown in blue) occurs at time zero, the non-equilibrium pair (shown in red) is still upstream of the collider [Fig.~\ref{fig:tdb_phase}(a)].
In the other state, $|\Psi_f\rangle$,  when the tunneling at the collider occurs at a later moment $t$, this pair instead already moves past the collider [Fig.~\ref{fig:tdb_phase}(b)].

To contract operators with time arguments $s_1$ and $s_2$ ($s_1 \to s_2$), these operators need to be moved next to each other, when evaluating $\langle \Psi_f| \Psi_i \rangle$.
As a consequence, one has to exchange the ``non-equilibrium'' operators at time moment $s_2$ with the collider operators at time $t$, leading to the exchanging phase $- i\pi\nu$, half in amplitude of the ``braiding'' phase, $\pm i 2\pi \nu$. 
The other phase contribution comes from the phase difference $\theta(\eta_1,\eta_2)$ 
between $\langle \Psi_f | \Psi_i\rangle$ and the Keldysh-ordered correlation function. For clarity, in Figs.~\ref{fig:tdb_phase}(c) to \ref{fig:tdb_phase}(f), we show operators with different Keldysh indexes, and the resulting values of $\theta$.
In these figures, when the operator in the bra-state is upstream of that in the ket-state, exchanging them leads to a phase factor $-i\pi\nu$, when both operators are creation or annihilation operators. 
The phase is opposite if the operator in the bra-state is downstream of that in the ket-state.
Following these figures, the total phase equals $-i\pi\nu (\eta_1 - \eta_2)$, as reported in, e.g., Ref.~\cite{SMorelPRB22}.

\color{black}

\section*{S2. Finite-temperature correlations in an out-of-equilibrium anyonic channel}

In Eq.~(1) of the main text, we show correlation functions of operators in channel A, for finite temperatures $T_\text{A}$ or $T_\text{SA}$ (see Fig.1 of the main text for the schematics of the setup).
In this section, we explain how to obtain these expressions. 
The correction to the equilibrium correlation function  of the tunneling operators at position $x=L$ in channel A, which arises at the first order in the diluter transmission $\mathcal{T}_\text{A}$, involves averaging in the presence of tunneling operators at the diluter ($x=0$) in channels A and SA:
\begin{align}
\big\langle \psi^\dagger_\text{A} (L,t^-) \psi_\text{A} (L,0^+)\big\rangle_\text{neq}^{(1)} &=  \mathcal{T}_\text{A} \int_{-\infty}^\infty  ds_1 \int_{-\infty}^\infty ds_2 \sum_{\eta_1\eta_2} \eta_1\eta_2 
\, 
e^{-i\nu e V (s_1 - s_2)}
\notag 
\\
&\times \big\langle \psi^\dagger_\text{A} (L,t^-) \psi_\text{A} (L,0^+) \psi^\dagger_\text{A} (0,s_1^{\eta_1}) \psi_\text{A} (0,s_2^{\eta_2}) \big\rangle_0 \langle \psi_\text{SA} (0,s_1^{\eta_1}) \psi^\dagger_\text{SA} (0,s_2^{\eta_2}) \big\rangle_0 .
\label{eq:1st_order_expansion}
\end{align}
Here, $\eta_1=\pm $ and $\eta_2 =\pm $ are the two Keldysh indexes,
the subscript ``neq'' denotes the averaging over the non-equilibrium state, the superscript ``(1)'' indicates the power of $\mathcal{T}_\text{A}$, and the subscript ``0'' indicates that correlations are evaluated at $\mathcal{T}_A = \mathcal{T}_\text{C} = 0$.
We first consider the correlation functions when the temperature of the source reservoir $T_\text{SA}$ is finite but the ambient temperature of channel A is zero, $T_\text{A} = 0$, and then generalize the results to the case when both $T_\text{A}$ and $T_\text{SA} $ are finite.

\textbf{Case I: \ $T_\text{SA}$ finite, and $T_\text{A} = 0$}

We begin by considering the case $T_\text{A} = 0$, when  $T_\text{SA}$ is the only finite temperature in the setup.
In this case, the correlation functions involving operators $\psi^\dagger_\text{A}$ and $\psi_\text{A}$ 
are power-law functions of time, whereas the correlations of $\psi^\dagger_\text{SA}$ and $\psi_\text{SA}$ produce a power of the hyperbolic function. Equation~\eqref{eq:1st_order_expansion} 
takes then the form:
\begin{align}
    \text{Eq.~(S1)}=& \frac{\mathcal{T}_A}{(2\pi \tau_c)^3} \, \frac{\tau_c^\nu}{(\tau_c + it)^\nu} \sum_{\eta_1\eta_2} \eta_1\eta_2 
    \int_{-\infty}^\infty  ds_1 \int_{-\infty}^\infty ds_2 
    \, e^{-i\nu e V (s_1 - s_2)}
    \notag\\
   &\times  \frac{\tau_c^\nu}{ \tau_c + i(s_1 - s_2) \chi_{\eta_1\eta_2} (s_1 - s_2) ]^\nu} 
   \,  \frac{[\tau_c + i(t-s_1 - L) \chi_{-\eta_1} (t - s_1)]^\nu [ \tau_c + (-s_2) \chi_{+\eta_2} (-s_2)]^\nu}{[\tau_c + i(t-s_2 - L) \chi_{-\eta_2} (t - s_2)]^\nu [ \tau_c + (-s_1) \chi_{+\eta_1} (-s_1)]^\nu}
     \notag
  \\
    & \times
   \left(\frac{\pi T_\text{SA} \tau_c}{
   \sin \left\{\pi T_\text{SA} [ \tau_c + i (s_1 - s_2) \chi_{\eta_1\eta_2} (s_1 - s_2) ] \right\}}\right)^\nu,
\label{eq:ta_0}
\end{align}
where $\chi_{\eta \eta'}(s>0) = \eta'$ and $\chi_{\eta \eta'}(s<0) =-\eta$, i.e., $$\chi_{\eta,\eta'}(s)=\eta'\, \Theta(s)-\eta\, \Theta(-s),$$
where $\Theta(s)$ is the Heaviside step function. 

Without loss of generality, we first consider $t > 0$.
The contribution of time-domain braiding to the correlation function is determined by the singularity in the integrand of Eq.~\eqref{eq:ta_0} at (cf. Ref.~\cite{SMorelPRB22})
$$s_1 \to s_2.
$$  
For $0<s_1=s_2<t$, we obtain
\begin{equation}
    \frac{[\tau_c + i(t-s_1 - L) \chi_{-\eta_1} (t - s_1)]^\nu [ \tau_c + (-s_2) \chi_{+\eta_2} (-s_2)]^\nu}{[\tau_c + i(t-s_2 - L) \chi_{-\eta_2} (t - s_2)]^\nu [ \tau_c + (-s_1) \chi_{+\eta_1} (-s_1)]^\nu}\ \to\ \exp[i\pi\nu (\eta_1 - \eta_2)].
    \label{eq:fraction}
\end{equation}
When $s_1=s_2$ is outside the segment $[0,t]$, the fraction in Eq.~\eqref{eq:fraction} becomes unity,
i.e., it does not depend on $\eta_1$ and $\eta_2$. 
Introducing $s=s_1-s_2$ in Eq.~\eqref{eq:ta_0}, it is instructive to re-write the time-domain-braiding contribution to the correlation function by performing explicitly the summation over the Keldysh indexes:
\begin{align}
  \text{Eq.~(S1)}_\text{\,TDB}=  
  &\frac{\mathcal{T}_A \, t}{(2\pi)^3} \frac{\tau_c^{3\nu-3}}{(\tau_c + it)^\nu} \sum_{\eta_1\eta_2} \eta_1\eta_2\,
  \, e^{i\pi\nu (\eta_1 - \eta_2)} \int_{-\infty}^\infty  \!\!ds \frac{e^{-i\nu e V s}}{[\tau_c + i s \chi_{\eta_1\eta_2} (s)]^\nu}\, \frac{(\pi T_\text{SA})^\nu}{\sin^\nu \left\{\pi T_\text{SA} [ \tau_c + i s \chi_{\eta_1\eta_2} (s) ] \right\}^\nu}
  \notag
  \\
    =& \frac{\mathcal{T}_A \, t}{(2\pi)^3} \frac{\tau_c^{3\nu-3}}{(\tau_c + it)^\nu} 
    \left\{\! \int_0^\infty \!\!ds \frac{e^{-i\nu V s}}{(\tau_c + i s )^\nu} \frac{(\pi T_\text{SA})^\nu}{\sin^\nu \left[\pi T_\text{SA} ( \tau_c + i s ) \right]^\nu} 
    +\!  \int_{-\infty}^0 \!\!ds \frac{e^{-i\nu e V s}}{(\tau_c - i s )^\nu} \frac{(\pi T_\text{SA})^\nu}{\sin^\nu \left[\pi T_\text{SA} ( \tau_c - i s ) \right]^\nu}\right.
    \notag 
    \\
+ &  \int_0^\infty ds \frac{e^{-i\nu e V s}}{(\tau_c - i s )^\nu} \frac{(\pi T_\text{SA})^\nu}{\sin^\nu \left[\pi T_\text{SA} ( \tau_c - i s ) \right]^\nu} 
+  \int_{-\infty}^0 ds \frac{e^{-i\nu e V s}}{(\tau_c + i s )^\nu} \frac{(\pi T_\text{SA})^\nu}{\sin^\nu \left[\pi T_\text{SA} ( \tau_c + i s ) \right]^\nu}
\notag 
\\
- & \left.  \int_{-\infty}^\infty ds \frac{e^{-i\nu e V s}}{(\tau_c + i s )^\nu} \frac{(\pi T_\text{SA})^\nu}{\sin^\nu \left[\pi T_\text{SA} ( \tau_c + i s ) \right]^\nu} e^{-2i\pi\nu}  
-  \int_{-\infty}^\infty ds \frac{e^{-i\nu e V s}}{(\tau_c - i s )^\nu} \frac{(\pi T_\text{SA})^\nu}{\sin^\nu \left[\pi T_\text{SA} ( \tau_c - i s ) \right]^\nu} e^{2i\pi\nu} \right\},
\label{eq-eta1-eta2}
\end{align}
Here, we set $s_1=s_2$ in the fraction (\ref{eq:fraction}) and took into account the configurations that produce a nontrivial phase factor
$e^{i\pi\nu (\eta_1 - \eta_2)}$ (otherwise the summation over $\eta_{1,2}$ yields zero). 
The integration over the center-of-mass time $(s_1+s_2)/2$ produces then the total time $t$, while the integration over the time difference $s$ can be safely extended to infinite limits, since the integral over $s$ is dominated by $s\lesssim \text{min}[1/(\nu e V),1/T_\text{SA}]\ll t$. Equation \eqref{eq-eta1-eta2}
can be written in the form emphasizing the two tunneling processes---tunneling from A to SA ($I_{\text{A} \to \text{SA}}$) and tunneling from SA to A ($I_{\text{SA} \to \text{A}}$):
\begin{align}
  \text{Eq.~(S1)}_\text{\,TDB}= &\frac{\mathcal{T}_A\, t}{(2\pi)^3} \frac{\tau_c^{3\nu-3}}{(\tau_c + it)^\nu}  \,\int_{-\infty}^\infty ds \frac{e^{-i\nu e V s}}{(\tau_c + i s )^\nu} \frac{(\pi T_\text{SA})^\nu}{\sin^\nu \left[\pi T_\text{SA} ( \tau_c + i s ) \right]^\nu} \left( 1-e^{-2i\pi\nu} \right) 
\notag
\\
+ & \frac{\mathcal{T}_A\, t}{(2\pi)^3} \frac{\tau_c^{3\nu-3}}{(\tau_c + it)^\nu} \,\int_{-\infty}^\infty ds \frac{e^{-i\nu e V s}}{(\tau_c - i s )^\nu} \frac{(\pi T_\text{SA})^\nu}{\sin^\nu \left[\pi T_\text{SA} ( \tau_c - i s ) \right]^\nu} \left( 1-e^{2i\pi\nu} \right) .
\label{eq:ta_0_integral}
\end{align}
Written in terms of the partial currents, Eq.~\eqref{eq:ta_0_integral} becomes
\begin{equation}
\begin{aligned}
   \text{Eq.~(S1)}_\text{\,TDB} &= \frac{1}{2\pi \tau_c} \frac{\tau_c^\nu}{(\tau_c + it)^\nu} \left[ -\frac{I_{\text{SA} \to \text{A}}}{\nu e} \left( 1-e^{2i\pi\nu} \right)t -\frac{I_{\text{A} \to \text{SA}}}{\nu e} \left( 1-e^{-2i\pi\nu} \right)t  \right]\\
   & = \frac{1}{2\pi \tau_c} \frac{\tau_c^\nu}{(\tau_c + it)^\nu} \left[ -\frac{S_\text{A}}{\nu^2 e^2} [1 - \cos (2\pi\nu) ]\, t + i \frac{I_\text{A}}{\nu e} \sin (2\pi\nu)\, t \right],
\end{aligned}
\label{eq:ta_0_result}
\end{equation}
where $S_A = \nu e ( I_{\text{SA} \to \text{A}} + I_{\text{A} \to \text{SA}} )$ is the noise in the non-equilibrium channel A, after the diluter.
It is defined as $S_\text{A} \equiv \int dt [\langle \hat{I}_\text{A}(t) \hat{I}_\text{A}(0) \rangle - I_\text{A}^2]$, with $\hat{I}_\text{A}$ the current operator, $\langle \hat{I}_\text{A} \rangle = I_\text{A}$. At zero temperature, $T_\text{SA}=0$, Eq.~\eqref{eq:ta_0_result} becomes the leading-order correlation functions of Refs.~\cite{SLeeSimNC22, SMorelPRB22, SschillerPRL23}, after representing the tunneling noise as $S_\text{A} = \nu e I_\text{A}$.
For negative times, $t < 0$, one replaces $-S_A [1 - \cos(2\pi\nu)]\, t/ \nu^2 e^2$ in Eq.~\eqref{eq:ta_0_result} with $ S_A [1 - \cos(2\pi\nu)]\, t/ \nu^2 e^2$.

Now we go to higher-order expansions of $\mathcal{T}_\text{A}$.
For the zero-temperature situation, one can sum up contributions to all orders of dilutions, if $\nu e V_\text{SA} > I_\text{A}/\nu e$.
For a finite temperature, the requirement of the resummation becomes modified, into $\nu e V_\text{SA} > S_\text{A}/\nu^2 e^2$.
With this requirement satisfied, we perform resummation over all orders of dilutions.
More specifically, assuming self-contraction (meaning independence of anyonic pairs) of $n$ pairs of non-equilibrium anyons, the corresponding correlation becomes
\begin{equation}
\begin{aligned}
    & \frac{1}{2\pi \tau_c} \frac{\tau_c^\nu}{(\tau_c + it)^\nu} \frac{C_{2n}^2 C_{2n-2}^2 \cdot \cdot \cdot C_2^2 2^n }{2n!} \frac{1}{n!} \left[ -\frac{S_\text{A}}{\nu^2 e^2} [1 - \cos (2\pi\nu) ] t + i \frac{I_\text{A}}{\nu e} \sin (2\pi\nu) t \right]^n\\
    &=  \frac{1}{2\pi \tau_c} \frac{\tau_c^\nu}{(\tau_c + it)^\nu} \frac{1}{n!} \left[ -\frac{S_\text{A}}{\nu^2 e^2} [1 - \cos (2\pi\nu) ] t + i \frac{I_\text{A}}{\nu e} \sin (2\pi\nu) t \right]^n,
\end{aligned}
\label{eq:nth_expansion}
\end{equation}
where $C_m^n = m!/[n! (m-n)!]$: the number of options to pick out $n$ items out of in-total $m$ ones. In Eq.~\eqref{eq:nth_expansion}, the factor $C_{2n}^2 C_{2n-2}^2 \cdot \cdot \cdot C_2^2 2^n$ refers to the number of self-contraction options of non-equilibrium anyons, and the factor $1/2n!$ is the prefactor of performing $2n$th order Keldysh expansion, and finally, the factor $1/n!$ removes the number of options on the picking order of $n$ anyonic pairs.
Resummation over $n$ then leads to
\begin{equation}
\langle \psi_A^\dagger (t) \psi_A (0)\rangle_\text{neq} \Big|_{T_\text{A}=0,T_\text{SA} } = \frac{1}{2\pi \tau_c} \frac{\tau_c^{\nu} }{(\tau_c + i t)^\nu   } \exp \left\{ -\frac{S_\text{A}}{\nu^2 e^2} [1 - \cos (2\pi\nu) ] | t|  + i \frac{I_\text{A}}{\nu e} \sin (2\pi\nu) t \right\}.
\end{equation}

\textbf{Case II: \ $T_\text{SA}$ and $T_\text{A} $ both finite}

In this case, Eq.~\eqref{eq:ta_0} is modified by using the hyperbolic functions of time for all the two-point correlators involved:  
\begin{equation}
\begin{aligned}
    \text{Eq.~(S1)}=& \frac{\mathcal{T}_A}{(2\pi \tau_c)^3} \frac{(\pi T_\text{SA} \tau_c^3)^\nu (\pi T_\text{A} \tau_c)^{2\nu}}{\sin^\nu[\pi T_\text{A}(\tau_c + it)]} \iint ds_1 ds_2\sum_{\eta_1\eta_2} \frac{\eta_1\eta_2 e^{-i\nu e V (s_1 - s_2)}}{ \sin^\nu \left\{\pi T_\text{sA} [ \tau_c + i (s_1 - s_2) \chi_{\eta_1\eta_2} (s_1 - s_2) ] \right\} }\\
    & \times \frac{1}{\sin^\nu \left\{\pi T_\text{A} [ \tau_c + i (s_1 - s_2) \chi_{\eta_1\eta_2} (s_1 - s_2) ] \right\}}\\
    & \times \frac{\sin^\nu \left\{ \pi T_\text{A} [\tau_c + i(t-s_1 - L) \chi_{-\eta_1} (t - s_1)]\right\} \sin^\nu \left\{\pi T_\text{A} [ \tau_c + (-s_2) \chi_{+\eta_2} (-s_2)]\right\}}{\sin^\nu \left\{ \pi T_\text{A} [\tau_c + i(t-s_2 - L) \chi_{-\eta_2} (t - s_2)]\right\}\sin^\nu \left\{\pi T_\text{A} [ \tau_c + (-s_1) \chi_{+\eta_1} (-s_1)]\right\}}.
\end{aligned}
\label{eq:ta_finite}
\end{equation}
The further derivation is analogous to that for $T_A=0$.When considering the time-domain-braiding process, we again take $s_1\to s_2$, with which the last line of Eq.~\eqref{eq:ta_finite} produces the exchange phase, leading to (assuming $t > 0$ first)
\begin{equation}
    \begin{aligned}
         \text{Eq.~(S1)}_\text{ TDB}
         &=  \frac{\mathcal{T}_A}{(2\pi \tau_c)^3} \frac{(\pi T_\text{SA} \tau_c^3)^\nu (\pi T_\text{A} \tau_c)^{2\nu}}{\sin^\nu[\pi T_\text{A}(\tau_c + it)]} t 
         \\
         &\times \left\{ \int_{-\infty}^\infty ds \frac{ e^{-i\nu e V s}}{ \sin^\nu \left[\pi T_\text{SA} ( \tau_c + i s  ) \right] } \frac{1}{\sin^\nu \left[\pi T_\text{A} ( \tau_c + i s ) \right]} \left( 1 - e^{-2i\pi\nu} \right) \right.\\
         &\left. +\int_{-\infty}^\infty ds \frac{ e^{i\nu e V s}}{ \sin^\nu \left[\pi T_\text{SA} ( \tau_c + i s  ) \right] } \frac{1}{\sin^\nu \left[\pi T_\text{A} ( \tau_c + i s ) \right] } \left( 1 - e^{2i\pi\nu} \right) \right\}
    \end{aligned}
    \label{eq:ta_finite_integral}
\end{equation}

Again, the expression above can be simplified, by noticing that to the leading order in $\mathcal{T}_A$, the partial currents read as
\begin{equation}
    \begin{aligned}
        I_{\text{SA}\to \text{A}} & = \nu e \frac{\mathcal{T}_A}{(2\pi \tau_c)^2} (\pi^2 T_\text{A} T_\text{SA} \tau_c^2)^\nu \int_{-\infty}^\infty ds \frac{ e^{i\nu e V s}}{ \sin^\nu \left[\pi T_\text{sA} ( \tau_c + i s  ) \right] } \frac{1}{\sin^\nu \left[\pi T_\text{A} ( \tau_c + i s ) \right] },\\
        I_{\text{A}\to \text{SA}} & = \nu e \frac{\mathcal{T}_A}{(2\pi \tau_c)^2} (\pi^2 T_\text{A} T_\text{SA} \tau_c^2)^\nu \int_{-\infty}^\infty ds \frac{ e^{-i\nu e V s}}{ \sin^\nu \left[\pi T_\text{sA} ( \tau_c + i s  ) \right] } \frac{1}{\sin^\nu \left[\pi T_\text{A} ( \tau_c + i s ) \right] },
    \end{aligned}
    \label{eq:sa_ia_identities}
\end{equation}
for tunneling from SA to A ($I_{\text{SA}\to \text{A}}$), and for the opposite direction of tunneling, A to SA ($I_{\text{A}\to \text{SA}}$).
With Eq.~\eqref{eq:sa_ia_identities}, Eq.~\eqref{eq:ta_finite_integral} becomes
\begin{equation}
\begin{aligned}
   \text{Eq.~(S1)}_\text{\, TDB} &= \frac{1}{2\pi \tau_c} \frac{(\pi T_\text{A} \tau_c)^\nu}{\sin^\nu [\pi T_\text{A}(\tau_c + it)]} \left[ -\frac{I_{\text{SA} \to \text{A}}}{\nu e} \left( 1-e^{2i\pi\nu} \right)t -\frac{I_{\text{A} \to \text{SA}}}{\nu e} \left( 1-e^{-2i\pi\nu} \right)t  \right]\\
   & = \frac{1}{2\pi \tau_c} \frac{(\pi T_\text{A} \tau_c)^\nu}{\sin^\nu [\pi T_\text{A}(\tau_c + it)]} \left[ -\frac{S_\text{A}}{\nu^2 e^2} [1 - \cos (2\pi\nu) ] t + i \frac{I_\text{A}}{\nu e} \sin (2\pi\nu) t \right],
\end{aligned}
\label{eq:ta_finite_result}
\end{equation}
which, in comparison to Eq.~\eqref{eq:ta_0_result}, contains a $T_\text{A}$ dependent prefactor. In addition, the value of $S_\text{A}$ is also 
implicitly modified by $T_\text{A}$.
For negative times $t<0$, one again replaces $-S_A [1 - \cos(2\pi\nu)] t/\nu^2 e^2$ by $S_A [1 - \cos(2\pi\nu)] t/\nu^2 e^2$.

Now we go to higher orders in the diluter transmission. In comparison to that of $T_\text{A} = 0$, where only $\nu e V_\text{SA} > S_\text{A}/\nu^2 e^2$ is required, here we further require $\nu e V_\text{SA} > T_\text{A}$. Indeed, if $T_\text{A}$ becomes large enough, the time $t$ between two tunneling events at the collider is bounded by the temperature inverse $1/T_\text{A}$, thus presenting resummation over higher-order contributions.
With both requirements satisfied, we obtain (following the same steps as for the case $T_A=0$):
\begin{equation}
\begin{aligned}
&\langle \psi_A^\dagger (t) \psi_A (0)\rangle_\text{neq} \Big|_{T_\text{A},T_\text{SA} } = \frac{1}{2\pi \tau_c} \frac{(\pi T_\text{A} \tau_c)^\nu}{\sin^\nu [\pi T_\text{A}(\tau_c + it)]} \exp \left\{ -\frac{S_\text{A}}{\nu^2 e^2} [1 - \cos (2\pi\nu) ] | t|  + i \frac{I_\text{A}}{\nu e} \sin (2\pi\nu) t \right\},\\
&\langle \psi_A (t) \psi^\dagger_A (0)\rangle_\text{neq} \Big|_{T_\text{A},T_\text{SA} } = \frac{1}{2\pi \tau_c} \frac{(\pi T_\text{A} \tau_c)^\nu}{\sin^\nu [\pi T_\text{A}(\tau_c + it)]} \exp \left\{ -\frac{S_\text{A}}{\nu^2 e^2} [1 - \cos (2\pi\nu) ] | t| - i \frac{I_\text{A}}{\nu e} \sin (2\pi\nu) t \right\},
\end{aligned}
\label{eq:psi_a_general_correlation}
\end{equation}
which is Eq.~(1) of the main text. 

An inspection of Eq.~\eqref{eq:psi_a_general_correlation} discloses an important message: the particle correlation $\langle \psi_A^\dagger (t) \psi_A (0)\rangle_{T_\text{A},T_\text{SA} }$, and the hole correlation $\langle \psi_A (t) \psi^\dagger_A (0)\rangle_{T_\text{A},T_\text{SA} }$ only differ by the imaginary phase in the exponential factor, even at finite temperatures.
This feature, remarkably, leads to the symmetry between the particle and hole distributions, with respect to the energy $I_\text{A} \sin(2\pi\nu)/\nu e$, see more details in Sec.~S3 below. 
Noteworthily, this symmetry between the particle and hole distributions is present if taking into account only time-domain-braiding processes.
Indeed, this symmetry is sabotaged, if allowing non-equilibrium anyons to participate in the tunneling correlation functions directly (i.e., when an anyon appears at the position of the collider, at moment 0 or/and $t$).
Nevertheless, non-equilbirium anyons have little direct influence on correlation functions in the strongly diluted limit---the limit addressed in the current work.
We conclude that the above symmetry between particle and hole distributions is an anyonic-specific feature, because in fermionic systems time-domain braiding is absent.

\section*{S3. Distribution functions, for equilibrium and non-equilibrium channels}

In the main text, we describe non-equilibrium anyonic channels by defining particle and hole distribution functions
\begin{equation}
\begin{aligned}
    & n_{\text{p},\alpha} (\epsilon) \equiv \int dt\, e^{-i\epsilon t} \langle \psi_\alpha^\dagger (t) \psi_\alpha(0)\rangle_a,\\
    & n_{\text{h},\alpha} (\epsilon) \equiv \int dt\, e^{i\epsilon t} \langle \psi_\alpha (t) \psi^\dagger_\alpha (0)\rangle_a,
    \label{eq:distributions_si}
\end{aligned}
\end{equation}
for three channels $\alpha =\, $SA, A and B with $a=0$, neq, and 0, respectively.
In this section, we 
present and compare distribution functions for different situations. 
When considering tunneling between, e.g., channels A and B, one encounters an integral of the form
\begin{equation}
\begin{aligned}
\mathcal{T}_\text{C} \int dt \langle \psi_\text{A}^\dagger (t) \psi^{}_\text{B} (t) \psi_\text{B}^\dagger (0) \psi^{}_\text{A} (0)\rangle_\text{neq} & = \mathcal{T}_\text{C} \int dt \langle \psi^\dagger_A ( t) \psi_A ( 0) \rangle_\text{neq}  \langle \psi_{B} ( t) \psi_{B}^\dagger (0) \rangle_0  \\
&=  \frac{\mathcal{T}_\text{C}}{(2\pi)^2} \int dt \iint d\epsilon_1 d\epsilon_2\, e^{-i(\epsilon_1 - \epsilon_2) t} n_{\text{p,A}} (\epsilon_1)  n_{\text{h,B}} (\epsilon_2 )\\
&=  \frac{\mathcal{T}_\text{C}}{2\pi} \int d\epsilon \, n_{\text{p,A}}(\epsilon)\, n_{\text{h,B}} (\epsilon ),
\end{aligned}
\label{eq:current_integral_form}
\end{equation}
where, unlike Eq.~(3) of the main text, for generality, we do not show explicitly the position at which tunneling occurs.
For a given value of energy $\epsilon$, Eq.~\eqref{eq:current_integral_form} can be interpreted as a product of two distribution functions defined in Eq.~\eqref{eq:distributions_si}: the anyonic particle distribution in A and the anyonic hole distribution in B.

We emphasize that the distributions defined by Eq.~\eqref{eq:current_integral_form} apply to all points between the diluter and the collider, for $\alpha =\, $A, B alike, not only to the location of the collider, at which the tunneling current is evaluated. At first glance, this seems counter-intuitive. Indeed, the non-equilibrium distribution functions appearing in Eq.~\eqref{eq:current_integral_form} are strongly affected by the phenomenon of time-domain braiding, which is generated by the presence of the tunneling bridge (the collider) connecting channels A and B at a single point in space. How can we get the same distribution function at the position upstream of the collider when the time-domain braiding is induced by the collider?
To understand this for $\alpha=\,$A, we interpret Eq.~\eqref{eq:current_integral_form} as the correlation between removing a particle at time 0 and position $x$ (that is, between the diluter and the collider), and adding another one at time $t$ at the same position, $x$. In order to probe the single-anyon distribution function $n_{\text{p,A}}(\epsilon)$ at position $x$, one should introduce a ``measurement device'' coupled to the single-anyon operator in channel $A$. This probe
can be introduced by adding an auxiliary channel, with the same filling fraction as that in A, which is tunnel-coupled to channel A at position $x$. 
With this auxiliary channel, time-domain braiding will take place between nonequilibrium anyons in channel A and the spontaneously generated particle-hole pair induced by the auxiliary channel at the point $x$.
Any measurement of the single-anyon distribution function inevitably introduces time-domain braiding at the measurement position. As a result, the non-equilibrium distribution function at the position of the probe will be affected by the time-domain braiding in exactly the same way as it is affected at the collider and, hence, will have the same form. This is why correlation functions of A channel operators calculated at the collider, where time-domain braiding occurs in the experimental setup, also apply to positions upstream of the collider.
Any probe of the distribution function with an auxiliary channel could be considered analogous to the collider that couples channel A with equilibrium channel B.

Importantly, the effect of time-domain braiding on the distribution function does not depend, to the lowest order, on the transparency of the collider probing the distribution function [cf. Eq.~\eqref{eq:current_integral_form}, where $\mathcal{T}_\text{C}\ll 1$ serves as a prefactor in the expression for the tunneling current and does not enter the distribution functions].
Thus, in the limit of weak tunneling, the probe will measure the same distribution function at any position downstream of the diluter, also including the locations downstream of the collider. The strength of the coupling to the probe will not affect the measured distribution function in the weak-tunneling limit, whereas time-domain braiding will be fully introduced by the probe. 
It is worth reiterating that the probe should be coupled to single-anyon operators. If, for example, we couple a $\nu=1/3$ channel A to a probe based on an integer edge ($\nu=1$), such that the tunneling between the channel and the probe is only possible for a triplet of anyons forming an electron, we would not able to measure the single-anyon distribution function. In this case, time-domain braiding is not effective (see discussion in Sec. S11 below) as electrons do not braid with anyons.
This implies that the coherent superposition of three charge $e/3$ anyons cannot be described in terms of single-anyon distribution functions: quantum correlations in such a triplet are crucially important.

%%%%%%%%%%%%%%%%%%%%%%%%%
\begin{figure}
  \includegraphics[width= 1 \linewidth]{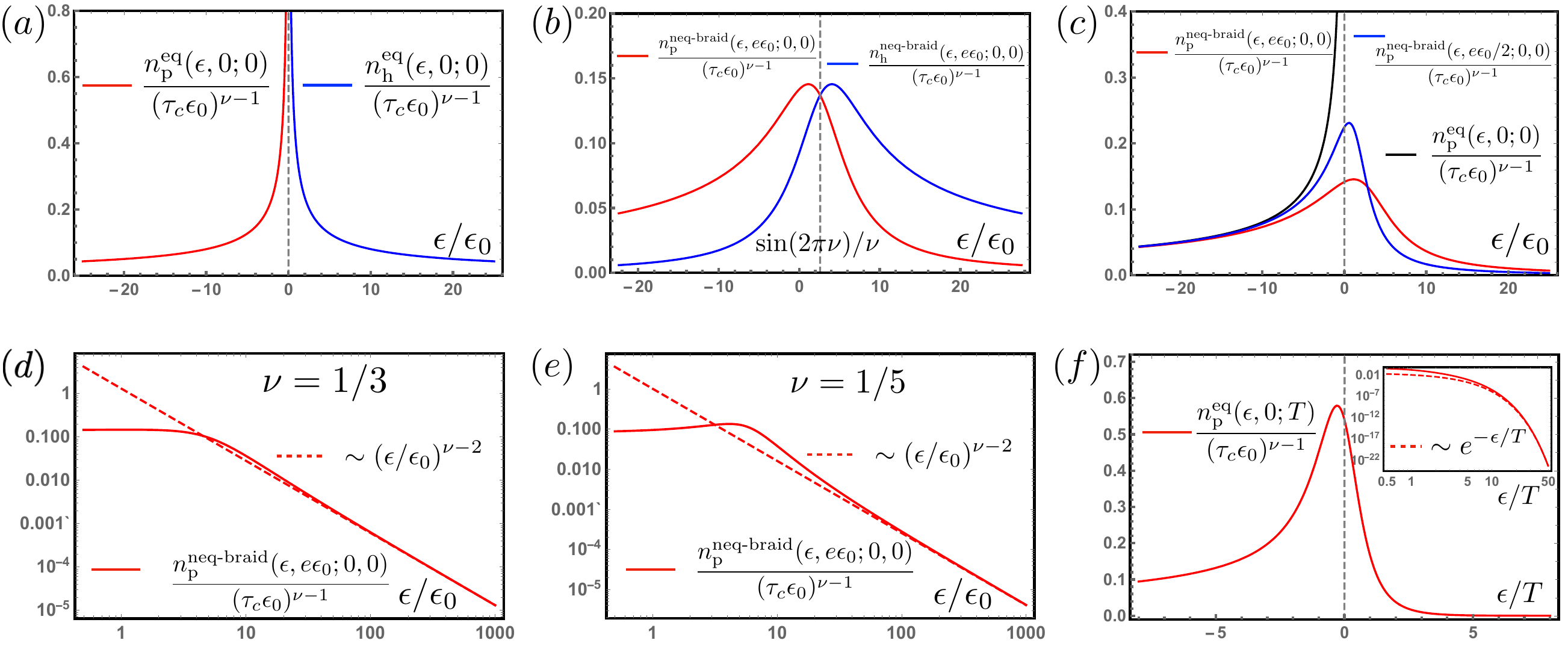}
  \caption{
  \textbf{Distribution functions for an anyonic channel.} An energy unit $\epsilon_0>0$ is used as a reference energy to make all quantities dimensionless, both
  for equilibrium and non-equilibrium distributions.
  Functions $n_\text{p,A}^\text{neq-braid}$ and $n_\text{h,A}^\text{neq-braid}$ refer, respectively, to the particle and hole distributions of non-equilibrium channel A. Their superscript ``neq-braid'' indicates that only time-domain-braiding processes are involved when evaluating the influence of non-equilibrium anyons on correlation functions.
  Functions $n_\text{p,B}^\text{eq}$ and $n_\text{h,B}^\text{eq}$ refer to distributions of equilibrium channel B.
  In panels (a) - (e), all distributions are obtained at zero temperatures, $T_\text{A} = T_\text{SA} = T_\text{B}=0$. The thermal effects will be discussed later, in Sec.~S6.
  \textbf{Panel (a)}: The particle and hole distributions for an equilibrium anyonic channel are only finite for states with energies below and above the Laughlin surface, respectively. \textbf{Panel (b)}: With a finite non-equilibrium current $I_A=e \epsilon_0$, both distributions are nonzero for all energies. The two distributions are related to each other by mirror symmetry with respect to the energy $\epsilon 
  = \epsilon_0 \sin (2\pi\nu)/\nu e$ (the gray dashed line). 
  \textbf{Panel (c)}: The non-equilibrium particle distributions (red and blue lines) approach the equilibrium one (black line) when the non-equilibrium current ($e\epsilon_0$ for the red curve and $e\epsilon_0/2$ for the blue curve) decreases. \textbf{Panels (d) and (e)}: Particle distributions for a non-equilibrium channel with $\nu = 1/3$ and $\nu =1/5$, respectively.
  In both cases, particle distributions decrease as power law $\sim (\epsilon/\epsilon_0)^{\nu - 2}$, for large energies $\epsilon\gg \epsilon_0$. \textbf{Panel (f)}: Particle distribution for an equilibrium anyonic channel with temperature $T$. Inset: the log-log plot of the particle distribution. The distribution decays exponentially $\sim \exp (-\epsilon/T)$, at large energies.
  }
  \label{fig:distributions}
\end{figure}
%%%%%%%%%%%%%%%%%%%%%%%%%

One can use Eqs.~\eqref{eq:distributions_si} and \eqref{eq:current_integral_form} to describe features of tunneling between different anyonic channels in the way conventionally employed for transport of fermions~\cite{SBlanterButtikerPhysRep00}.
At zero temperature $T = 0$, the equilibrium anyonic distribution functions of channel B or SA are given by
\begin{equation}
\begin{aligned}
n_\text{p}^\text{eq} (\epsilon, V; T = 0) & = \frac{ \tau_c^{\nu - 1} (\nu e V - \epsilon)^{\nu - 1}}{\Gamma (\nu)}\, [1 -  \Theta (\epsilon - \nu e V)]\\
n_\text{h}^\text{eq} (\epsilon, V; T = 0) & = \frac{ \tau_c^{\nu - 1} ( \epsilon - \nu e V)^{\nu - 1}}{\Gamma (\nu)}\,  \Theta (\epsilon - \nu e V),
\end{aligned}
\label{eq:voltage_distributions}
\end{equation}
where superscript ``eq'' emphasizes that channels B and SA are at equilibrium, $V$ equals $V_\text{B}$ or $V_\text{SA}$, $\Gamma (\nu)$ is the gamma function, and $\Theta$ is the step function.
The voltage bias is defined with respect to the chemical potential of channel A before the diluter.

The presence of Heaviside $\Theta$-functions in Eq.~\eqref{eq:voltage_distributions} means that the particle and hole distributions for a zero-$T$ equilibrium anyonic channel are only finite for states with energies below and above the Laughlin-sea surface, respectively. This feature is analogous to the property of fermionic distribution functions. 
However, in contrast to the fermionic case, zero-$T$ equilibrium anyonic distributions are not flat when they are nonzero. A related unique feature of anyonic distributions is that the sum of particle and hole distributions is not a constant. In great contrast, for fermions, the sum of particle and hole distributions equals one for all energies, a feature arising from the fermionic anticommutation relation.
The plots of Eq.~\eqref{eq:voltage_distributions} are shown in Fig.~\ref{fig:distributions}(a). Following Eq.~\eqref{eq:voltage_distributions} and Fig.~\ref{fig:distributions}(a), the particle (hole) distribution is finite when $\epsilon < \nu e V$, i.e., beneath (above) the Laughlin surface at the energy $\nu e V$, and equals zero for energies above the anyonic-sea surface.
When approaching the Laughlin surface, both the particle and hole distributions diverge as a power law.

Now we move to the non-equilibrium channel A with zero-temperature ($T_\text{A} = T_\text{SA} = 0$) distribution functions
\begin{equation}
\begin{aligned}
n_{\text{p,A}}^\text{neq-braid} (\epsilon, I_\text{A};  T_\text{A} = 0, T_\text{SA} = 0) & = \frac{\nu^{1-\nu}\Gamma (1 - \nu)}{\pi} \tau_c^{ \nu - 1} \text{Re} \left\{ e^{i\pi\nu/2} \left[ \frac{I_\text{A}}{e} \left(1 - e^{-2i\pi\nu} \right) - i \nu \epsilon \right]^{\nu-1} \right\}, \\
n_{\text{h,A}}^\text{neq-braid} (\epsilon, I_\text{A}; T_\text{A} = 0, T_\text{SA} = 0 ) & = \frac{\nu^{1-\nu}\Gamma (1 - \nu)}{\pi} \tau_c^{ \nu - 1} \text{Re} \left\{ e^{-i\pi\nu/2} \left[ \frac{I_\text{A}}{e} \left(1 - e^{-2i\pi\nu} \right) - i \nu \epsilon \right]^{\nu-1} \right\},
\end{aligned}
\label{eq:current_distributions}
\end{equation}
where superscript ``neq-braid'' indicates that only time-domain braiding (addressed by ``braid'') processes are involved when evaluating the influence of non-equilibrium (addressed by ``neq'') anyons on correlation functions. In contrast to Eq.~\eqref{eq:voltage_distributions} where we used the voltage (chemical potential) to parametrize the distribution functions, here we use the non-equilibrium current $I_\text{A}$ as the second argument: now we fix the chemical potential and tune the amplitude of $I_\text{A}$.
Plots of these non-equilibrium distribution functions are shown in Fig.~\ref{fig:distributions}(b).
These non-equilibrium distribution functions are further compared, in Fig.~\ref{fig:distributions}(c), with equilibrium ones. Clearly, particle distribution gradually approaches the equilibrium one (the black curve), when decreasing the non-equilibrium current ($I_\text{A}/e = \epsilon_0$ for the red curve and  $I_\text{A}/e = \epsilon_0/2$ for the blue curve).

With a non-equilibrium current in channel A, the zero-temperature [$T_\text{A} = T_\text{SA} = 0$, cf. Eq.~\eqref{eq:current_distributions}] particle distribution in this channel is finite even for states with positive energies, meaning that non-equilibrium current excites particles into states with higher energies, mimicking the feature of fermionic distributions induced by thermal fluctuations.
This is exactly the reason to define an effective temperature in a non-equilibrium channel.
In addition, as shown by Figs.~\ref{fig:distributions}(d) and \ref{fig:distributions}(e), particle distributions decrease in a power-law manner with increasing energy [as $\epsilon^{\nu - 2}$, see Fig.~\ref{fig:effective_vs_real}].
This power-law energy dependence, remarkably, distinguishes the anyonic distribution characterized by an effective temperature introduced by the non-equilibrium current, from an equilibrium thermal distribution.
Indeed, for a real temperature $T_\text{B}$, equilibrium distribution functions of channel B read as
\begin{equation}
\begin{aligned}
n_{\text{p,B}}^\text{eq} (\epsilon,V_\text{B}; T_\text{B}) & = \frac{1}{2 \pi\Gamma (\nu)} \left( \frac{1}{2 \pi T_\text{B} \tau_c} \right)^{1 - \nu} \exp\left(-\frac{\epsilon - \nu V_\text{B}}{2 T_\text{B}}\right)  \Bigg| \Gamma \left( \frac{\nu}{2} + i \frac{\epsilon - \nu V_\text{B}}{2 \pi T_\text{B}} \right) \Bigg|^2,\\
n_{\text{h,B}}^\text{eq} (\epsilon,V_\text{B}; T_\text{B}) & = \frac{1}{2 \pi\Gamma (\nu)} \left( \frac{1}{2 \pi T_\text{B} \tau_c} \right)^{1 - \nu} \exp\left(\frac{\epsilon - \nu V_\text{B}}{2 T_\text{B}}\right) \Bigg| \Gamma \left( \frac{\nu}{2} + i \frac{\epsilon - \nu V_\text{B}}{2 \pi T_\text{B}} \right) \Bigg|^2,
\end{aligned}
\label{finite-TB-eq}
\end{equation}
with the plot of $n_{\text{p,B}}^\text{eq}$ shown in Fig.~\ref{fig:distributions}(f) for $V_\text{B} = 0$.
Following this figure, for positive energies, the thermal distribution function $n_{\text{p,B}}^\text{eq}$ decays exponentially away from the Laughlin surface, $\sim \exp (-\epsilon/T)$ (cf. Fig.~\ref{fig:effective_vs_real}), much faster than Eq.~\eqref{eq:current_distributions} that is induced by a non-equilibrium current.
A more detailed comparison is provided in Sec.~S9, to distinguish between non-equilibrium particle distributions with effective parameters and equilibrium ones with real chemical potential and temperature.

Equation~\eqref{eq:current_distributions} is obtained when $S_\text{A} = \nu e I_\text{A}$, corresponding to the situation where $T_\text{SA} = T_\text{A} = 0$.
Before ending this section, we provide the general expression for the distribution functions when $S_\text{A} \neq \nu e I_\text{A}$: 
\begin{equation}
\begin{aligned}
n_{\text{p,A}}^\text{neq-braid} (\epsilon, I_\text{A}, S_\text{A}) & = \frac{\nu^{1-\nu}\Gamma (1 - \nu)}{\pi} \tau_c^{ \nu - 1} \text{Re} \left( e^{i\pi\nu/2} \left\{ \frac{S_\text{A}}{\nu e^2} \left[1 - \cos(2\pi\nu) \right] + i\frac{I_\text{A}}{e} \sin (2\pi\nu) - i \nu \epsilon \right\}^{\nu-1} \right), \\
n_{\text{h,A}}^\text{neq-braid} (\epsilon, I_\text{A}, S_\text{A}) & = \frac{\nu^{1-\nu}\Gamma (1 - \nu)}{\pi} \tau_c^{ \nu - 1} \text{Re} \left( e^{-i\pi\nu/2} \left\{ \frac{S_\text{A}}{\nu e^2} \left[1 - \cos(2\pi\nu) \right] + i\frac{I_\text{A}}{e} \sin (2\pi\nu) - i \nu \epsilon \right\}^{\nu-1} \right).
\end{aligned}
\label{eq:current_distributions_ia_sa}
\end{equation}
Based on Eq.~\eqref{eq:current_distributions_ia_sa}, shapes of distribution functions are determined by the non-equilibrium current noise, $S_\text{A}$.
The non-equilibrium current, $I_\text{A}$, induces a shift of distribution functions in energy. Thus, the distribution functions for diluted anyonic beams are parametrized by the two quantities: effective potential $V_\text{eff}$ (related to $I_\text{A}$) and effective temperature $T_\text{eff}$ (related to $S_\text{A}$). This is similar to conventional fermionic or bosonic functions at equilibrium. However, the shape of the out-of-equilibrium anyonic distributions is drastically different from the Fermi (or Bose) functions.

%%%%%%%%%%%%%%%%%%%%%%%%%
\begin{figure}
  \includegraphics[width= \linewidth]{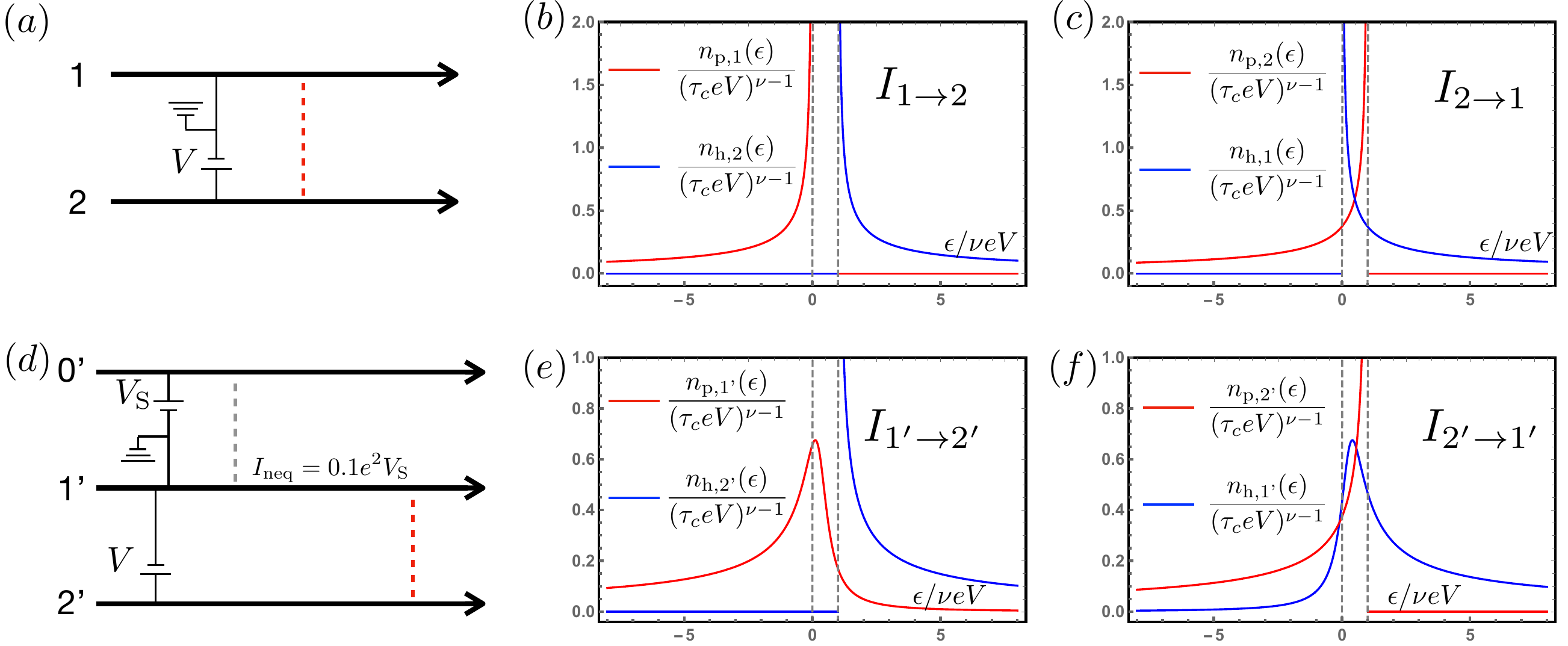}
  \caption{
  \textbf{Representation of zero-temperature inter-channel tunneling in terms of anyonic distribution functions.} The particle and hole distribution functions are shown by red and blue solid curves, respectively.
  A finite tunneling rate requires particle and hole distributions to be both finite for each tunneling event.
  \textbf{Panel (a)}: First setup -- two equilibrium channels communicate through a quantum point contact (red dashed line).
  \textbf{Panel (b)}: Tunneling from channel 1 to channel 2 (yielding a partial tunneling current $I_{1\to 2}$) is determined by the product of particle ($n_\text{p,1}$ of channel 1) and hole ($n_\text{h,2}$ of channel 2) distributions. $I_{1\to 2} = 0$ since either particle or hole distribution equals zero, for all energies.
  \textbf{Panel (c)}: Tunneling from channel 2 to channel 1 ($I_{2\to 1}$) is determined by the product of particle ($n_\text{p,2}$ of channel 2) and hole ($n_\text{h,1}$ of channel 1) distributions. $I_{2\to 1}$ receives contributions for energies $0<\epsilon < \nu e V$, where both particle and hole distributions are finite.
 \textbf{Panel (d)}: Second setup -- a non-equilibrium channel (1') and an equilibrium one (2'). The gray dashed line denotes a diluter, through which a non-equilibrium current is provided to channel 1'. We choose, without loss of generality, $I_\text{neq} = 0.1 e^2 V_\text{S}$, thus fixing the renormalized transmission of the diluter (which is, in fact, also a function of $V_\text{S}$).  
  \textbf{Panel (e)}: Particle ($n_\text{p,1'}$ of channel 1') and hole ($n_\text{h,2'}$ of channel 2') distributions that are relevant to the tunneling from 1' to 2' ($I_{1'\to 2'}$). The partial tunneling current now receives contributions from energies $\epsilon > \nu e V$.
  In great contrast to the equilibrium case (b), the non-equilibrium current $I_\text{neq}$ in channel 1' enables tunneling of anyons from 1' to 2'.
  \textbf{Panel (f)}: Particle ($n_\text{p,2'}$ of channel 2') and hole ($n_\text{h,1'}$ of channel 1') distributions that determine tunneling from 2' to 1' ($I_{2'\to 1'}$). The partial tunneling current receives contributions from energies $\epsilon < \nu e V$, where both distributions are nonzero.
  }
  \label{fig:tunnelings}
\end{figure}
%%%%%%%%%%%%%%%%%%%%%%%%%

\section*{S4. Interpreting zero-temperature tunneling current with distribution functions}

In this section, we analyze the zero-temperature tunneling current in the two setups [shown in Fig.~\ref{fig:tunnelings}(a) and \ref{fig:tunnelings}(d)] by considering the corresponding particle and hole distributions (we will specify distributions by the label of corresponding channels).
We will simplify the notation for distribution functions, by assigning them only one argument: the energy $\epsilon$.
Before addressing each of the two setups in detail, we notice, based on Eq.~\eqref{eq:current_integral_form}, that a finite tunneling current requires (i) finite particle distributions in the channel where particles are from, and (ii) finite hole distributions in the channel where particles tunnel into.

Now we move to two specific systems.
The first setup is depicted in Fig.~\ref{fig:tunnelings}(a), where two equilibrium channels are bridged by a tunneling quantum point contact (the red dashed line). Channel 2 is biased by voltage $V$ with respect to the grounded channel (channel 1).
Tunneling between these two channels involves two competing processes: tunneling of particles from channel 1 to channel 2 [represented by the current $I_{1\to 2}$ in Fig.~\ref{fig:tunnelings}(b)] and tunneling of particles from channel 2 to channel 1 [represented by the current $I_{2\to 1}$ in Fig.~\ref{fig:tunnelings}(c)].
Based on Figs.~\ref{fig:tunnelings}(b) and \ref{fig:tunnelings}(c), for equilibrium cases, zero-temperature tunneling is only possible from the channel with a larger bias to the one with a smaller one.
Indeed, in this case, states with energies $0<\epsilon < \nu eV$ have finite particle and hole distributions [Fig.~\ref{fig:tunnelings}(c)] and, hence, the product $n_{p,2}(\epsilon) n_{h,1}(\epsilon)$ is finite in this energy interval, enabling tunneling from 2 to 1.
The tunneling current in the opposite direction, from 1 to 2, is totally forbidden, as the product of particle or hole distribution $n_{p,1}(\epsilon) n_{h,2}(\epsilon)$ vanishes for all energies [Fig.~\ref{fig:tunnelings}(b)].

In the setup with a non-equilibrium state in one of the channels [Fig.~\ref{fig:tunnelings}(d)], the situation is different. Specifically, here, channel 1' is grounded but it carries a non-equilibrium (diluted) current $I_\text{neq}$ supplied (through the diluter) by the channel 0' biased by voltage $V_\text{S}$. 
With this non-equilibrium current, channel 1' has a finite particle distribution for all energies. Even for positive energies, the particle and hole distributions in channel 1' are both finite, decreasing as a power law, 
$\epsilon^{\nu - 2}$ [Figs.~\ref{fig:distributions}(d) and \ref{fig:tunnelings}(e)]. Consequently, tunneling from 1' to 2', i.e., $I_{1'\to 2'}$, is non-zero as long as the hole distribution $n_{h,2'}(\epsilon)$ in channel 2' is finite, i.e., for $\epsilon > \nu e V$. Remarkably, $n_{p,1'}(\epsilon)$ remains finite even for energies larger than $\nu e V_\text{S}$ -- the energy corresponding to the Laughlin surface of the source channel 0'. This means that the partial tunneling current $I_{1'\to 2'}$ is non-zero for $V_\text{S}<V$, i.e., when the source channel that produces the non-equilibrium state has a smaller voltage than the equilibrium channel into which particles tunnel.
Similarly, tunneling from channel 2' to channel 1', which yields  $I_{2'\to 1'}$, now involves states with a larger range of energy than in the first setup with equilibrium channels: $\epsilon < \nu e V$ instead of $0<\epsilon < \nu eV$.

\section*{S5. Additional criteria to define the effective chemical potential}

In the main text, the non-equilibrium channel A is calibrated by coupling it to an equilibrium channel B. More specifically, the effective chemical potential $V_\text{eff}$ of A is defined to be equal to the potential $V_\text{B}$ applied to B, $V_\text{eff} = V_\text{B}$, such that no \textit{net} charge current tunnels between A and B for this value of $V_\text{B}$.
Following this definition, 
\begin{equation}
    V_\text{eff} = I_\text{A} \sin (2\pi\nu)/\nu^2 e^2 
    \label{VeffIA}
\end{equation}
depends on the non-equilibrium current $I_\text{A}$ in channel A and the filling fraction $\nu$.
To support this conclusion, in this section, we provide two more criteria to define the effective potential, which yields the same value of $V_\text{eff}$
given by Eq.~(\ref{VeffIA}).

\subsection*{S5A. The neutrality requirement}

Before introducing the neutrality requirement, we first define ``modified'' particle and hole distributions
\begin{equation}
\begin{aligned}
\delta n_{\text{p,A}}^\text{neq-braid} ( \epsilon, I_\text{A}; T_\text{A}, T_\text{SA}) \equiv n_{\text{p,A}}^\text{neq-braid} (\epsilon, I_\text{A}; T_\text{A}, T_\text{SA}) - n_{\text{p,A}}^\text{eq} (\epsilon - \nu e V_\text{eff}, 0; T_\text{A}, T_\text{SA}),\\
\delta n_{\text{h,A}}^\text{neq-braid} (\epsilon, I_\text{A}; T_\text{A}, T_\text{SA}) \equiv n_{\text{h,A}}^\text{neq-braid} (\epsilon, I_\text{A}; T_\text{A}, T_\text{SA}) - n_{\text{h,A}}^\text{eq} (\epsilon - \nu e V_\text{eff}, 0 ; T_\text{A}, T_\text{SA}),
\end{aligned}
\label{eq:modified_distributions}
\end{equation}
as the difference between non-equilibrium distributions, and equilibrium ones,
after a shift in energy $\nu e V_\text{eff}$. Following Eq.~\eqref{eq:modified_distributions}, we define the neutrality requirement as
\begin{equation}
\begin{aligned}
&\int_{\nu e V_\text{eff}}^\infty d\epsilon\  \delta n_{\text{p,A}}^\text{neq-braid} ( \epsilon, I_\text{A}; T_\text{A}, T_\text{SA}) = \int_{-\infty}^{\nu e V_\text{eff}} d\epsilon\  \delta n_{\text{h,A}}^\text{neq-braid} (\epsilon, I_\text{A}; T_\text{A}, T_\text{SA}),\\
&\int_{\nu e V_\text{eff}}^\infty d\epsilon\  \delta n_{\text{h,A}}^\text{neq-braid} (\epsilon, I_\text{A}; T_\text{A}, T_\text{SA}) = \int_{-\infty}^{\nu e V_\text{eff}} d\epsilon\  \delta n_{\text{p,A}}^\text{neq-braid} ( \epsilon, I_\text{A}; T_\text{A}, T_\text{SA}).
\end{aligned}
\label{eq:equalities}
\end{equation}
With equation above, we are requiring two equalities: (i) between the extra particle distribution above the energy $\nu e V_\text{eff}$, and the extra hole distribution below the energy $\nu e V_\text{eff}$; and likewise, (ii) between the extra particle distribution below the energy $\nu e V_\text{eff}$, and the extra hole distribution above the energy $\nu e V_\text{eff}$.
Alternatively, Eq.~\eqref{eq:equalities} requires the particle and hole distributions to be symmetric to each other, with respect to the energy $\nu V_\text{eff}$: the position of the shifted anyonic surface.

Both equalities are satisfied by choosing $V_\text{eff} = I_\text{A} \sin (2\pi\nu)/\nu^2 e^2$, which is exactly Eq.~\eqref{VeffIA}.
To illustrate this, we show in Figs.~\ref{fig:potential_criteria}(a)-\ref{fig:potential_criteria}(c) the values of $\delta n^\text{neq-braid}_\text{p} (\epsilon, I_\text{A}; T_\text{A}, T_\text{SA})$ and $\delta n^\text{neq-braid}_\text{h} ( \epsilon, I_\text{A}; T_\text{A}, T_\text{SA})$, for different temperatures.
Clearly, $\delta n^\text{neq-braid}_\text{p} ( \epsilon-I_\text{A} \sin (2\pi\nu)/\nu e, I_\text{A}; T_\text{A}, T_\text{SA}) = \delta n^\text{neq-braid}_\text{h} (  -\epsilon+I_\text{A} \sin (2\pi\nu)/\nu e, I_\text{A}; T_\text{A}, T_\text{SA})$, meaning $\delta n^\text{neq-braid}_\text{p}$ and $\delta n^\text{neq-braid}_\text{h}$ are indeed symmetric with respect to the energy $\epsilon_\mu = I_\text{A} \sin (2\pi\nu)/\nu e$.
Equalities of Eq.~\eqref{eq:equalities} are clearly satisfied, with the symmetry mentioned above.

%%%%%%%%%%%%%%%%%%%%%%%%%
\begin{figure}[h!]
  \includegraphics[width= 0.95 \linewidth]{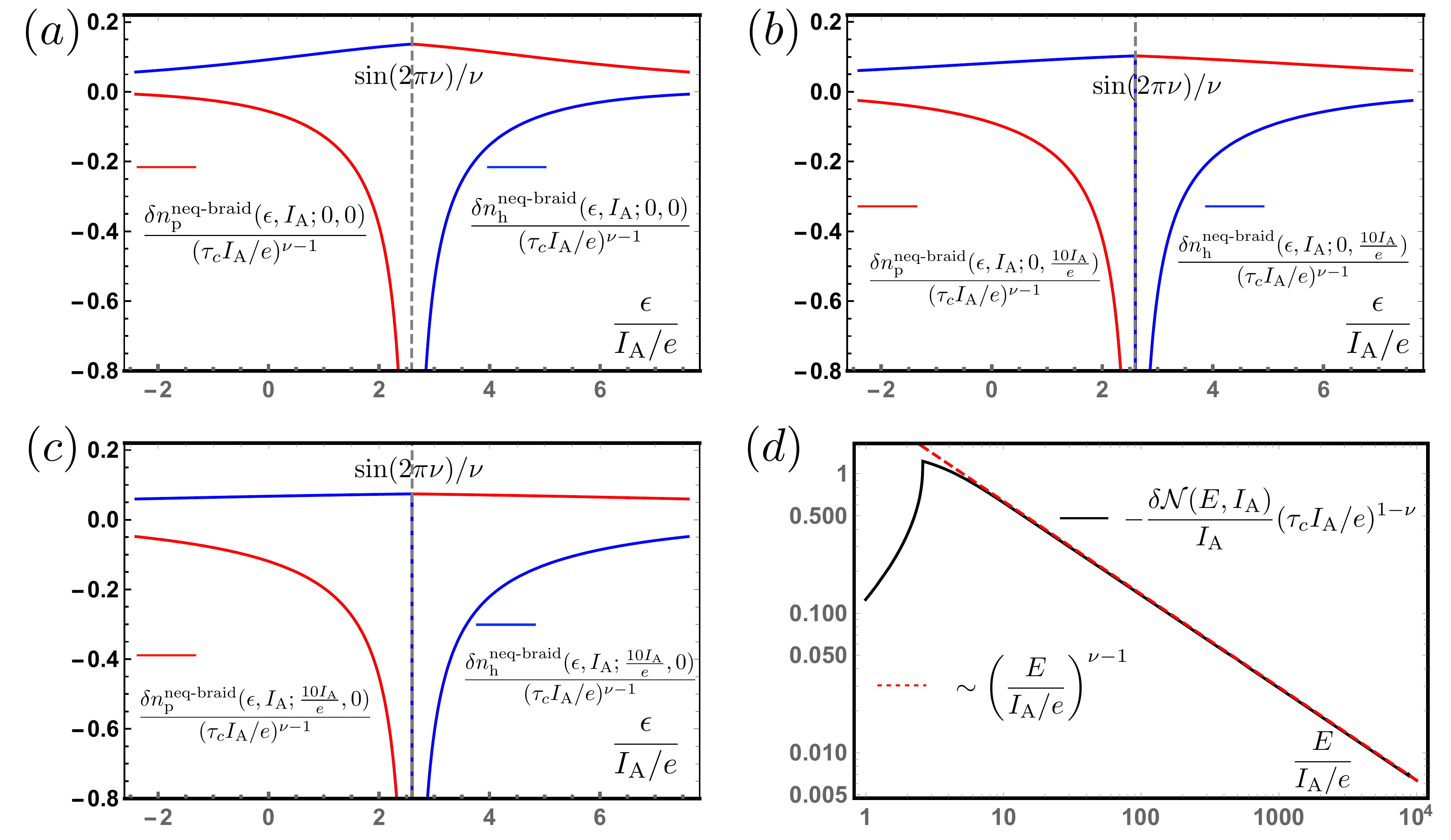}
  \caption{
  \textbf{Two extra criteria to define the effective chemical potential.} \textbf{Panel (a)}: Plots of the modified particle and hole distribution functions defined in Eq.~\eqref{eq:modified_distributions} for $T_\text{A} = T_\text{SA} = 0$. \textbf{Panels (b) and (c)}: Modifications of the particle and hole distributions for $T_\text{A} = 0$, $T_\text{SA} = 10 I_\text{A}/e$ and $T_\text{A} = 10 I_\text{A}/e$, $T_\text{SA} = 0$. For all three figures, the modified functions for particle and hole states are symmetric to each other with respect to the energy $I_\text{A} \sin (2\pi\nu )/\nu e$ (the gray dashed vertical line). \textbf{Panel (d)}: The log-log plot of the particle number within an energy strip (integrated distribution),  Eq.~\eqref{eq:modified_number}. The integral decreases in a power-law way and thus vanishes when the integration limits approach infinity, $E \to \infty$.
 }
  \label{fig:potential_criteria}
\end{figure}
%%%%%%%%%%%%%%%%%%%%%%%%%

\subsection*{S5B. Equality between ``modified'' and initial particle numbers}

As another criterion, if the effective chemical potential $ V_\text{eff}$ is reasonably defined, we would like the particle numbers in the channel with a non-equilibrium current $I_\text{A}$ to be equal to that of an equilibrium channel, that has a relative shift $V_\text{eff}$ in the chemical potential, i.e.,
\begin{equation}
\begin{aligned}
\int_{-\infty}^\infty d\epsilon \delta n^\text{neq-braid}_\text{p} (  \epsilon, I_\text{A}; 0,0) = \int_{-\infty}^\infty d\epsilon \delta n^\text{neq-braid}_\text{h} (\epsilon, I_\text{A} ; 0,0) = 0,
\end{aligned}
\label{eq:second_requirement}
\end{equation}
for zero temperatures. To verify this requirement, we define another integral
\begin{equation}
    \delta \mathcal{N} (E,I_\text{A}) \equiv e \int_{-E}^{E} d\epsilon \delta n^\text{neq}_\text{p} (\epsilon,I_\text{A};0 ,0),
    \label{eq:modified_number}
\end{equation}
as the modification, introduced by the non-equilibrium current, of particle number within the energy strip $(-E,\,E)$ compared to that of an equilibrium channel.
When $V_\text{eff} = I_\text{A}\sin (2\pi\nu)/\nu^2 e^2$, we check the requirement Eq.~\eqref{eq:second_requirement} in Fig.~\ref{fig:potential_criteria}(b). Following Fig.~\ref{fig:potential_criteria}(b), the integral in Eq.~\eqref{eq:modified_number} decays to zero as a power law, i.e., $\sim E^{-2/3}$.
When $E \to \infty$, $\delta \mathcal{N} \to 0$, indicating that Eq.~\eqref{eq:second_requirement} is satisfied by taking $V_\text{eff} = I_\text{A}\sin (2\pi\nu)/\nu^2 e^2$.

We can further complement the above consideration by performing a Fourier transformation:
\begin{equation}
\begin{aligned}
    \lim_{E \to \infty} \delta \mathcal{N} (E,I_\text{A}) & = e \int_{-\infty}^\infty d \epsilon \ [ \delta n^\text{neq}_\text{p} (\epsilon,I_\text{A};0 ,0)]\\
    & = \lim_{t\to 0} [ \langle \psi_A^\dagger (t) \psi_A (0)\rangle_{T_\text{A},T_\text{SA} } - \langle \psi_B^\dagger (t) \psi_B (0)\rangle_{T_\text{B}}  ]\\
    & = \lim_{t\to 0} \frac{\tau_c^{\nu - 1}}{2\pi} \frac{\left(-\frac{S_\text{A}}{\nu^2 e^2}\right) [1 - \cos (2\pi\nu) ] | t| }{( i t)^\nu} \to 0,
\end{aligned}
\label{eq:anyonic_number_difference}
\end{equation}
where we have set $V_\text{B} = V_\text{eff} \equiv I_\text{A} \sin (2\pi\nu)/\nu^2 e^2$ and used the fact that the difference between the two correlation functions is not singular at $t \to 0$ (so that one can take $\tau_c = 0$ in the denominator).

\subsection*{S5C. Role of direct tunneling of non-equilibrium anyons}

Following previous section, when considering only the time-domain-braiding contributions, the three criteria for defining the effective chemical potential, namely (i) zero tunneling current, (ii) neutrality condition, and (ii) equality between modified and initial total particle numbers, are consistent.
In this section, we exclude time-domain braiding, but include only direct tunneling of non-equilibrium anyons. Here, we assume that channel A is out of equilibrium and channel B is at equilibrium, as in the setup shown in Fig. S2(d).
More specifically, assuming channel A distributions are modified by particles with energies between $0$ and $\nu e V_\text{SA}$,
relevant correlation functions now become
\begin{equation}
\begin{aligned}
    \big\langle \psi^\dagger_\text{A} (t) \psi_\text{A} (0) \big\rangle_\text{direct} & \approx \frac{\tau_c^{\nu - 1}}{2\pi (\tau_c + i t)^\nu} \left[ 1 + \mathcal{T} \left(e^{i\nu e V_\text{SA} t} -1\right)\right],\\
    \big\langle \psi_\text{B} (t) \psi^\dagger_\text{B} (0) \big\rangle_\text{direct} & \approx \frac{\tau_c^{\nu - 1}}{2\pi (\tau_c + i t)^\nu} e^{-i\nu e V_\text{B} t},
\end{aligned}
\end{equation}
with $\mathcal{T}$ the renormalized transmission probability through the diluter.
Let us now apply the above criteria to these correlation functions.

\textbf{Criterion I: \ Zero tunneling current}

As the first criterion, the effective chemical potential can be defined after requiring a vanishing tunneling current, leading to
\begin{equation}
\begin{aligned}
    (V_\text{SA} - V_\text{B})^{2\nu - 1} \mathcal{T}  = V_\text{B}^{2\nu - 1} (1 - \mathcal{T}) \qquad
    \Rightarrow\qquad  V_\text{eff} = V_\text{B} = V_\text{SA} \left[ 1 + \left( \frac{\mathcal{T}}{1 - \mathcal{T}}\right)^{1/(1 - 2\nu)} \right]^{-1},
\end{aligned}
\label{eq:direct_tunneling_veff}
\end{equation}
at zero temperature.
In the first line of Eq.~\eqref{eq:direct_tunneling_veff}, the left- and right-hand sides refer to tunneling from channel A to channel B and from B to A, respectively.

Interestingly, in the strongly diluted limit, $\mathcal{T} \ll 1$, $V_\text{eff}$ behaves differently depending on whether $\nu$ is larger or smaller than $1/2$.
Indeed, if $\nu > 1/2$ (including the fermionic case, $\nu = 1$), $V_\text{eff} /V_\text{SA} \ll 1$ is a small number.
In contrast, if $\nu < 1/2$ (including Laughlin states), $V_\text{eff} \approx V_\text{SA}$.
Actually, in the latter case, the tunneling from A to B is greatly amplified by the large time cutoff $\propto (V_\text{SA} - V_\text{eff})^{2\nu - 1}$, which compensates for the smallness of the transmission probability.

\textbf{Criterion II: Equality of modified particle distributions}
\\
Now, we move to the second criterion. Following derivations of Eq.~\eqref{eq:anyonic_number_difference}, we get
\begin{equation}
\begin{aligned}
   & \lim_{t \to 0} [ \big\langle \psi^\dagger_\text{A} (t) \psi_\text{A} (0) \big\rangle_\text{direct} - \big\langle \psi^\dagger_\text{B} (t) \psi_\text{B} (0) \big\rangle_\text{direct}]
   \approx  \lim_{t \to 0} \frac{i\tau_c^{\nu - 1}}{2 \pi} \frac{(\mathcal{T}  \nu e V_\text{SA}-\nu e V_\text{B} ) t}{(it)^\nu},
\end{aligned}
\end{equation}
which yields zero, if $\nu < 1$, for any finite value of $V_\text{B}$.
Similarly to the previous consideration for anyons with time-domain braiding, the second criterion, i.e., zero modification of total particle number, is always satisfied if $\nu < 1$.
When $\nu \ge 1$, the second criterion gives instead $V_\text{eff} = \mathcal{T} V_\text{SA}$, which agrees with that from the first criterion, Eq.~\eqref{eq:direct_tunneling_veff}, only if $\nu = 1$.

\section*{S6. Heat current}
\label{sec:heat current}

This section provides details of calculating the tunneling heat current between channels A and B.
We define the partial heat current from channel A to channel B as
$$J_{\text{A}\to \text{B}} = \int d\epsilon \mathcal{J}_{\text{A}\to \text{B}} (\epsilon),$$ where
\begin{equation}
    \mathcal{J}_{\text{A}\to \text{B}} (\epsilon) \equiv \mathcal{T}_\text{C} \, n^\text{neq-braid}_\text{p,A} (\epsilon, I_\text{A};T_\text{A},T_\text{SA})\, n^\text{eq}_\text{h,B} (\epsilon, V_\text{B};  T_\text{B}) \left(\epsilon - \nu e V_\text{B} \right)
    \label{eq:heat}
\end{equation}
is the differential heat (defined as the energy with respect to the Laughlin surface $\nu e V_\text{B}$, of B) carried by states with the energy $\epsilon$.
Here, $n^\text{neq-braid}_\text{p,A}$ and $n^\text{eq}_\text{h,B}$ refer to the non-equilibrium particle distribution in channel A and the equilibrium hole distribution in B, respectively.
Equation~\eqref{eq:heat} can be obtained
after Fourier transformation of the tunneling heat current derived in, e.g., Ref.~\cite{SEbisuSchillerOregPRL22}.
The superscript of $n^\text{neq-braid}_\text{p,A}$
(``neq-braiding'') indicates that the non-equilibrium distribution function involves only time-domain braiding contribution. Direct tunneling of non-equilibrium particles is not included. 
In addition, $n^\text{neq-braid}_\text{p,A}$ contains four arguments:  the particle energy $\epsilon$, the non-equilibrium current $I_\text{A}$, and two temperatures $T_\text{A}$ and $T_\text{SA}$.
The other distribution function, $n^\text{eq}_\text{h,B}$, depends on three variables: the particle energy $\epsilon$, the bias $V_\text{B}$ (defined with respect to the chemical potential of channel A before the diluter), and the temperature in channel B, $T_\text{B}$.
Below, we focus on the situation where $\nu e V_\text{B} = \nu e V_\text{eff} = I_\text{A} \sin (2\pi\nu)/\nu e$, and study the heat current for different temperatures.

Before moving to specific cases, we begin by rewriting the heat current, Eq.~\eqref{eq:heat}, in another form that involves an integral over time.
Let us consider a general case where two channels, 1 and 2, with equal chemical potentials $V_0$ (either effective or real potential), communicate through a quantum point contact with transmission $\mathcal{T}_\text{C}$. 
In this case, the partial heat current from channel 1 to channel 2, $J_{1 \to 2}$ (not the net tunneling heat current), becomes
\begin{equation}
\begin{aligned}
    J_{1 \to 2}& = \mathcal{T}_\text{C} \int d\epsilon\  (\epsilon - \nu e V_0)\  n_\text{p,1} (\epsilon) n_\text{h,2} (\epsilon) \ = \mathcal{T}_\text{C} \int d\epsilon\ \epsilon\   n_\text{p,1} (\epsilon + \nu e V_0) n_\text{h,2} (\epsilon + \nu e V_0) \\
    & = \mathcal{T}_\text{C} \int d\epsilon \int dt_1 \int dt_2\  \epsilon \ e^{-i\epsilon (t_1 - t_2)} f_1 (t_1) f_2 (t_2) = -i\pi \mathcal{T}_\text{C} \int dt \left\{ \left[\frac{\partial}{\partial t} f_1 (t) \right] f_2 (t) - f_1 (t) \frac{\partial}{\partial t} f_2 (t) \right\},
\end{aligned}
\label{eq:heat_time_integral}
\end{equation}
where $n_\text{p,1}$ and $n_\text{h,2}$ are the particle distribution in channel 1 and the hole distribution in channel 2, respectively.
In Eq.~\eqref{eq:heat_time_integral}, $\nu e V_0$ refers to the energy with respect to which the heat is defined. For cases where the chemical potentials (effective or real) of the communicating channels are equal, $V_0$ is chosen such that tunneling heat current comes solely from thermal fluctuations (stemming from an effective or real temperature).
Functions $f_1$ and $f_2$ in Eq.~\eqref{eq:heat_time_integral} are Fourier transformed functions of these two distributions, after a shift $\nu e V_0$ in energy. Specifically,
\begin{equation}
\begin{aligned}
    f_1 (t) & \equiv \frac{1}{2\pi} \int d\epsilon e^{i\epsilon t} n_\text{p,1} (\epsilon + \nu e V_0) = \frac{e^{-i\nu eV_0 t}}{2\pi} \int d\epsilon e^{i\epsilon t} n_\text{p,1} (\epsilon), \\
    f_2 (t) & \equiv \frac{1}{2\pi} \int d\epsilon e^{-i\epsilon t} n_\text{h,2} (\epsilon + \nu e V_0) = \frac{e^{i\nu e V_0 t}}{2\pi} \int d\epsilon e^{-i\epsilon t} n_\text{h,2} (\epsilon).
\end{aligned}
\end{equation}
During the derivation Eq.~\eqref{eq:heat_time_integral}, we have used the identity
\begin{equation}
\int d\epsilon \ \epsilon
\ e^{-i\epsilon t_0 } = 2\pi i \frac{\partial}{\partial t_0} \delta (-t_0),
\end{equation}
and integration by parts.
One can similarly calculate the tunneling heat current from 2 to 1, $J_{2\to 1}$, with which the net tunneling heat current between channels 1 and 2 is given by  
$$J_Q = J_{1\to 2} - J_{2\to 1}.$$

\subsection*{S6A. Zero ambient temperatures}

To begin with, we show details of the calculation of the heat current between channels A and B, when $V_\text{eff} = I_\text{A} \sin (2\pi\nu)/\nu^2 e^2$, at zero temperatures, $T_\text{A} = T_\text{SA} = T_\text{B} = 0$.
In this case, we simplify the notation by writing
$n^\text{neq-braid}_\text{p,A} (\epsilon,I_\text{A};T_\text{A} = 0,T_\text{SA}=0) \to n_\text{p}^\text{neq-braid} (\epsilon,I_\text{A})$ and $n^\text{eq}_\text{h,B} ( \epsilon,V_\text{B}; T_\text{B}) \to n^\text{eq}_\text{h} (\epsilon,V_\text{B}) = n^\text{eq}_\text{h} (\epsilon - \nu e V_\text{B},0)$.
The heat current can be evaluated with Eq.~\eqref{eq:heat_time_integral}, where the functions $f_1 (t)$ and $f_2 (t)$ are given by
\begin{equation}
\begin{aligned}
f_1 (t) & \equiv \frac{1}{2\pi} \int d\epsilon\,  e^{i\epsilon t}\, n^\text{neq-braid}_\text{p} \left( \epsilon + I_\text{A} \frac{\sin(2\pi\nu)}{\nu e},I_\text{A}\right) =  \langle \psi^\dagger_\text{A} (t) \psi_\text{A} (0) \rangle_\text{neq} \exp\left[ - i I_\text{A} \frac{\sin(2\pi\nu)}{\nu e} t \right] \\
& = \frac{1}{2\pi} \frac{\tau_c^{\nu-1}}{(\tau_c + it)^\nu} 
\exp\left[-\frac{I_\text{A}}{\nu e} [1 - \cos (2 \pi \nu )] |t| \right], \\
f_2 (t) & \equiv \frac{1}{2\pi} \int d\epsilon\,
e^{- i\epsilon t}\,  n^\text{eq}_\text{h} (\epsilon , 0) = \frac{1}{2\pi} \frac{\tau_c^{\nu-1}}{(\tau_c + it)^\nu}.
\end{aligned}
\label{eq:f1_f2}
\end{equation}

The heat current, Eq.~\eqref{eq:heat_time_integral}, contains the time derivatives of both $f_1 (t)$ and $f_2(t)$.
These derivatives act, in particular, on the function $(\tau_c + it)^{-\nu}$. It turns out that such contributions however cancel out in Eq.~\eqref{eq:heat_time_integral}; the only term that survives in Eq.~\eqref{eq:heat_time_integral} originates from the time derivative of the exponential factor in $f_1(t)$.
Note that this exponential factor depends on the absolute value $|t|$.
As a consequence, the tunneling heat current from A to B takes the form
\begin{equation}
\begin{aligned}
J_{\text{A} \to \text{B}} & =  i\pi \frac{I_\text{A}}{\nu e} [1 - \cos (2 \pi\nu)] \mathcal{T}_\text{C}\, \frac{\tau_c^{2\nu-2}}{(2\pi)^2} \int dt \frac{e^{-\frac{I_\text{A}}{\nu e} [1 - \cos (2 \pi\nu)]|t|}}{(\tau_c + i t)^{2\nu}} \text{sgn}(t)\\
& = \pi \frac{I_\text{A}}{\nu e} [1 - \cos (2 \pi\nu)]  \frac{\tan(\pi\nu)}{\nu } \frac{I_{\text{A} \to \text{B}}}{e} = \pi \tan^2 (\pi\nu)\, V_\text{eff}\, I_{\text{A} \to \text{B}},
\end{aligned}
\end{equation}
where $I_{\text{A} \to \text{B}}$ stands for the tunneling charge current from channel A (with non-equilibrium current $I_\text{A}$) to the equilibrium channel B characterized by a finite potential $V_\text{B}$.
We see now that the factor $$\frac{I_\text{A}}{\nu e} [1 - \cos (2 \pi\nu)] \sim e V_\text{eff}$$ plays the role of the effective temperature.
Indeed, for two unbiased channels, one with temperature $T$ and the other at zero temperature, the heat current should be proportional to $T I_{\text{A} \to \text{B}}$.

Now, we consider the heat current produced by the tunneling processes from channel B to channel A: 
\begin{equation}
    \begin{aligned}
        J_{\text{B}\to \text{A}} & = \mathcal{T}_\text{C} \int d\epsilon\,  n_\text{h}^\text{neq-braid} (I_\text{A}, \epsilon) \, n_\text{p}^\text{eq} \left( \epsilon - I_\text{A} \frac{\sin 2\pi\nu}{\nu e}, 0\right) \left[\epsilon - I_\text{A} \frac{\sin(2\pi\nu)}{\nu e} \right]\\
        & =  \mathcal{T}_\text{C} \int d\epsilon\ \epsilon\   n^\text{neq-braid}_\text{h} \left( \epsilon + I_\text{A} \frac{\sin(2\pi\nu)}{\nu e},I_\text{A}\right)  n^\text{eq}_\text{p} \left( \epsilon , 0 \right).
    \end{aligned}
    \label{eq:heat_inverse}
\end{equation}
Noticing that [cf. Fig.~\ref{fig:distributions}(b)]
$$n_\text{h}^\text{neq-braid} ( \epsilon+I_\text{A} \sin(2\pi\nu)/\nu e,I_\text{A}) = n_\text{p}^\text{neq-braid} ( -\epsilon+I_\text{A} \sin(2\pi\nu)/\nu e,I_\text{A})$$ 
and 
$$ n^\text{eq}_\text{p} \left(\epsilon,0 \right) =  n^\text{eq}_\text{h} \left( -\epsilon,0 \right),$$ Eq.~\eqref{eq:heat_inverse} can be rewritten as
\begin{equation}
    \begin{aligned}
        J_{\text{B}\to \text{A}}  &=  \mathcal{T}_\text{C} \int d\epsilon\ \epsilon\   n^\text{neq-braid}_\text{h} \left( \epsilon + I_\text{A} \frac{\sin(2\pi\nu)}{\nu e}, I_\text{A}\right)  n^\text{eq}_\text{p} \left( \epsilon ,0\right)\\
        & = \mathcal{T}_\text{C} \int d\epsilon\ \epsilon\   n^\text{neq-braid}_\text{p} \left( -\epsilon + I_\text{A} \frac{\sin(2\pi\nu)}{\nu e},I_\text{A}\right)  n^\text{eq}_\text{h} \left( - \epsilon , 0 \right)\\
        & = - \mathcal{T}_\text{C} \int d\epsilon\ \epsilon\   n^\text{neq-braid}_\text{p} \left( \epsilon + I_\text{A} \frac{\sin(2\pi\nu)}{\nu e}, I_\text{A}\right)  n^\text{eq}_\text{h} \left( \epsilon , 0\right) = -  J_{\text{A}\to \text{B}}.
    \end{aligned}
    \label{eq:heat_a_to_b}
\end{equation}
The total heat current $J_Q$ between channels A and B then reads
\begin{equation}
    J_Q = J_{\text{A} \to \text{B}} - J_{\text{B} \to \text{A}} = 2\pi \tan^2 (\pi\nu) \,V_\text{eff}\, I_{\text{A} \to \text{B}}.
\end{equation}

We plot the heat current carried by the state with energy $\epsilon$ in Fig.~\ref{fig:different_temperatures}. Here, the first two plots, Fig.~\ref{fig:different_temperatures}(a) and (b), show the results for $T_\text{A} = T_\text{SA} = T_\text{B} = 0$.
The particle distributions are symmetric to the corresponding hole distributions, with respect to the energy of the effective chemical potential $\nu e V_\text{eff}$.
As another important feature, for the zero-temperature situation, $\mathcal{J}_{\text{A} \to \text{B}}$ and $-\mathcal{J}_{\text{B} \to \text{A}}$ are both positive for all energies, indicating that the heat current is only allowed to transport from A to B, even for nonzero charge transport from B to A.
This is because heat refers to energy defined with respect to the Laughlin surface of the equilibrium channel B.

%%%%%%%%%%%%%%%%%%%%%%%%%
\begin{figure}[h!]
  \includegraphics[width= 1.0 \linewidth]{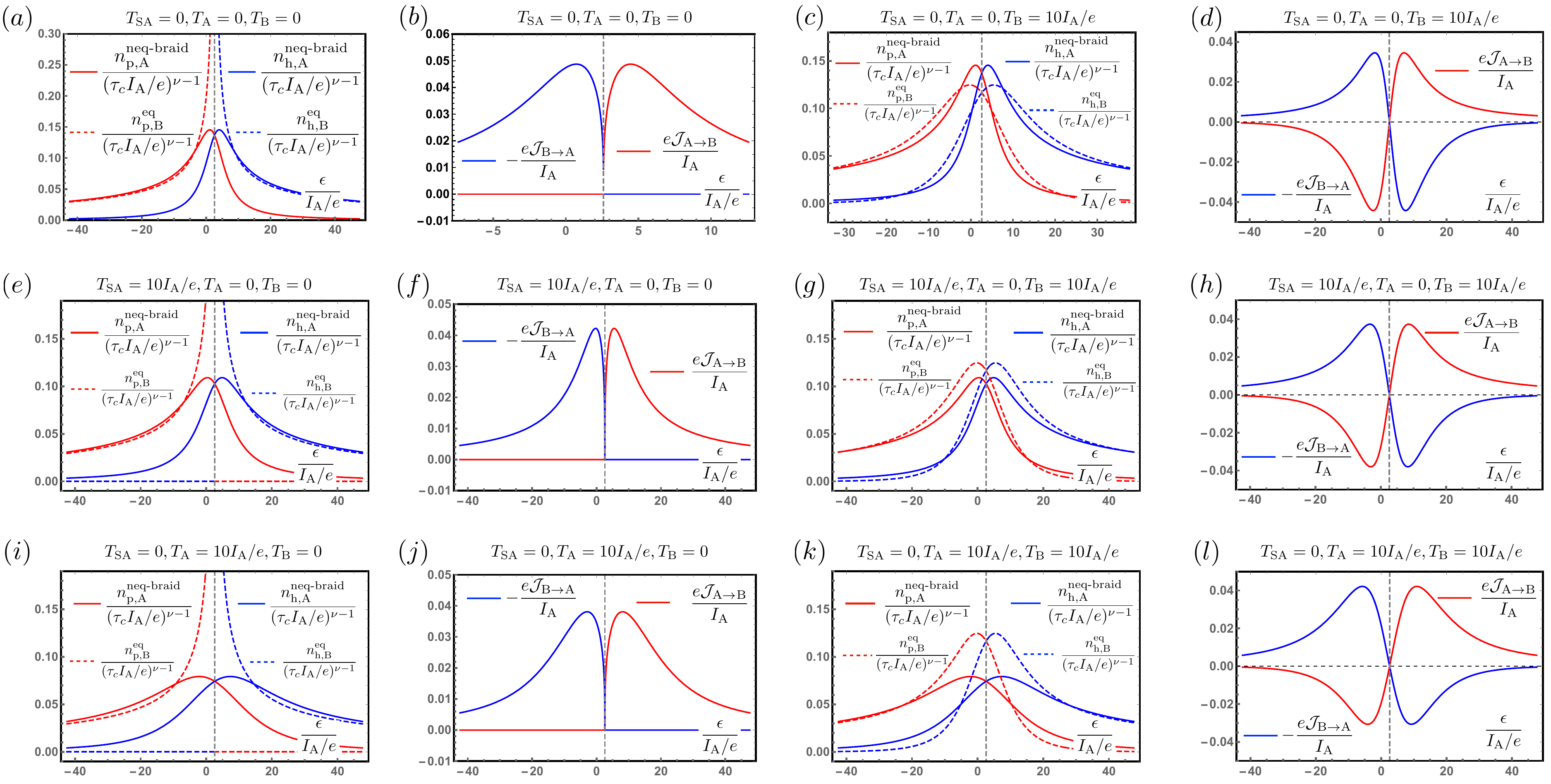}
  \caption{
  \textbf{Heat flow for different arrangements of temperatures in channels A and B.}
  Functions $n_\text{p,A}^\text{neq-braid}$ and $n_\text{h,A}^\text{neq-braid}$ refer, respectively, to the particle and hole distributions in the non-equilibrium channel A. Their superscript ``neq-braid'' indicates that only time-domain braiding processes are included when evaluating the influence of non-equilibrium anyons on correlation functions.
  Functions $n_\text{p,B}^\text{eq}$ and $n_\text{h,B}^\text{eq}$ refer to the corresponding distributions in the equilibrium channel B. 
  For convenience, temperatures are labeled on top of each panel and neglected in legends.
  \textbf{Panel (a)}: The particle distribution functions for $T_\text{SA} = T_\text{A} = T_\text{B} = 0$.
  The red curves represent the particle distributions in channels A (solid red curve) and B (dashed red curve). The blue curves represent the hole distributions in channels A (solid blue curve) and B (dashed blue curve).
  \textbf{Panel (b)}: Plots of the differential heat flows $\mathcal{J}_{\text{A} \to \text{B}}$ (red curve) and $-\mathcal{J}_{\text{B} \to \text{A}}$ (blue curve). Both curves are positive, meaning that heat only flows from A to B, for $T_\text{SA} = T_\text{A} = T_\text{B} = 0$.
  \textbf{Panels (c) and (d)} show the corresponding results for $T_\text{SA} = T_\text{A} = 0$, but $T_\text{B}$ finite. Now, the differential heat flows can be either positive or negative.
  \textbf{Panels (e) - (h)}: Curves correspond to those in panel (a) - (d) but with a finite temperature of channel SA, $T_\text{SA} = 10 I_\text{A}/e$.
  \textbf{Panels (i) - (l)}: Curves correspond to those in (a) - (d) but with a finite temperature of channel A, $T_\text{A} = 10 I_\text{A}/e$. In panels (e) - (l), we take $V_\text{SA} = 30 I_\text{A}/e^2$.
 }
  \label{fig:different_temperatures}
\end{figure}
%%%%%%%%%%%%%%%%%%%%%%%%%

\subsection*{S6B. Finite $T_\text{B}$ and $T_\text{SA} = T_\text{A} = 0$}

When the equilibrium channel B is at finite temperature $T_B\neq 0$, its correlation functions are given by Eq.~(\ref{finite-TB-eq}).
Again, we assume that channel B is biased such that no charge current tunnels through the collider:  $V_\text{B} = V_\text{eff}=I_\text{A} \sin (2\pi\nu)/\nu^2 e^2$.
For finite $T_B$, it is hard to obtain a compact analytical form of the heat current. Therefore, in Fig.~\ref{fig:different_temperatures}, we present results of the numerical integration performed in Eq.~\eqref{eq:heat_time_integral} with the known distribution functions given by Eqs.~\eqref{eq:current_distributions} and \eqref{finite-TB-eq}.
In Fig.~\ref{fig:different_temperatures}(c), we show the particle (red curves) and hole (blue curves) distribution functions for channels A (solid curves) and B (dashed curves).
Similar to the zero-temperature case [Fig.~\ref{fig:different_temperatures}(a)], particle and hole distributions at finite $T_\text{B}$, shown in Fig.~\ref{fig:different_temperatures}(c), remain symmetric to each other with respect to the Laughlin surface (at $\epsilon = \nu e V_\text{eff}$) of channel B.
Finite temperature $T_\text{B}$ ``softens'' the equilibrium particle distributions compared to Fig.~\ref{fig:different_temperatures}(a). As a consequence, the corresponding differential heat current [Fig.~\ref{fig:different_temperatures}(d)] can have negative values at low energies, in great contrast to that in the zero-temperature case [Fig.~\ref{fig:different_temperatures}(b)].

Note that, in Fig.~\ref{fig:different_temperatures}(d), the temperature in channel B, $T_\text{B} = 10I_\text{A}/e$, is chosen to be smaller than the temperature [$\approx 17.3 I_\text{A} $, see Fig.~1(d) of the main text] at which the total heat current $$J_Q = \int d\epsilon\, [\mathcal{J}_{\text{A}\to \text{B}} (\epsilon) - \mathcal{J}_{\text{B}\to \text{A}}(\epsilon)] $$ equals zero. 
Nevertheless, at this temperature, $\mathcal{J}_{\text{A}\to \text{B}}(\epsilon) < \mathcal{J}_{\text{B}\to \text{A}}(\epsilon)$ for $\epsilon \leq 15 I_\text{A}/e$, indicating heat flow from B to A at low energies.
This heat current from B to A, is balanced and exceeded by the reverse heat current, from A to B, at large energies.
Indeed, in Fig.~\ref{fig:different_temperatures}(d) [and likewise, Figs.~\ref{fig:different_temperatures}(h) and \ref{fig:different_temperatures}(l)], curves with positive values decay much slower than their negative counterparts, indicating that the total heat flow is dominated by the partial heat flow from A to B at large energies.

\subsection*{S6C. Finite $T_\text{SA}$ and $T_\text{A} = 0$}

Next, we consider the situation where $T_\text{SA}$ is finite and $T_\text{A} = 0$.
In this case, Eq.~\eqref{eq:psi_a_general_correlation} leads to
\begin{equation}
\begin{aligned}
\left.\langle \psi_A^\dagger (t) \psi_A (0)\rangle_{\text{neq}}
\right|_{T_\text{A} = 0,\,T_\text{SA} } & = \frac{1}{2\pi \tau_c } \frac{\tau_c^{\nu}}{(\tau_c + it)^\nu} \exp \left\{ -\frac{S_\text{A}}{\nu^2 e^2} [1 - \cos (2\pi\nu) ] | t|  + i \frac{I_\text{A}}{\nu e} \sin (2\pi\nu) t \right\},\\
\left.
\langle \psi_A (t) \psi^\dagger_A (0)\rangle_{\text{neq}}\right|_{T_\text{A} = 0,\,T_\text{SA} }  &  = \frac{1}{2\pi \tau_c } \frac{\tau_c^{\nu}}{(\tau_c + it)^\nu} \exp \left\{ -\frac{S_\text{A}}{\nu^2 e^2} [1 - \cos (2\pi\nu) ] | t|  - i \frac{I_\text{A}}{\nu e} \sin (2\pi\nu) t \right\},
\end{aligned}
\label{eq:psi_a_ta0}
\end{equation}
where $T_\text{SA}$ is implicitly contained in the noise of the diluted channel
$$S_\text{A} \equiv \int dt [\langle \hat{I}_\text{A} (0,t) \hat{I}_\text{A} (0,0)\rangle - I_\text{A}^2].$$
Indeed, $S_\text{A} = \nu e I_\text{A}$ when $T_\text{A} = T_\text{SA} = 0$, and deviates from this value otherwise.

When $T_\text{B} = 0$, the heat current from A to B can be similarly obtained following Eqs.~\eqref{eq:heat_time_integral} and \eqref{eq:f1_f2}, after replacing $I_\text{A}$ in Eq.~\eqref{eq:f1_f2} with $ S_\text{A}/\nu$, leading to
\begin{equation}
    J_Q = 2 \pi \frac{S_\text{A}}{\nu^3 e^3} [1 - \cos (2 \pi\nu)] \,\tan(\pi\nu) I_{\text{A} \to \text{B}} .
\end{equation}
Thus, when $T_\text{SA} $ is finite, the effective temperature becomes proportional to the current noise $S_\text{A}$ in the diluted channel A, which now combines the non-equilibrium and thermal parts of the tunneling noise of the diluter.

For a more comprehensive understanding, we plot the distribution functions and the differential heat flow, in Figs.~\ref{fig:different_temperatures}(e) and \ref{fig:different_temperatures}(f), respectively, for $T_\text{SA} = 10 I_\text{A}/e$ and $T_\text{A} = T_\text{B} = 0$.
Clearly, when $T_\text{B} = 0$, the spectral heat flows $\mathcal{J}_{\text{A} \to \text{B}}$ and $-\mathcal{J}_{\text{B} \to \text{A}}$ are both positive for all energies, indicating positive heat flow from A to B.
This is in line with the fact that in channel B, particle distributions are zero for energies larger than $\nu e V_\text{B}$.

When $T_\text{B}$ is finite, particle and hole distributions in channel B are again softened by thermal fluctuation.
This fact is shown in \ref{fig:different_temperatures}(g) for $T_\text{SA} = T_\text{B} = 10 I_\text{A}/e$, where particle distribution in channel B remains finite for states with heat ($\epsilon - \nu e V_\text{B}$) within the window of the order of $T_\text{B}$.
As a consequence, depending on the state energy, the heat can either flow from channel A to channel B or from B to A, as shown in Fig.~\ref{fig:different_temperatures}(h).
Note that, for energies above the Laughlin surface of channel B, the particle distribution in B (the red dashed curve) is influenced by temperature $T_\text{B}$ and decays much faster than the particle distribution in channel A (the red solid curve), which is induced by a non-equilibrium current.
This is an important difference between a real temperature and an effective one that is induced by a non-equilibrium current.

\subsection*{S6D. Finite $T_\text{A}$ and $T_\text{SA} = 0$}

Finally, we consider the situation where $T_\text{A}$ is finite but $T_\text{SA} =0$.
In this case, the correlation functions in channel A read
\begin{equation}
\begin{aligned}
\left.\langle \psi_A^\dagger (t) \psi_A (0)\rangle_{\text{neq} }\right|_{T_\text{SA} = 0,\,T_\text{A} } & = \frac{1}{2\pi \tau_c} \frac{(\pi T_\text{A} \tau_c)^\nu}{\sin^\nu [\pi T_\text{A}(\tau_c + it)]} \exp \left\{ -\frac{S_\text{A}}{\nu^2 e^2} [1 - \cos (2\pi\nu) ] | t|  + i \frac{I_\text{A}}{\nu e} \sin (2\pi\nu) t \right\},\\
\left.\langle \psi_A (t) \psi^\dagger_A (0)\rangle_{\text{neq} }\right|_{T_\text{SA} = 0,\,T_\text{A} }&  = \frac{1}{2\pi \tau_c} \frac{(\pi T_\text{A} \tau_c)^\nu}{\sin^\nu [\pi T_\text{A}(\tau_c + it)]} \exp \left\{ -\frac{S_\text{A}}{\nu^2 e^2} [1 - \cos (2\pi\nu) ] | t|  - i \frac{I_\text{A}}{\nu e} \sin (2\pi\nu) t \right\},
\end{aligned}
\label{eq:psi_sa_ta0}
\end{equation}
which contains two $T_\text{A}$-dependent factors: the equilibrium correlation function in the prefactor, which explicitly depends on $T_\text{A}$, and the noise $S_\text{A}$ that is implicitly influenced by $T_\text{A}$.

When $T_\text{B} = T_\text{SA} = 0$, the corresponding distribution functions are provided in Fig.~\eqref{fig:different_temperatures}(i), for $T_\text{A} = 10 I_\text{A}/e$. In comparison to those at finite $T_\text{SA}$ [Fig.~\ref{fig:different_temperatures}(e)], distribution functions of channel A in Fig.~\eqref{fig:different_temperatures}(i) are flatter, indicating a stronger influence of thermal fluctuation resulting from finite $T_\text{A}$. This result agrees with the observation above: $T_\text{A}$ influences the correlation function in a two-fold way (explicitly and implicitly), while $T_\text{SA}$ influences the correlation function only implicitly, through the noise $S_\text{A}$.
We also show in Fig.~\eqref{fig:different_temperatures}(j) the spectral content of the heat current. Again, since $T_\text{B} = 0$, the energy-resolved heat current at all energies flow is from channel A to channel B.

When $T_\text{B}$ is finite, we plot distribution functions in Fig.~\ref{fig:different_temperatures}(k). Again, here distribution functions in A are flatter, in comparison to that in Fig.~\ref{fig:different_temperatures}(g), indicating a stronger thermal influence from $T_\text{A}$.
The corresponding differential heat flow is shown in Fig.~\ref{fig:different_temperatures}(l).

\subsection*{S6E. Effective temperature $T_\text{eff}$ of the non-equilibrium channel A at $T_A = 0$}

Before ending this section, we derive the relation between the noise $S_\text{A}$ of the non-equilibrium anyonic channel A, and its effective temperature $T_\text{eff}$.
Following Eq.~\eqref{eq:heat_time_integral} and using correlation functions Eq.~\eqref{eq:psi_a_ta0} for  the anyon operators in channel A and
\begin{equation}
\begin{aligned}
    \langle \psi^\dagger_\text{B} (t) \psi_\text{B} (0)\rangle_0 &= \frac{1 }{2\pi \tau_c} \frac{(\pi T_\text{B} \tau_c)^\nu}{ \sin^\nu \left[ \pi T_\text{B} (\tau_c + i t) \right]  } e^{i\nu e V_\text{B}},\\
    \langle \psi_\text{B} (t) \psi_\text{B}^\dagger (0)\rangle_0 &= \frac{1 }{2\pi \tau_c} \frac{(\pi T_\text{B} \tau_c)^\nu}{ \sin^\nu \left[ \pi T_\text{B} (\tau_c + i t) \right]  } e^{-i\nu e V_\text{B}},
\end{aligned}
\end{equation}
for the anyon operators in channel B, 
after setting $V_\text{B} = V_\text{eff} = I\sin(2\pi\nu)/\nu^2 e^2$, we write the condition of zero heat current for $T_\text{B} = T_\text{eff}$ as
\begin{equation}
   \int \! dt \! \left\{ \frac{\partial}{\partial t} \left[ \frac{e^{-\frac{S_\text{A}}{\nu^2 e^2} [1 - \cos(2\pi\nu)] |t| }}{(\tau_c + it)^\nu} \right] \frac{1}{\sin^\nu[\pi T_{\text{eff}} (\tau_c + i t)]} -  \frac{e^{-\frac{S_\text{A}}{\nu^2 e^2} [1 - \cos(2\pi\nu)] |t| }}{(\tau_c + it)^\nu}  \frac{\partial}{\partial t} \left[\frac{1}{\sin^\nu[\pi T_{\text{eff}} (\tau_c + i t)]} \right] \right\} = 0.
   \label{eq:zero_heat_current_origin}
\end{equation}
The equation above simplifies into
\begin{equation}
    \int dt' \frac{e^{-\frac{s_c}{\pi} [1 - \cos(2\pi\nu)] |t'|}}{(\tau_c' + it')^\nu \sin^\nu (\tau_c' + it')} \left\{ -\frac{i\nu}{\tau_c' + it'} + \frac{i\nu}{\tan (\tau_c' + i t')}-\frac{s_c}{\pi} [1 - \cos(2\pi\nu)] \text{sgn}(t')  \right\} = 0,
    \label{eq:eff_temp_definition}
\end{equation}
where $\tau_c' \equiv \pi T_{\text{eff},\alpha} \tau_c$,\ $t' \equiv \pi T_{\text{eff}} t$, and
\begin{equation}
   s_c \equiv \frac{S_\text{A}}{\nu^2 e^2 T_{\text{eff}}}.
   \label{eq:sc}
\end{equation}
is a \emph{dimensionless constant} (given $\nu$ fixed), at which the integral in Eq.~\eqref{eq:eff_temp_definition} vanishes.
For the situation displayed in Fig.~1 of the main text, $\nu = 1/3$, we have $s_c \approx 0.174$.
In that case, the factor $s_c[1 - \cos(2\pi\nu)]/\pi = 0.083$ is a numerically small quantity, and we expand the exponential in Eq.~\eqref{eq:eff_temp_definition} to leading order in $s_c[1 - \cos(2\pi\nu)]/\pi$, which 
yields an approximate expression for $s_c$:
\begin{equation}
    s_c \approx 
    \frac{ \int dt'\dfrac{i \pi\nu}{(\tau_c' + i t')^{\nu +1}\sin^\nu (\tau_c' + i t') } 
    \Big[ (\tau_c' + i t')\cot (\tau_c' + i t') -1\Big]}
    {\int dt' \dfrac{1 - \cos(2 \pi\nu)}{(\tau_c' + i t')^{\nu+1 }\sin^\nu (\tau_c' + i t') } 
    \Big[ \tau_c'\ \text{sgn} (t') + i (1 - \nu) |t'|  + i\nu |t'| (\tau_c' + i t')\cot (\tau_c' + i t')\Big]} ,
    \label{eq:sc_approx}
\end{equation}
which equals $0.152$ for $\nu = 1/3$, close to the result $0.174$, obtained by numerically solving Eq.~\eqref{eq:eff_temp_definition}.
For smaller values of $\nu$, the factor $s_c[1 - \cos(2\pi\nu)]/\pi$ further decreases: following Eq.~\eqref{eq:sc_approx}, it approximately equals 0.064 and 0.052 for $\nu = 1/5$ and $1/7$, respectively. In comparison, their exact values, obtained by numerically solving Eq.~\eqref{eq:eff_temp_definition}, are equal to 0.087 and 0.076, respectively. 
Clearly, following Eqs.~\eqref{eq:eff_temp_definition} and \eqref{eq:sc}, the effective temperature
\begin{equation}
    T_{\text{eff}} = \frac{S_\text{A}}{\nu^2 e^2 s_c},
    \label{eq:teff_sc}
\end{equation}
is proportional to the noise $S_\text{A}$ of the non-equilibrium channel A.

Notice that the parameter $s_c$ enters Eq.~\eqref{eq:eff_temp_definition} only through the product $s_c[1-\cos(2\pi\nu)]$. Therefore, 
in the fermionic case, when $\nu = 1$ and hence $1-\cos(2\pi\nu)=0$, the left-hand-side of Eq.~\eqref{eq:eff_temp_definition} can only equal zero, when $s_c$ becomes infinitely large. Indeed, if $s_c$ is a finite constant at $\nu\to 1$, the left-hand-side of Eq.~\eqref{eq:eff_temp_definition} becomes $s_c$-independent, with a constant integral outcome $\approx -1.05 i$. Thus, 
\begin{equation}
    s_c\propto \frac{1}{1-\cos(2\pi \nu)}.
    \label{sc-propto}
\end{equation}
The divergence of $s_c$ for a fermionic system can also be inferred from its approximate expression, Eq.~\eqref{eq:sc_approx}, where the denominator vanishes when $\nu = 1$.
As a consequence, $T_\text{eff}$ vanishes for fermions at $\nu\to 1$, 
\begin{equation}
    T_\text{eff}\propto (1-\nu)^2,\quad \nu\to 1,
\end{equation}
as follows from Eq.~\eqref{eq:teff_sc} with $s_c\to \infty$ from Eq.~(\ref{sc-propto}).
The fact indicates that effective temperature, $T_\text{eff}$, originates exclusively from braiding of anyons, in full analogy with the braiding-induced effective chemical potential $V_\text{eff}$.

\section*{S7. Tunneling-current noise and witness function $\mathcal{W}$}

Experimentally, to measure the value of $\mathcal{W}$ [defined by Eq.~(10) of the main text], one needs to know the tunneling-current noise at the collider,
$$S_\text{T}=\int dt\, \langle \delta I_\text{T} (t) \delta I_\text{T} (0) \rangle,$$
where $\delta I_\text{T} \equiv \hat{I}_\text{T} - I_\text{T}$ describes fluctuations of the tunneling current.
In general, measuring $S_\text{T}$ is a demanding task. Here, we show, however, that in the considered setups [Fig.~1(a) of the main text], $S_\text{T}$ can be experimentally obtained by measuring the auto-correlation noise in channel B.
To demonstrate this, we start by decomposing the tunneling current noise into several components (assuming the collider is located at the position $x = L$):
\begin{equation}
\begin{aligned}
& \int dt\, \langle \delta I_\text{T} (t) \delta I_\text{T} (0) \rangle = \int dt\, \langle [ \delta I_\text{B} (x>L, t) - \delta I_\text{B} (x<L, t) ] 
\,[ \delta I_\text{B} (x>L, 0) - \delta I_\text{B} (x<L, 0) ] \rangle\\
& =  \int dt \, \langle \delta I_\text{B} (x>L, t) \, \delta I_\text{B} (x>L, 0) \rangle +  \int dt\, \langle \delta I_\text{B} (x<L, t) \, \delta I_\text{B} (x<L, 0) \rangle \\
 &-  \int dt\, \langle \delta I_\text{B} (x>L, t)\,  \delta I_\text{B} (x<L, 0) \rangle-  \int dt\, \langle \delta I_\text{B} (x<L, t) \, \delta I_\text{B} (x>L, 0) \rangle,
\end{aligned}
\label{eq:noise_expressions}
\end{equation}
where $\delta I_\text{B} \equiv \hat{I}_\text{B} - I_\text{B}$ denotes fluctuations of the current $I_\text{B}$ in channel B.
In the second line of Eq.~\eqref{eq:noise_expressions}, the first term is the current auto-correlation in channel B downstream of the collider and 
the second term is the noise upstream of the collider. We neglect this term, as it depends only on the temperature $T_\text{B}$.

Now we move to the last line of Eq.~\eqref{eq:noise_expressions}, to show that it is zero at \textit{zero temperature}.
We can simplify the first term of the last line of Eq.~\eqref{eq:noise_expressions} as follows:
\begin{equation}
\begin{aligned}
& \int dt\, \langle \delta I_\text{B} (x>L, t) \, \delta I_\text{B} (x<L, 0) \rangle & = \nu^2 e^2 \int dt \langle \delta \hat{n}_\text{p,B} (x>L, t)\,  \delta \hat{n}_\text{p,B}(x<L, 0) \rangle,
\end{aligned}
\end{equation}
where $\delta \hat{n}_\text{p,B}$ is the operator (hence, marked with a ``hat'') that refers to fluctuation (indicated by $\delta$) of the particle distribution in channel B. 
This correlation function can be further cast in the form of an integral over the energy:
\begin{equation}
\int d\epsilon\, \left[\langle  \hat{n}_\text{p,B} (x>L, \epsilon)\,  \hat{n}_\text{p,B} (x<L, \epsilon) \rangle - \langle  \hat{n}_\text{p,B} (x>L, \epsilon) \rangle\, \langle  \hat{n}_\text{p,B} (x<L, \epsilon) \rangle\right].
\label{eq:b_fluctuation}
\end{equation}
At zero temperature, Eq.~\eqref{eq:b_fluctuation} is only nonzero if $\epsilon < \nu e V_B$, i.e., below the Laughlin surface of channel B.
Indeed, otherwise (when $\epsilon > \nu e V_B$) both the first and second terms of Eq.~\eqref{eq:b_fluctuation} vanish, because there are no particles with corresponding energies in channel B.
However, when $\epsilon < \nu e V_B$, tunneling from A to B vanishes, since the hole distribution in B vanishes, i.e., $n_\text{h,B}(x<L, \epsilon) = \langle \hat{n}_\text{h,B}(x<L, \epsilon) \rangle = 0$.
As a consequence, at the collider, only tunneling from B to A modifies the particle distribution $n_\text{p,B}=\langle\hat{n}_\text{p,B}\rangle$, leading to
\begin{equation}
\langle  \hat{n}_\text{p,B} (x>L, \epsilon)  \, \hat{n}_\text{p,B}(x<L, \epsilon) \rangle = n_\text{p,B}(x<L, \epsilon)\, \big[ n_\text{p,B} (x<L, \epsilon) - \tilde{\mathcal{T}}_\text{C} \,n_\text{p,B} (x<L, \epsilon)\, n_\text{h,A} (x<L, \epsilon) \big],
\label{eq:average_op_prod}
\end{equation}
for $\epsilon < \nu e V_B$.
Here, $\tilde{\mathcal{T}}_\text{C}$ refers to the (renormalized) transmission observed in experiment, not $\mathcal{T}_\text{C}$ in the tunneling Hamiltonian.

On the other hand, we have
\begin{equation}
\begin{aligned}
 \langle \hat{n}_\text{p,B} (x<L, \epsilon) \rangle &= n_\text{p,B}(x<L, \epsilon),\quad \text{ if }\epsilon < \nu e V_B,\\
 \langle \hat{n}_\text{p,B} (x<L, \epsilon) \rangle &= 0\quad \text{otherwise};\\
\langle \hat{n}_\text{p,B} (x>L, \epsilon) \rangle &= n_\text{p,B} (x<L, \epsilon) - \tilde{\mathcal{T}}_\text{C}\, n_\text{p,B}(x<L, \epsilon)\, n_\text{h,A}(x<L, \epsilon), \quad \text{ if }\epsilon< \nu e V_B, \\
\langle \hat{n}_\text{p,B} (x>L, \epsilon) \rangle  &= \tilde{\mathcal{T}}_\text{C}\,n_\text{h,B}(x<L, \epsilon)\, n_\text{p,A}(x<L, \nu \epsilon)\quad \text{otherwise}.
\end{aligned}
\label{eq:averages}
\end{equation}
Following Eq.~\eqref{eq:averages}, 
the product $\langle  \hat{n}_\text{p,B} (x>L, \epsilon) \rangle\, \langle  \hat{n}_\text{p,B} (x<L, \epsilon) \rangle$ is only finite for $\epsilon < \nu e V_B$, where it is equal to the average of the product of operators.
We have thus arrived at the conclusion that, at zero temperature,
\begin{equation}
\int dt\, \langle \delta I_\text{B} (x>L, t) \, \delta I_\text{B} (x<L, 0) \rangle = 0.
\label{eq:ib_int}
\end{equation}
We emphasize that the vanishing of Eq.~\eqref{eq:ib_int} resorts to the fact that channel B is equilibrium upstream of the collider; otherwise, Eqs.~\eqref{eq:average_op_prod} and \eqref{eq:ib_int} will not be exactly equal. Following similar steps, we can prove that the last term of the last line of Eq.~\eqref{eq:noise_expressions} also vanishes:
\begin{equation}
\int dt\, \langle \delta I_\text{B} (x<L, t)\,  \delta I_\text{B} (x>L, 0) \rangle = 0.
\end{equation}
As a consequence, when $V_\text{B} $ is much larger than the experimental temperature, the tunneling-current noise in the setup shown in Fig.~1(a) of the main text can be obtained from measurements of the current auto-correlations in channel B [second line of Eq.~\eqref{eq:noise_expressions}].

\section*{S8. Distribution functions with non-equilibrium particle tunneling}

In both the main text and the Supplementary Information (see Secs.~S3 and S6; in particular,  Figs.~\ref{fig:distributions} and \ref{fig:different_temperatures}), we have shown a remarkable property of anyonic distributions. Specifically, if only time-domain braiding processes are included in the correlation functions, the particle and hole distribution functions are symmetric to each other with respect to the energy $I_\text{A} \sin(2\pi\nu)/\nu e$ that corresponds to the effective chemical potential $\nu e V_\text{eff}$.

%%%%%%%%%%%%%%%%%%%%%%%%%
\begin{figure}[h!]
  \includegraphics[width= 0.8 \linewidth]{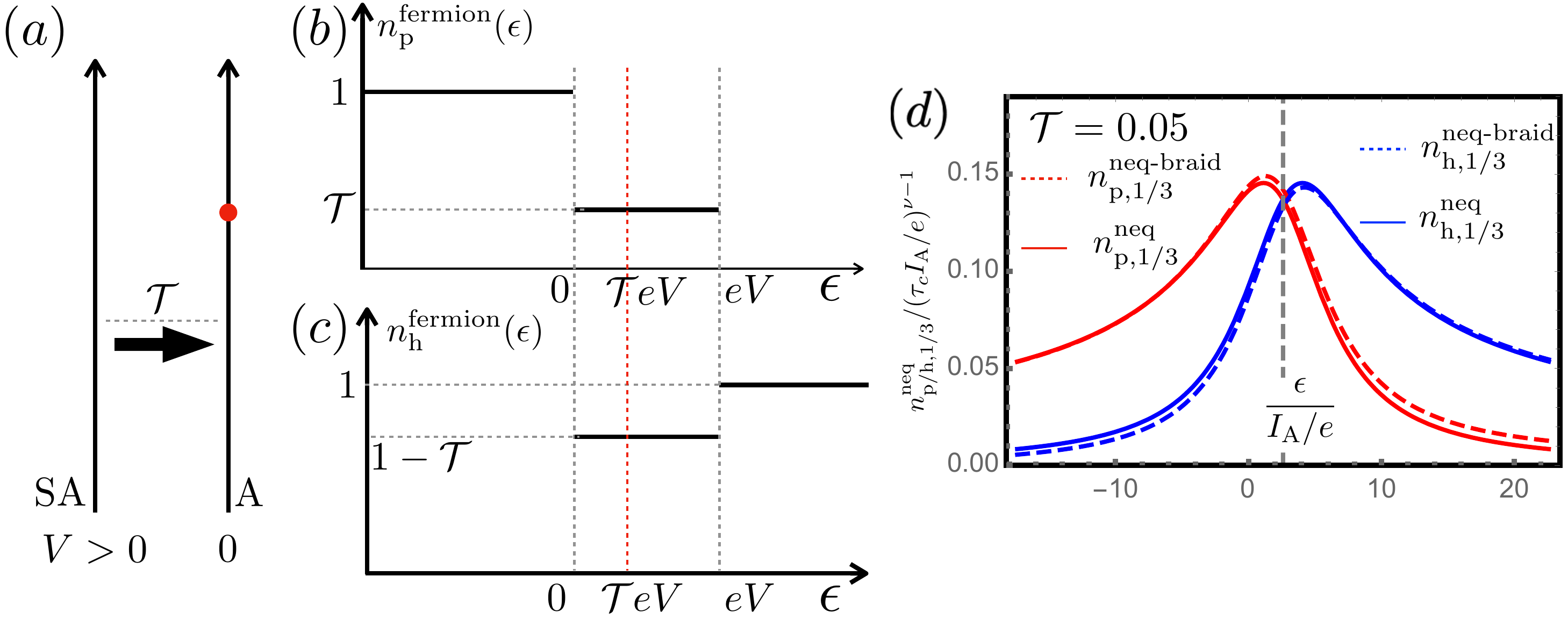}
  \caption{
  \textbf{Symmetry between particle and hole distribution functions.}
  \textbf{Panel (a)}: The structure we consider. The source SA has a finite voltage bias with respect to that of A. Channel A becomes non-equilibrium downstream of the diluter with transmission $\mathcal{T}_\text{A}$. We present the distribution functions at the position of the red dot after the diluter.
  \textbf{Panels (b) and (c)}: Particle and hole distribution functions in the fermionic case at zero temperature.
  The red dashed line highlights the modification of the effective chemical potential induced by an extra amount of non-equilibrium fermions. 
  \textbf{Panel (d)}: Particle and hole distribution functions of the $\nu=1/3$ anyonic channel for $\mathcal{T}_\text{A} = 0.05$.
  Here, the solid and dashed lines represent $n_{\text{p/h},\nu}^\text{neq}$ (all processes included) and $n_{\text{p/h},\nu}^\text{neq-braid}$ (only time-domain-braiding processes included), respectively.
  When $\mathcal{T}_\text{A}\ll 1$, the two curves agree very well.
 }
  \label{fig:asymmetric_distributions}
\end{figure}
%%%%%%%%%%%%%%%%%%%%%%%%%

This symmetry is, however, not a fundamental one: it is sabotaged by the processes of direct tunneling of non-equilibrium particles.
More specifically, 
the total distribution functions in the out-of-equilibrium channel [channel A in Fig.~\ref{fig:asymmetric_distributions}(a)] can be written as
\begin{equation}
\begin{aligned}
    n_{\text{p/h}}^\text{neq} \approx n_{\text{p/h}}^\text{neq-braid}
    +n_{\text{p/h}}^\text{neq-direct},
\end{aligned}
\label{eq:two_distribution_contributions}
\end{equation}
where the first term, $n_{\text{p/h},\nu}^\text{neq-direct}$, refers to the modification of particle/hole distributions directly by injection of non-equilibrium particles. 
when a particle state (fermionic or anyonic) with energy $\epsilon$ has probability $\mathcal{T}_\text{A}$ to receive one non-equilibrium particle, the corresponding distribution is modified simply by $n_{\text{p/h}}^\text{neq-direct} = \mathcal{T}_\text{A}$, i.e., the expected number of received non-equilibrium particles.

This modification is neglected in the evaluation of the current and noise in Refs.~\cite{SRosenowLevkivskyiHalperinPRL16, SLeeSimNC22, SLeeNature23}, where direct tunneling of anyons (between two non-equilibrium channels or between non-equilibrium and equilibrium ones) is considered to be less important than the tunneling of anyon-hole pairs (leading to time-domain braiding) in the situation of strong dilution.
During the time-domain-braiding processes, non-equilibrium particles supplied through diluters are not allowed to directly tunnel at the collider. Instead, they indirectly assist tunneling at the collider, by establishing nonlocal correlations between particle-hole tunneling events at the collider.
These anyon-hole tunneling processes produce the second second term, $n_{\text{p/h},\nu}^\text{neq-braid}$, in the nonequilibrium distribution function~\eqref{eq:two_distribution_contributions}. 
In our work, we also focus on the effect of such processes after assuming strong enough dilution.
Crucially, the symmetry between particle and hole distributions, thoroughly discussed in this work (in e.g., Figs.~\ref{fig:distributions} and \ref{fig:different_temperatures}), is only valid for  $n_{\text{p/h}}^\text{neq-braid}$. 
In this section, we explicitly show that the other contribution $n_{\text{p/h}}^\text{neq}$ will sabotage this symmetry.
For simplicity, we will address here the zero-temperature situation.

It is instructive to start with the fermionic case, $\nu = 1$, for which $n_{\text{p/h}}^\text{neq-braid} = 0$, since fermions do not produce a nontrivial braiding phase. The results for the particle and hole distributions at zero temperature (double-step distributions) are shown in Fig.~\ref{fig:asymmetric_distributions}(b) and Fig.~\ref{fig:asymmetric_distributions}(c), respectively. In these figures, the effective chemical potential is highlighted by the red dashed line. Clearly, for the fermionic case, 
the particle and hole distributions are, in general, not symmetric to each other with respect to the effective chemical potential. Actually, only for the special case $\mathcal{T}_\text{A} = 0.5$, the particle and hole distributions become symmetric to each other with respect to the red dashed line.

Now, we move to the anyonic case, where we include the direct contribution of non-equilibrium anyons to the distribution functions (again, for simplicity, at zero ambient temperature). For an anyonic channel A with non-equilibrium (diluted) current $I_\text{A}$, the full correlation functions read~\cite{f1}:
\begin{align}
\big\langle T_K \psi^\dagger_A (L,t^-) \psi_A (L,0^+) \big\rangle_\text{neq} &=  \frac{\tau_c^{\nu - 1}}{2\pi( \tau_c + it)^\nu} \left\{  \underbrace{e^{- \frac{I_\text{A}}{\nu e} [1 - \cos(2\pi\nu) ] |t| + i \frac{I_\text{A}}{\nu e} \sin(2\pi\nu) t}}_{\textbf{braid}}\right.\notag \\
+& \left. \underbrace{\frac{I_\text{A}}{e^2 V_\text{SA}} C(\nu) \left[  e^{i\nu e V_\text{SA} t} + i \text{sgn} (t) \sin(\pi\nu)  e^{- \frac{I_\text{A}}{\nu e} [1 - \cos(2\pi\nu) ] |t| + i \frac{I_\text{A}}{\nu e} \sin(2\pi\nu) t} \right] }_{\textbf{direct}} \right\},
\label{eq:positive_t_correlations-1}\end{align}
\begin{align}
\big\langle T_K \psi_A (L,t^-) \psi^\dagger_A (L,0^+) \big\rangle_\text{neq} &= \frac{\tau_c^{\nu - 1}}{2\pi( \tau_c + it)^\nu} \left\{ \underbrace{ e^{- \frac{I_\text{A}}{\nu e} [1 - \cos(2\pi\nu) ] |t| - i \frac{I_\text{A}}{\nu e} \sin(2\pi\nu) t} }_{\textbf{braid}} \right.\notag \\
+&\left. \underbrace{ \frac{I_\text{A}}{e^2 V_\text{SA}} C(\nu) \left[  e^{-i\nu e V_\text{SA} t} - i \text{sgn} (t) \sin(\pi\nu)  e^{- \frac{I_\text{A}}{\nu e} [1 - \cos(2\pi\nu) ] |t| - i \frac{I_\text{A}}{\nu e} \sin(2\pi\nu) t}\right]}_{\textbf{direct}} \right\},
\label{eq:positive_t_correlations}
\end{align}
where $V_\text{SA}$ is the bias in the source channel from which non-equilibrium current tunnels and
\begin{equation}
C(\nu) \equiv \tan\left( \pi\nu \right) \frac{\Gamma (1 - \nu)^2}{\Gamma (1-2\nu)} \frac{1}{\nu^2}.
\label{eq:cnu}
\end{equation}
In Eqs.~\eqref{eq:positive_t_correlations-1} and \eqref{eq:positive_t_correlations}, following the decomposition introduced by Eq.~\eqref{eq:two_distribution_contributions}, 
the first and second lines of each of the correlation functions refer to time-domain braiding and direct tunneling of non-equilibrium anyons, respectively.

The Fourier transformation yields the energy-dependent distribution functions:
\begin{align}
    & n_{\text{p,A}}^\text{neq}(\epsilon)  =n_{\text{p,A}}^\text{neq-braid}(\epsilon)+n_{\text{p,A}}^\text{neq-direct}(\epsilon) \notag \\
    & = \underbrace{\frac{\nu^{1-\nu}\Gamma (1 - \nu)}{\pi} \tau_c^{ \nu - 1}
    \text{Re} \left\{ e^{i\pi\nu/2} \left[ \frac{I_\text{A}}{e} \left(1 - \cos(2\pi\nu) \right) + i\frac{I_\text{A}}{e} \sin(2\pi\nu) - i \nu \epsilon \right]^{\nu-1} \right\}}_{\textbf{braid}}\notag  \\
    & + \underbrace{ \mathcal{T}_\text{A} \left( \frac{ \tau_c^{\nu - 1} (\nu e V_\text{SA} - \epsilon)^{\nu - 1}}{\Gamma (\nu)}\,  \Theta (-\epsilon + \nu e V_\text{SA}) + \frac{\nu^{1-\nu}\Gamma (1 - \nu)}{\pi} \tau_c^{ \nu - 1} \text{Im} \left\{ e^{i\pi\nu/2} \left[ \frac{I_\text{A}}{e} \left(1 - e^{-2i\pi\nu} \right) - i \nu \epsilon \right]^{\nu-1} \right\} \right)}_{\textbf{direct}},  
     \label{eq:full_neq_distributions-p}
    \\
    & n_{\text{h},A}^\text{neq}(\epsilon) = n_{\text{h},A}^\text{neq-braid}(\epsilon)+n_{\text{h},A}^\text{neq-direct}(\epsilon) \notag
    \\
     & = \underbrace{\frac{\nu^{1-\nu}\Gamma (1 - \nu)}{\pi} \tau_c^{ \nu - 1} 
    \text{Re}
    \left\{ e^{i\pi\nu/2} \left[ \frac{I_\text{A}}{e} \left(1 - \cos (2\pi\nu) \right) - i \frac{I_\text{A}}{e} \sin (2\pi\nu) + i \nu \epsilon \right]^{\nu-1} \right\}}_{\textbf{braid}}\notag \\
    & + \underbrace{\mathcal{T}_\text{A}\left( \frac{ \tau_c^{\nu - 1} ( \epsilon - \nu e V_\text{SA})^{\nu - 1}}{\Gamma (\nu)}\,  \Theta (\epsilon - \nu e V_\text{SA}) - \frac{\nu^{1-\nu}\Gamma (1 - \nu)}{\pi} \tau_c^{ \nu - 1} \text{Im} \left\{ e^{i\pi\nu/2} \left[ \frac{I_\text{A}}{e} \left(1 - e^{2i\pi\nu} \right) + i \nu \epsilon \right]^{\nu-1} \right\} \right)}_{\textbf{direct}},
    \label{eq:full_neq_distributions-h}
\end{align}
where terms corresponding to time-domain braiding and direct tunneling of non-equilibrium anyons, like Eq.~\eqref{eq:positive_t_correlations}, are explicitly marked out.
Here $\mathcal{T}_\text{A} \equiv I_\text{A} C(\nu)/e^2 V_\text{SA}$ refers to the experimentally measured transmission coefficient of the diluter.
In Eqs.~\eqref{eq:full_neq_distributions-p} and \eqref{eq:full_neq_distributions-h}, the first lines of $n_{\text{p,A}}^\text{neq}$ and $n_{\text{h,A}}^\text{neq}$ refer to the time-domain braiding contributions,
$n_{\text{p,A}}^\text{neq-braid}$ and $n_{\text{h,A}}^\text{neq-braid}$ of Eq.~\eqref{eq:current_distributions_ia_sa}, at zero temperature (i.e., when $S_\text{A} = \nu e I_\text{A}$).
Notice that as $\nu < 1$ and $\mathcal{T}_\text{A}\ll 1$, these terms are much more significant in comparison to the rest terms (with the mark ``direct'').
For the fermionic case, $\nu = 1$, these terms however vanish, which is consistent with the fact that fermions do not braid.
As another feature, remarkably, particle and hole distribution functions contributed by time-domain braiding are symmetric to each other, with respect to $\epsilon = I_\text{A} \sin(2\pi\nu)/\nu e$.
In great contrast, terms marked by ``direct'' have different symmetries.
Indeed, when consider the first term of the direct-tunneling contribution, particle and hole distributions are symmetric with respect to a different energy, $\epsilon = \nu e V_\text{SA}$. When moving to the second term (induced by direct tunneling), the particle and hole distributions are never symmetric. Instead, now they are \textit{anti-symmetric} with respect to $\epsilon = I_\text{A} \sin(2\pi\nu)/\nu e$, due to the difference in sign (in the prefactor).

We plot the anyonic distributions given by Eqs.~~\eqref{eq:full_neq_distributions-p} and \eqref{eq:full_neq_distributions-h} for $\nu=1/3$ in Fig.~\ref{fig:asymmetric_distributions}(d).
In this figure, the dashed lines represent distributions $n_{\text{p/h}}^\text{neq-braid}$,
where only time-domain braiding is included.
The solid lines represent the full distributions  $n_{\text{p/h}}^\text{neq}$, for a very small value of the diluter's transmission, $\mathcal{T}_\text{A} = 0.05$. 
Clearly, the inclusion of $n_{\text{p/h}}^\text{neq-direct}$ introduces a weak asymmetry to the previously symmetric particle and hole distribution functions.
This asymmetry is however negligible under the condition of strong dilution, where the full distribution $n_{\text{p/h}}^\text{neq}$ agrees remarkably well with $n_{\text{p/h}}^\text{neq-braid}$ induced by time-domain braiding.

It is also worth demonstrating explicitly how Eqs.~\eqref{eq:full_neq_distributions-p} and \eqref{eq:full_neq_distributions-h} correctly reduce to the double-step fermionic distributions (shown in Fig.~\ref{fig:asymmetric_distributions}b) in the $\nu = 1$ limit.
Consider, for instance, the particle distribution, Eq.~\eqref{eq:full_neq_distributions-p}. Making use of the singularity $\Gamma(1-\nu)\propto 1/(1-\nu)$ at $\nu\to 1$, 
we can keep only the terms proportional to $1-\nu$ within the curly brackets; higher-order terms vanish after taking the $\nu \to 1$ limit. 
As a result, Eq.~\eqref{eq:full_neq_distributions-p}
simplifies in the $\nu = 1$ limit, as follows: 
\begin{equation}
\begin{aligned}
    n_\text{p,A}^\text{neq}(\epsilon) \Big|_{\nu\to 1}& = \frac{1}{\pi}\lim_{\nu\to 1}\frac{1}{1-\nu}\text{Re} \left[ e^{i\pi\nu/2} e^{(\nu - 1)\ln (-i\nu \epsilon)}\right] + \mathcal{T}_\text{A} \Theta (-\epsilon + e V_\text{SA})\\
    &= \frac{1}{\pi}\lim_{\nu\to 1}\frac{1}{1-\nu}\text{Re} \left\{ \left[ -\frac{\pi}{2} (\nu - 1 ) + i \right] \left[ 1 + (\nu -1) \ln (-i\nu \epsilon) \right] \right\}+ \mathcal{T}_\text{A} \Theta (-\epsilon + e V_\text{SA})\\
    &= \frac{1}{\pi}\lim_{\nu\to 1}\frac{\text{Re} \left\{ \left[ -\frac{\pi}{2} (\nu - 1 ) + i \right] \left[ 1 + (\nu -1) \Theta (\epsilon) \left( - i\frac{\pi}{2} + \ln |\nu \epsilon|\right) + (\nu -1) \Theta (-\epsilon) \left( i\frac{\pi}{2} + \ln |\nu \epsilon|\right) \right] \right\}}{1-\nu}\\
    & \quad + \mathcal{T}_\text{A} \Theta (-\epsilon + e V_\text{SA})\\
    & =  \Theta (-\epsilon) + \mathcal{T}_\text{A} \Theta (-\epsilon + e V_\text{SA}),
\end{aligned}
\label{eq:particle_distribution_nu_1}
\end{equation}
reproducing the distribution function of the fermionic case. Here, the first and second terms correspond to the equilibrium and non-equilibrium contributions, respectively.
Notice that in Eq.~\eqref{eq:particle_distribution_nu_1}, $\ln (-i \nu \epsilon) = i\text{Arg} ( \nu \epsilon ) + \ln |\nu \epsilon|$ takes the principle value, i.e., with $\text{Arg} ( \nu \epsilon )$ between $-\pi$ and $\pi$. This requirement comes from the fact that in distributions of Eq.~\eqref{eq:full_neq_distributions-h}, the phase of expressions within curly brackets is assumed to be within the range $(-\pi,\pi)$.
Following Eq.~\eqref{eq:particle_distribution_nu_1}, the general formulas of distributions remain valid also for $\nu=1$. 
However, in the analysis of the fermionic case, it is, of course, natural to start with the well-known double-step distribution functions (Fig.~\ref{fig:asymmetric_distributions}b) or the correlation functions [Eqs.~\eqref{eq:positive_t_correlations-1} and \eqref{eq:positive_t_correlations}], instead of taking the limit $\nu\to 1$ in the general formulas.

We can now approximately evaluate the effective chemical potential, to the leading order in small $\mathcal{T}_\text{A}$, in the presence of direct tunneling of non-equilibrium anyons.
For simplicity, we assume zero ambient temperature for all channels.
This approximation is taken, as $V_\text{eff}$ is temperature-independent if including only time-domain braiding.
Strictly speaking, the inclusion of tunneling of non-equilibrium anyons also modifies the effective temperature---a conclusion that will be presented in detail in a coming work by the authors.
With distribution functions \eqref{eq:full_neq_distributions-p} and \eqref{eq:full_neq_distributions-h}, we obtain tunneling currents of Laughlin quasiparticles at the central collider,
\begin{align}
    I_{\text{A} \to \text{B}} & \simeq \nu e \mathcal{T}_\text{C} \frac{\tau_c^{2\nu-2}}{4\pi^2 } \Gamma (1 - 2\nu)
      \left(\underbrace{ 2\text{Re} \left\{ e^{i\pi\nu} \left[  2\sin^2(\pi\nu) \frac{I_\text{A}}{\nu e}+ i  \sin (2\pi\nu)\frac{I_\text{A}}{\nu e } -i \nu e V_\text{B} \right]^{2\nu-1} \right\} }_{\textbf{braid}}  \right.\notag\\
    & +\underbrace{ 2 \mathcal{T}_\text{A} \left( \nu e V_\text{SA} -\nu e V_\text{B} \right)^{2\nu -1}  \sin (2\pi\nu) \Theta (V_\text{SA} - V_\text{B})}_{\textbf{direct}}\notag\\[-0.5cm]
    & \left. \underbrace{ + 2 \mathcal{T}_\text{A} \sin (\pi\nu) \text{Im} \left\{ e^{i\pi\nu} \left[ 2\sin^2(\pi\nu) \frac{I_\text{A}}{\nu e}  + i \sin (2\pi\nu)\frac{I_\text{A}}{\nu e } -i \nu e V_\text{B} \right]^{2\nu-1} \right\}}_{\textbf{direct}}\right),
     \label{eq:general_currents_AB}
    \end{align}
    \begin{align}
    I_{\text{B} \to \text{A}} & \simeq \nu e \mathcal{T}_\text{C} \frac{\tau_c^{2\nu-2}}{4\pi^2 } \Gamma (1 - 2\nu)  \left(\underbrace{ 2\text{Re} \left\{ e^{i\pi\nu} \left[  2\sin^2(\pi\nu) \frac{I_\text{A}}{\nu e} - i \sin (2\pi\nu)\frac{I_\text{A}}{\nu e } + i \nu e V_\text{B} \right]^{2\nu-1} \right\}}_{\textbf{braid}} \right.\notag \\
    & + \underbrace{2 \mathcal{T}_\text{A} \left(- \nu e V_\text{SA} + \nu e V_\text{B} \right)^{2\nu -1}  \sin (2\pi\nu)  \Theta ( V_\text{B} - V_\text{SA})}_{\textbf{direct}}\notag \\[-0.5cm]
    & \left. \underbrace{ - 2 \mathcal{T}_\text{A} \sin (\pi\nu) \text{Im} \left\{ e^{i\pi\nu} \left[  2\sin^2(\pi\nu) \frac{I_\text{A}}{\nu e} - i \sin (2\pi\nu)\frac{I_\text{A}}{\nu e } + i \nu e V_\text{B} \right]^{2\nu-1} \right\}}_{\textbf{direct}} \right),
    \label{eq:general_currents_BA}
\end{align} 
from current from A to B ($I_{\text{A} \to \text{B}}$), and
that of the opposite direction ($I_{\text{B} \to \text{A}}$), respectively.
Again, here we explicitly mark out contributions induced by time-domain braiding, and direct tunneling of non-equilibrium anyons, respectively.
It is worth mentioning that Eqs.~\eqref{eq:general_currents_AB} and \eqref{eq:general_currents_BA}
can be alternatively obtained directly from the correlation functions in the time domain [i.e., Eq.~\eqref{eq:positive_t_correlations} for the non-equilibrium channel A, and Eq.~(2) of the main text (after taking $T_\text{B} = 0$)].
Notice that Eq.~\eqref{eq:general_currents_BA} is strictly only valid for $\nu < 1/2$, as terms $\propto (\tau_c I_\text{A}/\nu e)^{1-2\nu}$ have been neglected in the integral over $\epsilon$. Discussion on the fermionic case will be provided at the end of this section.

Noticeably, 
Eqs.~\eqref{eq:general_currents_AB} and \eqref{eq:general_currents_BA}, obtained for Laughlin quasiparticles (i.e., for $\nu \le 1/3$), display features that are distinct from those in the fermionic case [cf. discussions around Eq.~\eqref{eq:positive_t_fermionic_correlations}].
Firstly, for Laughlin quasiparticles, inter-channel tunneling is dominated by the modification of distribution functions by time-domain braiding processes.
Clearly, the contributions of the first lines to both $I_{\text{A} \to \text{B}}$ and $I_{\text{B} \to \text{A}}$ are much larger than those of the third lines, as $\mathcal{T}_\text{A} \ll 1$.
The comparison between amplitudes of the first and second lines is more involved.
For Laughlin quasiparticles, $2 \nu - 1 < 0$ (e.g., for $\nu = 1/3$).
When $V_\text{B} \approx I_\text{A} \sin (2\pi\nu)/\nu^2 e^2$ (corresponding to $V_\text{eff}$ induced by braiding), the first line becomes much larger than the second line, because  $(I_\text{A}/e)^{2\nu - 1} \gg (eV_\text{SA})^{2\nu - 1}$ for systems in the strongly diluted limit (this is in addition to a small prefactor $\mathcal{T}_\text{A}\ll 1$ in the second line).
Consequently, for Laughlin quasiparticles, the condition for finding the effective potential is dominated by the current that is induced by time-domain braiding.

To gain a more intuitive understanding, we expand the currents in 
$$\delta V_\text{B} \equiv V_\text{B} - I_\text{A} \sin (2\pi\nu)/\nu^2 e^2,$$ which refers to the correction to $V_\text{eff}$ that is introduced by tunneling of non-equilibrium anyons. Since $\mathcal{T}_\text{A}\ll 1$, we take $V_\text{B} < V_\text{SA}$. For Laughlin quasiparticles, Eqs.~\eqref{eq:general_currents_AB} and \eqref{eq:general_currents_BA}
then become
\begin{equation}
\begin{aligned}
    I_{\text{A} \to \text{B}} & \approx \nu e \mathcal{T}_\text{C} \frac{\tau_c^{2\nu-2}}{4\pi^2 } \Gamma (1 - 2\nu)\left( 2\text{Re} \left\{ e^{i\pi\nu} \left[ 2\sin^2(\pi\nu)\frac{I_\text{A}}{\nu e} \right]^{2\nu-1} \left[ 1 -  \frac{i (2\nu-1)\nu e}{2 \sin^2 (\pi\nu)I_\text{A} }\, \nu e \delta V_\text{B} \right] \right\}   \right.\\
    & + 2 \mathcal{T}_\text{A} \left( \nu e V_\text{SA} -\nu e V_\text{eff} \right)^{2\nu -1}  \sin (2\pi\nu)  \\
    & \left. + 2 \mathcal{T}_\text{A} \sin (\pi\nu) \text{Im} \left\{ e^{i\pi\nu} \left[ 2\sin^2(\pi\nu)\frac{I_\text{A}}{\nu e} \right]^{2\nu-1}\left[  1 -  \frac{i (2\nu-1)\nu e}{ 2 \sin^2 (\pi\nu)I_\text{A}}  \, \nu e \delta V_\text{B}\right] \right\} \right),\\
    I_{\text{B} \to \text{A}} & \approx\nu e \mathcal{T}_\text{C} \frac{\tau_c^{2\nu-2}}{4\pi^2 }\Gamma (1-2\nu) \left( 2\text{Re} \left\{ e^{i\pi\nu} \left[ 2\sin^2(\pi\nu)\frac{I_\text{A}}{\nu e} \right]^{2\nu-1} \left[ 1 +   \frac{i (2\nu-1) \nu e}{ 2 \sin^2 (\pi\nu)I_\text{A}} \, \nu e \delta V_\text{B}\right] \right\}  \right.\\
    & \left. - 2 \mathcal{T}_\text{A} \sin (\pi\nu) \text{Im} \left\{ e^{i\pi\nu} \left[ 2\sin^2(\pi\nu)\frac{I_\text{A}}{\nu e} \right]^{2\nu-1}\left[  1 +  \frac{i (2\nu-1) \nu e}{ 2 \sin^2 (\pi\nu)I_\text{A}}\, \nu e \delta V_\text{B}  \right] \right\} \right).
\end{aligned} 
\label{eq:approximate_general_currents}
\end{equation}

By requiring $I_{\text{A} \to \text{B}} = I_{\text{B} \to \text{A}}$, and keeping only leading order in $\mathcal{T}_\text{A}$ and $\nu^2 e^2\delta V_\text{B}/[I_\text{A} \sin(2\pi\nu)]$, we obtain the approximate value of $\delta V_\text{B}$  via
\begin{equation}
\begin{aligned}
    & 2\sin (\pi\nu) (1 - 2\nu ) \left[  2 \sin^2 (\pi\nu) \frac{I_\text{A}}{\nu e}\right]^{2\nu - 1} \frac{\nu^2 e^2 \delta V_\text{B}}{2 \sin^2 (\pi\nu)I_\text{A}} \\
    &\approx  \mathcal{T}_\text{A} (\nu e V_\text{SA})^{2\nu - 1} \sin (2\pi\nu) + 2 \mathcal{T}_\text{A} \sin (\pi\nu)^2 \left[  2 \sin^2 (\pi\nu) \frac{I_\text{A}}{\nu e}\right]^{2\nu - 1}\\ &\approx  2 \mathcal{T}_\text{A} \sin (\pi\nu)^2 \left[  2 \sin^2 (\pi\nu) \frac{I_\text{A}}{\nu e} \right]^{2\nu - 1},
\end{aligned}
\end{equation}
where, at the last step, we have again assumed $2\nu - 1 <0$ and $V_\text{SA} \gg 2\sin^2(\pi\nu) I_\text{A}/\nu^2e^2$.
As a result, to leading order in $\mathcal{T}_\text{A} $,
\begin{equation}
   \delta V_\text{B}  \simeq \mathcal{T}_\text{A} \frac{2\sin^3(\pi\nu) I_\text{A}}{(1-2\nu)\nu^2 e^2 } = \mathcal{T}_\text{A} V_\text{eff}\frac{\sin(\pi\nu) \tan(\pi\nu)}{1-2\nu} \propto \mathcal{T}_\text{A} V_\text{eff},
   \label{eq:anyonic_delta_vb}
\end{equation}
which indicates that $\delta V_\text{B}$ is indeed much smaller than $V_\text{eff} = I_\text{A} \sin(2\pi\nu)/\nu^2 e^2$ in the $\mathcal{T}_\text{A} \ll 1$ limit, as long as $\nu$ does not approach $1/2$ (which is the case for Laughlin filling fractions).

Now, for the sake of comparison, we return to the fermionic case, $\nu = 1$, where the correlation functions \eqref{eq:positive_t_correlations} read as
\begin{equation}
\begin{aligned}
\big\langle T_K \psi^\dagger_A (L,t^-) \psi_A (L,0^+) \big\rangle_\text{neq}\Big|_{\nu=1} =&  \frac{1}{2\pi( \tau_c + it)} \left( 1 +  2\pi \frac{I_\text{A}}{e^2 V_\text{SA}}   e^{i e V_\text{SA} t}  \right),\\
\big\langle T_K \psi_A (L,t^-) \psi^\dagger_A (L,0^+) \big\rangle_\text{neq}\Big|_{\nu=1}  =& \frac{1}{2\pi( \tau_c + it)} \left( 1 + 2\pi \frac{I_\text{A}}{e^2 V_\text{SA}}   e^{-i e V_\text{SA} t}  \right).
\end{aligned}
\label{eq:positive_t_fermionic_correlations}
\end{equation}
Here, the first term in the particle and hole distributions corresponds to the first lines of Eq.~\eqref{eq:positive_t_correlations}, after taking $\nu = 1$, which is an outcome of the absence of time-domain braiding for fermions.
As a consequence, the tunneling currents of different directions become (when channel B has zero temperature)
\begin{equation}
\begin{aligned}
    I_{\text{A} \to \text{B}} \big|_{\nu =  1} & = e \mathcal{T}_\text{C} \frac{I_\text{A}}{e^2}  ,\\
    I_{\text{B} \to \text{A}} \big|_{\nu =  1}& = e \mathcal{T}_\text{C}  \frac{1}{2\pi} V_\text{B}.
\end{aligned}
\label{eq:currents_fermionic}
\end{equation}
where we have taken $V_\text{B} > 0$ and $V_\text{SA} > V_\text{B}$.
In comparison to the anyonic counterpart  [Eq.~\eqref{eq:approximate_general_currents}], $I_{\text{A} \to \text{B}}$ contains now only the second line of the anyonic version, Eq.~\eqref{eq:approximate_general_currents}, corresponding to the tunneling of diluted non-equilibrium anyons (injected from the source SA) at the central collider. On the other hand, $I_{\text{B} \to \text{A}}$ corresponds to the first line of its anyonic counterpart. We then find 
\begin{equation}
    V_\text{eff}\big|_{\nu=1} = 2\pi I_\text{A}/e^2,
    \label{eq:veff_fermionic}
\end{equation}
a result proportional to $\mathcal{T}_\text{A}$.

Note that $V_\text{eff}|_{\nu=1}$, which originates from direct tunneling of diluted non-equilibrium particles, strongly differs from Eq.~\eqref{eq:anyonic_delta_vb}, where the inclusion of tunneling of non-equilibrium anyons merely introduces corrections, $\propto \mathcal{T}_\text{A}^2$, to $V_\text{eff}$.
This difference lies in the fact that for the anyonic case, i.e., in Eq.~\eqref{eq:approximate_general_currents}, both $I_{\text{A} \to \text{B}}$ and $I_{\text{B} \to \text{A}}$ are dominated by time-domain-braiding processes, which are absent for fermions.
Notice that Eq.~\eqref{eq:veff_fermionic} remains valid even under a finite temperature in channel B.
One can thus likewise define an effective temperature for a non-equilibrium fermionic channel, by enforcing the heat current to vanish.
When $\mathcal{T}_\text{A}$ is small, this effective temperature is given by
\begin{equation}
    T_\text{eff} \Big|_{\nu = 1} \approx \frac{\sqrt{3}}{\pi} \sqrt{\mathcal{T}_\text{A}} e V_\text{SA} = 2\sqrt{\frac{3}{\mathcal{T}_\text{A}}} \frac{I_\text{A}}{e},
\end{equation}
which depends not only on the non-equilibrium current $I_\text{A}$, but also on details of the diluter, i.e., the transmission $\mathcal{T}_\text{A}$.

In great contrast, as shown by Fig.~2 of the main text, if including only time-domain braiding for the anyonic case (where such processes are dominant), the effective temperature depends on the ambient temperature, and more remarkably, universally on the noise $S_\text{A}$ in the non-equilibrium channel.
This difference in effective temperatures is thus another crucial factor to distinguish a non-equilibrium anyonic channel from a fermionic one.

\section*{S9. Effective versus real temperatures and potentials}
\label{sec:effective_vs_real}

In the main text, we have shown that a non-equilibrium current $I_\text{A}$ introduces both an effective temperature and an effective shift of the chemical potential. Their manifestation in the distribution functions of non-equilibrium anyons is, however, not identical to that of real temperature and chemical potential.
In this section, we illustrate this point, by comparing the particle distributions in the two situations: (i) Distribution $n_\text{p,A}^\text{neq-braid} (\epsilon, I_\text{A}; T_\text{A} = 0, T_\text{SA} = 0,)$ of channel A for zero ambient temperature and finite non-equilibrium current $I_\text{A}$, characterized by the effective temperature $T_\text{eff}$ and an effective shift of the chemical potential $V_\text{eff}$. (ii) Distribution $n_\text{p,B}^\text{eq} (\epsilon,V_\text{B};T_\text{B})$ of an equilibrium channel B at a real temperature chosen to satisfy $T_\text{B} = T_\text{eff}$ and a real potential $V_\text{B} = V_\text{eff}$.

%%%%%%%%%%%%%%%%%%%%%%%%%
\begin{figure}[h!]
  \includegraphics[width= 0.5 \linewidth]{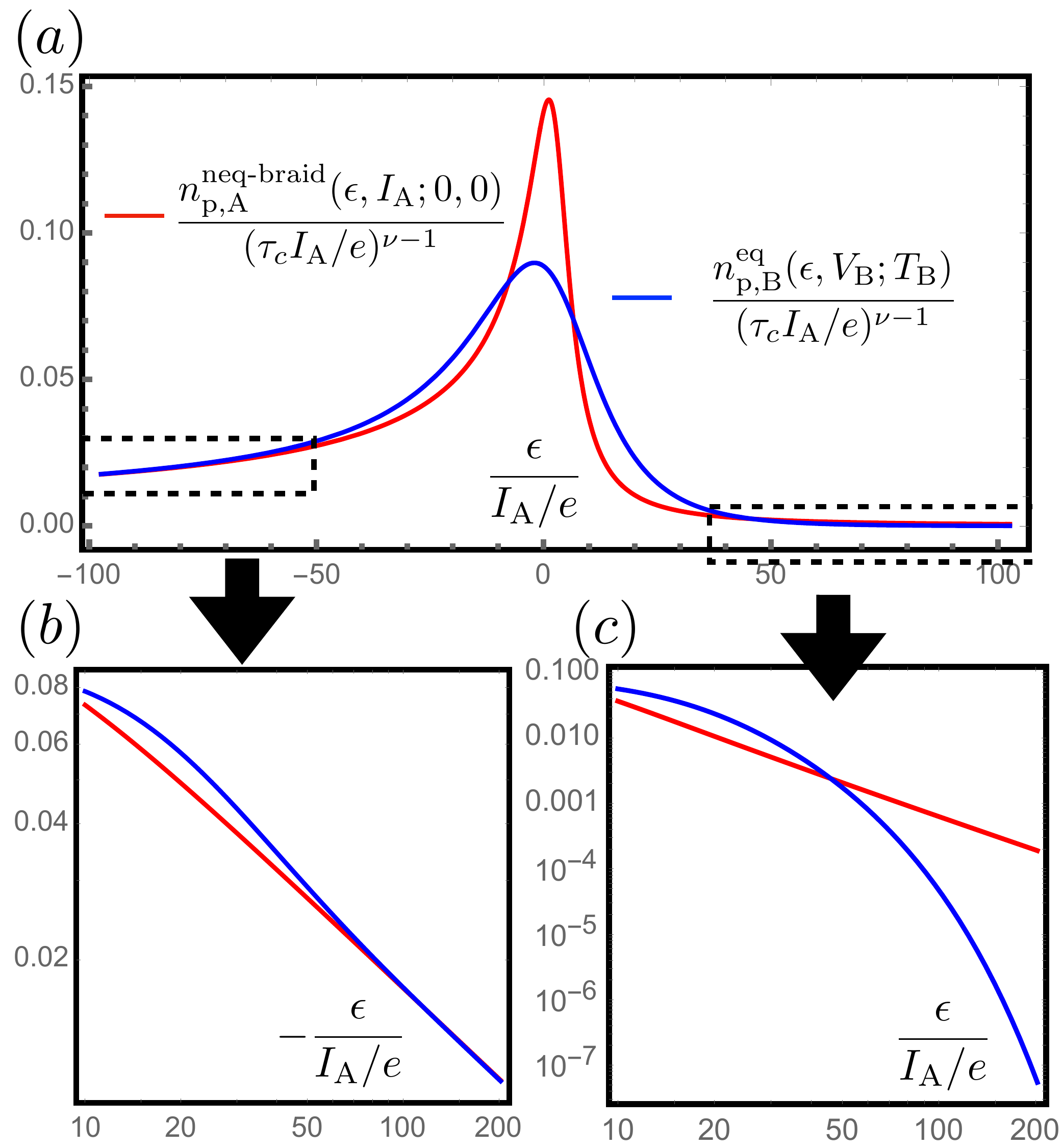}
  \caption{
  \textbf{Particle distribution functions for non-equilibrium (red curves) and equilibrium (blue curves) anyons.} Here, $V_\text{B}$ and $T_\text{B}$ of the equilibrium channel B are chosen to be equal to the effective potential $V_\text{eff}$ and effective temperature $T_\text{eff}$ of the non-equilibrium channel A. \textbf{Panel (a)}: The \textit{landscapes} of these two distributions. The asymptotic features are shown in \textbf{panels (b) and (c)} for (log-log plots of) negative and positive energies, respectively.
  The thermal particle distribution in channel B decays exponentially with increasing energy for $\epsilon>0$, while the non-equilibrium distribution in channel A decays in a power-law manner.
 }
  \label{fig:effective_vs_real}
\end{figure}
%%%%%%%%%%%%%%%%%%%%%%%%%

The plots of particle distributions in cases (i) and (ii) are shown in Fig.~\ref{fig:effective_vs_real}, with red and blue curves, respectively. Figure~\ref{fig:effective_vs_real}(a) shows the overall landscape of particle distribution functions, comparing the non-equilibrium distribution driven by a dilute current $I_A$ in channel A (red curve) with the equilibrium anyonic distribution in channel B (blue curve) at real potential and temperature $V_\text{B}$ and $T_\text{B}$ chosen to be equal to their effective counterparts in channel A.
The equilibrium distribution has a smaller peak height, in comparison to that of the non-equilibrium one.
Further, Figs.~\ref{fig:effective_vs_real}(b) and \ref{fig:effective_vs_real}(c) display the log-log plots of the asymptotic behavior of distribution functions, for negative and positive energies, respectively.
Based on these figures, although the two curves nicely agree with each other for negative energies, they are clearly distinct for positive energies. Indeed, the non-equilibrium distribution function decreases in a power-law manner with increasing energy [red curve in Fig.~\ref{fig:effective_vs_real}(c)] -- in great contrast to the equilibrium one that decreases exponentially [blue curve in Fig.~\ref{fig:effective_vs_real}(c)].

\section*{S10. Johnson-Nyquist noise}

In this section, we derive the tunneling current noise $S_\text{T}$
at the collider that connects the non-equilibrium channel A and the equilibrium channel B.
Previous discussions on $S_\text{T}$ in Sec.~S7 focuses on its experimental accessibility.
In this section, we write $S_\text{T}$ in the form of a Johnson-Nyquist noise, to show that $T_\text{eff}$ indeed has the function of temperature.

The expression for the tunneling-current noise involves the two products of correlation functions:
\begin{equation}
\begin{aligned}
    \langle \psi_\text{A}^\dagger (t) \psi_\text{A} (0)\rangle_\text{neq} \langle \psi_\text{B} (t) \psi_\text{B}^\dagger (0)\rangle_0 =& \frac{1}{4\pi^2} \frac{\tau_c^{2\nu-2} (\pi T_\text{B})^\nu}{(\tau_c + i t)^{\nu} \sin^\nu \left[ \pi T_\text{B} (\tau_c + i t) \right]  } \\
    \times & \exp \left\{ -\frac{S_\text{A}}{\nu^2 e^2} [1 - \cos (2\pi\nu) ] | t|  + i \left[ \frac{I_\text{A} }{\nu e} \sin (2\pi\nu) -\nu e V_\text{B}\right] t \right\},\\
    \langle \psi_\text{A} (t) \psi_\text{A}^\dagger (0)\rangle_\text{neq} \langle \psi_\text{B}^\dagger (t) \psi_\text{B} (0)\rangle_0 =& \frac{1}{4\pi^2} \frac{\tau_c^{2\nu-2} (\pi T_\text{B})^\nu}{(\tau_c + i t)^{\nu} \sin^\nu \left[ \pi T_\text{B} (\tau_c + i t) \right]  }\\
    \times & \exp \left\{ -\frac{S_\text{A}}{\nu^2 e^2} [1 - \cos (2\pi\nu) ] | t|  - i \left[ \frac{I_\text{A} }{\nu e} \sin (2\pi\nu) -\nu e V_\text{B}\right] t \right\}.
\end{aligned}
\label{eq:correlations_setup2}
\end{equation}
With Eq.~\eqref{eq:correlations_setup2}, the charge conductance [for $V_\text{B} = V_\text{eff} \equiv I_\text{A} \sin(2\pi\nu)/\nu^2 e^2$] reads
\begin{equation}
\begin{aligned}
    G_\text{T} & \!=\! \partial_{(V_\text{eff} - V_\text{B})} I_\text{T} \Big|_{V_\text{B} = V_\text{eff}} \!\!\!\!=\! 2 \mathcal{T}_\text{C} \nu^2 e^2 \!\!\int \! dt \frac{1}{4\pi^2} \frac{i t\, \tau_c^{2\nu-2}  (\pi T_\text{B})^\nu}{(\tau_c \!+\! i t)^{\nu} \sin^\nu \left[ \pi T_\text{B} (\tau_c \!+\! i t) \right]  } \exp \left\{ -\frac{S_\text{A}}{\nu^2 e^2} [1 \!-\! \cos (2\pi\nu) ] | t| \right\}.
\end{aligned}
\end{equation}
When, in addition, the real and effective temperatures are equal, $T_\text{B} = T_\text{eff} = S_\text{A}/(\nu^2 e^2 s_c)$, the conductance takes the form
\begin{equation}
\begin{aligned}
    G_\text{T}\Big|_{T_\text{B} = T_\text{eff}} & 
    = 2 \mathcal{T}_\text{C} \nu^2 e^2 \!\!\int \! dt \frac{1}{4\pi^2} \frac{i t\, \tau_c^{2\nu-2} (\pi T_\text{B})^\nu}{(\tau_c \!+\! i t)^{\nu} \sin^\nu \left[ \pi T_\text{B} (\tau_c \!+\! i t) \right]  } \exp \left\{ -T_\text{B} s_c [1 \!-\! \cos (2\pi\nu) ] | t| \right\}\\
    & = \mathcal{T}_\text{C} \nu^2 e^2 \frac{(\pi T_\text{B}\tau_c)^{2\nu - 2}}{2\pi^2}   \!\!\int \! dt' \frac{i t' }{(\tau_c' + i t')^{\nu} \sin^\nu (\tau_c' + i t')   } \exp \left\{ -\frac{s_c [1 - \cos (2\pi\nu) ]}{\pi} | t'| \right\}\\
    & = \nu^2 e^2\mathcal{T}_\text{C} \frac{(\pi T_\text{B} \tau_c)^{2\nu - 2}}{2\pi^2} g_1 (\nu),
\end{aligned}
\label{eq:conductance_integral}
\end{equation}
where 
\begin{equation}
    g_1(\nu)  \equiv \int dt' \frac{i t' }{(\tau_c' + i t')^{\nu} \sin^\nu (\tau_c' + i t')   } \exp \left\{ -\frac{s_c [1 - \cos (2\pi\nu) ]}{\pi} | t'| \right\},
\end{equation}
is a dimensionless constant that depends only on the filling factor $\nu$, and we introduced dimensionless variables $ \tau_c' \equiv \pi T_{\text{eff}} \tau_c$ and $ t' \equiv \pi T_{\text{eff}} t$. 
With Eq.~\eqref{eq:conductance_integral}, we can write the renormalized transmission as
\begin{equation}
    \mathcal{T}_\text{C}^\text{renorm} \equiv \mathcal{T}_\text{C} \frac{(\pi T_\text{B} \tau_c)^{2\nu - 2}}{\pi} g_1 (\nu),
    \label{eq:effective_transmission}
\end{equation}
so that the conductance can be written as 
\begin{equation}
    G_\text{T} = \nu^2 e^2 \mathcal{T}_\text{C}^\text{renorm}/(2\pi).
    \label{GT-TC}
\end{equation}

The tunneling-current noise is given by
\begin{equation}
\begin{aligned}
    S_\text{T} & = \mathcal{T}_\text{C} \nu^2 e^2 \int dt \big[\langle \psi_\text{A}^\dagger (t) \psi_\text{A} (0)\rangle_\text{neq} \langle \psi_\text{B} (t) \psi_\text{B}^\dagger (0)\rangle_0 + \langle \psi_\text{A} (t) \psi_\text{A}^\dagger (0)\rangle_\text{neq} \langle \psi_\text{B}^\dagger (t) \psi_\text{B}(0)\rangle_0 \big]
    \\
    &= 2 \mathcal{T}_\text{C} \nu^2 e^2 \int dt \frac{1}{4\pi^2} \frac{\tau_c^{2\nu-2} (\pi T_\text{B})^\nu}{(\tau_c + i t)^{\nu} \sin^\nu \left[ \pi T_\text{B} (\tau_c + i t) \right]  } \exp \left\{ -\frac{S_\text{A}}{\nu^2 e^2} [1 - \cos (2\pi\nu) ] | t| \right\}\\
    &=  \nu^2 e^2 \mathcal{T}_\text{C} \frac{(\pi T_\text{B} \tau_c)^{2\nu-2}}{2\pi} g_0 (\nu) T_\text{B},
\end{aligned}
\end{equation}
where
\begin{equation}
    g_0 (\nu) \equiv \int dt'  \frac{1}{(\tau_c' + i t')^{\nu} \sin^\nu \left( \tau_c' + i t' \right)  } \exp \left\{ -\frac{s_c [1 - \cos (2\pi\nu) ]}{\pi} | t'| \right\}
\end{equation}
is another dimensionless $\nu$-dependent quantity.
With the renormalized transmission $\mathcal{T}_\text{C}^\text{renorm}$ given by Eq.~\eqref{eq:effective_transmission}, the tunneling-current noise for $T_B=T_\text{eff}$
\begin{equation}
    S_\text{T} = \nu^2 \frac{g_0 (\nu)}{2 g_1 (\nu)} e^2 \mathcal{T}_\text{C}^\text{renorm}  T_\text{B} = \frac{\pi g_0 (\nu)}{g_1 (\nu)} G_\text{T} \, T_\text{eff},
\end{equation}
can be written in the conventional form for the Johnson-Nyquist noise, with the tunneling conductance $G_\text{T}$ given by Eq.~\eqref{GT-TC}.

\section*{S11. Tunneling between channels with different filling fractions}

In the main text, we show that the effective chemical potential of an out-of-equilibrium anyonic channel becomes meaningful only after specifying the charge of an individual tunneling quasiparticle.
This complexity arises because inter-edge tunneling can involve quasiparticles with different charges in anyonic systems.
This section extends the consideration of the main text and presents the effective chemical potential when channel A has a filling fraction $m_\text{A} \nu$. We also illustrate the statements of the main text with an example of tunneling between an edge with a Laughlin filling fraction and an integer edge.

\subsection*{S11A.~Tunneling between edges with filling factors $m_A\nu$ and $m_B \nu$}

We show in the main text that the effective chemical potential depends on the charge $m \nu e$ of particles tunneling between channels A and B that host quasiparticles with charge $\nu e$ and $m_\text{B} \nu e$, respectively.
In this section, we generalize this conclusion to the case when the filling fraction of channel A equals $m_\text{A} \nu $, with $m_\text{A}$ an integer, i.e., when the ratio of the filling factors of channels B and A is a rational (rather than integer) number.

In this case, the tunneling from A to B (see  Fig.~\ref{fig:different_fillings}) is described by the operator 
$$H_T\propto \mathcal{T}_\text{C} \exp [ -i( m \sqrt{\nu/m_\text{A}} \phi_\text{A} - m\sqrt{\nu/m_\text{B}} \phi_\text{B} ) ],$$
with $\phi_\text{A}$ and $\phi_\text{B}$ the bosonic fields in the two channels (in the main text, $m_\text{A}=1$). The correlation functions of the tunneling operators are now given by 
\begin{equation}
\begin{aligned}
    \Big\langle e^{-i\,m\sqrt{\frac{\nu}{m_\text{A}}} \phi_\text{A} (t) } \,e^{i\,m\sqrt{\frac{\nu}{m_\text{A}}} \phi_\text{A} (0)} \Big\rangle_\text{neq} &=
    \frac{1}{2\pi \tau_c} \left(\frac{\tau_c}{\tau_c + i t}\right)^{m^2\nu /  m_\text{A}}\!\!\!
    \exp\left(-\frac{I_\text{A} \left\{\left[ 1 - \cos\left(2m \pi\nu \right) \right] |t| - i \sin \left(2m \pi\nu \right) t  \right\}}{m_\text{A} \nu e} \right) ,\\
    \Big\langle e^{i\, m \sqrt{\frac{\nu}{m_\text{B}}} \phi_\text{B} (t) } \, e^{-i \,m \sqrt{\frac{\nu}{m_\text{B}}} \phi_\text{B} (0) }  \Big\rangle_0 &=
    \frac{1}{2\pi \tau_c} \frac{\left(\pi T_\text{B} \tau_c\right)^{m^2\nu /  m_\text{B}}}{\sin^{m^2\nu /  m_\text{B}} [\pi T_\text{B} (\tau_c + it) ] }
    \exp({-i\,m\,\nu\, e V_\text{B} \,t}),
\end{aligned}
\label{eq:different_filling_correlations}
\end{equation}
where we take the finite-temperature version (with temperature $T_\text{B}$) of correlation in channel B, for later convenience.

The net tunneling current between $\text{A}$ and $\text{B}$ vanishes, when (for all $T_\text{B}$ values)
\begin{equation}
    V_\text{B} = V_\text{eff} = I_\text{A} \sin\left( 2m\pi\nu \right)/(m m_\text{A} \nu^2 e^2),
    \label{eq:veff_different_fillings}
\end{equation}
corresponding to the vanishing of imaginary phases in the product of the two correlation functions in Eq.~\eqref{eq:different_filling_correlations}.
When $m_\text{A} = 1$, Eq.~\eqref{eq:veff_different_fillings} reduces to Eq.~(8) of the main text.
According to Eq.~\eqref{eq:veff_different_fillings},
$V_\text{eff}$ depends on the charge $m\nu e$ of quasiparticles that tunnel between channels A and B and on the filling fraction $m_A \nu$ of the non-equilibrium channel, but does not depend on the filling fraction $m_B \nu$ of the equilibrium channel.

%%%%%%%%%%%%%%%%%%%%%%%%%
\begin{figure}
  \includegraphics[width= 0.6 \linewidth]{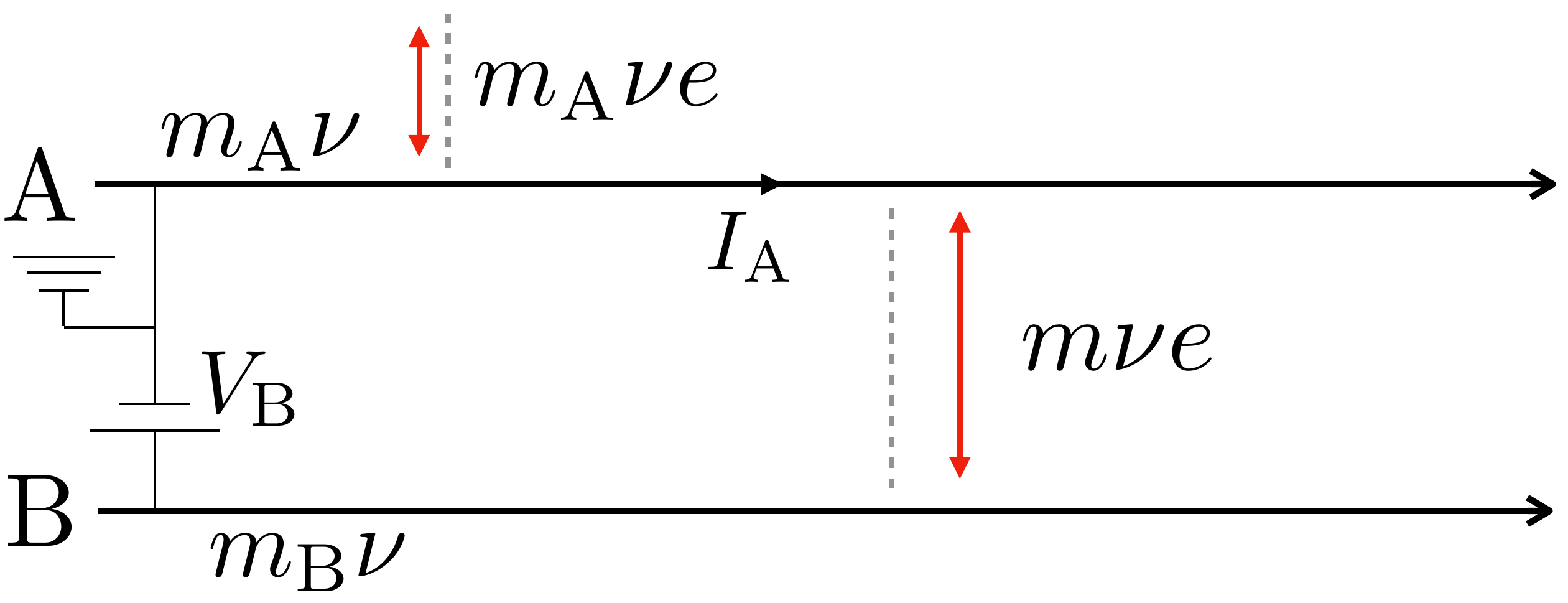}
  \caption{
  \textbf{Tunneling between two edges with different filling fractions.} The grounded anyonic channel A with filling fraction $m_\text{A} \nu$ carries a non-equilibrium current $I_\text{A}$ supplied by the diluter that transmits charge-$m_\text{A} \nu e$ quasiparticles from the not-shown source. The effective potential $V_\text{eff}$ of channel A is calibrated by an equilibrium channel B (with filling fraction $m_\text{B} \nu$ and bias $V_\text{B}$). These two channels communicate via a tunneling bridge transmitting quasiparticles of charge $m\nu e$.
  }
  \label{fig:different_fillings}
\end{figure}
%%%%%%%%%%%%%%%%%%%%%%%%%

As stated in the main text, the effective temperature also depends on the charge of quasiparticles that tunnel at the collider.
Indeed, with Eq.~\eqref{eq:different_filling_correlations} that accounts for different charges of quasiparticles, the requirement of zero heat current becomes [cf.~ Eq.~\eqref{eq:zero_heat_current_origin}]
\begin{equation}
\begin{aligned}
   \int \! dt \! \left\{  \frac{1}{\sin^{m^2 \nu/m_\text{B}}[\pi T_{\text{eff}} (\tau_c + i t)]} \frac{\partial}{\partial t} \left[ \frac{e^{-\frac{S_\text{A}}{m_\text{A}^2\nu^2 e^2} [1 - \cos(2\pi\nu)] |t| }}{(\tau_c + it)^{m^2\nu/m_\text{A}}} \right]\right.\\
   \left.-  \frac{e^{-\frac{S_\text{A}}{m_\text{A}^2\nu^2 e^2} [1 - \cos(2\pi\nu)] |t| }}{(\tau_c + it)^{m^2\nu/m_\text{A}}}  \frac{\partial}{\partial t} \left[\frac{1}{\sin^{m^2 \nu/m_\text{B}}[\pi T_{\text{eff}} (\tau_c + i t)]} \right] \right\} = 0,
\end{aligned}
\label{eq:effective_temperature_different_charge}
\end{equation}
where we have already taken $V_\text{B} = V_\text{eff} = I_\text{A} \sin\left( 2m\pi\nu \right)/(m m_\text{A} \nu^2 e^2)$.
Equivalently, we can write Eq.~\eqref{eq:effective_temperature_different_charge} in a dimensionless form, leading to
\begin{equation}
    \int dt' \frac{e^{-\frac{s_c}{m_\text{A}^2\pi} [1 - \cos(2\pi\nu)] |t'|}}{(\tau_c' + it')^{m^2\nu/m_\text{A}} \sin^{m^2\nu/m_\text{B}} (\tau_c' + it')} \left\{ -\frac{im^2\nu/m_\text{A}}{\tau_c' + it'} + \frac{im^2\nu/m_\text{B}}{\tan (\tau_c' + i t')}-\frac{s_c}{m_\text{A}^2\pi} [1 - \cos(2\pi\nu)] \text{sgn}(t')  \right\} = 0,
    \label{eq:eff_temp_definition-m}
\end{equation}
where we keep the definition of $s_c \equiv S_\text{A}/(\nu^2 e^2 T_\text{eff})$ [cf.~Eq.~\eqref{eq:sc}].
Equation~\eqref{eq:eff_temp_definition-m} discloses two features. First, it indicates that $T_\text{eff} \propto S_\text{A}$ remains valid, for given values of $m$, $m_\text{A}$ and $m_\text{B}$.
As the second feature, the value of $s_c$, obtained by solving Eq.~\eqref{eq:eff_temp_definition-m}, clearly depends on the values of $m$, $m_\text{A}$, and $m_\text{B}$.
As a consequence, the temperature-noise ratio $T_\text{eff}/S_\text{A}$ depends on the charge of quasiparticles, a feature shared by the effective chemical potential $V_\text{eff}$. Further, $T_\text{eff}$ also depends on the value of $m_\text{B}$, the charge carried by a single quasiparticle in channel B.
The latter dependence is, however, absent for $V_\text{eff}$, see  Eq.~\eqref{eq:veff_different_fillings}.

\color{black}

\subsection*{S11B.~Fermionic tunneling case}

In this section, we illustrate one of our key statements: the effective potential of the diluted anyonic channel depends on the type of quasiparticles that tunnel between the channel and the ``probe''.
Here, we consider a tunneling bridge characterized by $m\nu = 1$ (cf. Fig.~\ref{fig:different_fillings}). In this case, channels A and B communicate with each other by fermion tunneling. One can think, for example, of the tunneling between the channels through a ``vacuum'' (no fractional quasiparticles can be transmitted across such a barrier).
Further, when $m_A < 1/\nu$, this process corresponds to the so-called Andreev-like tunneling, where tunneling of a fermion of charge $e$ is accompanied by the reflection of another fractional charge.

Interestingly, when taking $m\nu =1$, the effective potential $V_\text{eff}$, given by  Eq.~\eqref{eq:veff_different_fillings}, equals zero. This fact indicates that chemical potential arising from time-domain braiding is zero for Andreev-like tunneling \cite{SSandler1998, SKaneFisherPRB03, SPierreNC23, SGuX2023Andreev}. This observation agrees with the fact that tunneling fermions do not ``braid'' with nonequilibrium anyons. Indeed, similar to non-interacting fermionic systems, the generation of an effective chemical potential now only originates from the direct tunneling of nonequilibrium particles at the collider.

The absence of the contribution of the time-domain braiding to the effective chemical potential is consistent with the discussion in Sec.~S3.
Indeed, although non-equilibrium distributions induced by time-domain braiding arise upstream of the collider, the detection of these distributions, as well as the induced effective chemical potential, would require (i) coupling the non-equilibrium channel to another channel with the same filling fraction $\nu$ (ii) via quasiparticles with charge $\nu e$. However, the distributions and effective potential induced by time-domain braiding are not captured with Andreev-like tunneling processes: the tunneling probe should be directly coupled to the quasiparticles whose non-equilibrium distribution is studied.

\section*{S12.~Composite edges: tunneling between $\nu = 1/3$ and $2/3$ edges}

In the main text, we show that the effective chemical potential of an out-of-equilibrium anyonic channel is uniquely defined only after specifying the charge of individual tunneling quasiparticles.
This is because inter-edge tunneling processes in anyonic systems can involve quasiparticles with different charges.

In this section, we discuss an example in which the edges A and B have filling fractions $\nu = 1/3$ and $2/3$, respectively. In this setup, both charge-$e/3$ and charge-$2e/3$ quasiparticles can participate in the inter-edge tunneling. We further discuss the preferred scheme that is more suitable to study the tunneling of charge-$2e/3$ quasiparticles. The geometry of the setup considered here for composite edges remains the same as in Fig. 1(a) of the main text, with a diluter and a collider connecting edges SA, A, and B (cf. Fig.~\ref{fig:different_fillings}).

An edge of a fractional quantum Hall system at filling fraction $\nu=2/3$ is a composite edge comprising counter-propagating channels~\cite{SAMacDonald90, SWenPRL90, SWenPRB91a, SWenPRB91b, SJohnsonMacDonald91, SKaneFisherPolchinskiPRL94}. 
We will denote with $\nu$ the filling fraction of the state whose physical edge is a composite one and use $\nu_c$ for the filling fraction of individual channels forming the edge.
The common starting point for the analysis of the structure of a $\nu=2/3$ edge is its representation in terms of two channels with $\nu_c = 1$ and $\nu_c=-1/3$. This corresponds to the so-called 
interaction-free fixed point.
Intra-edge interactions mix the bare $\nu_c = 1$ and $\nu_c=-1/3$ channels. For sufficiently strong interactions between the bare channels, the eigenmodes of the $\nu = 2/3$ edge are close to the $\nu_c = 2/3$ charge mode and the neutral mode, which are exact modes at the ``Kane-Fisher-Polchinski fixed point''~\cite{SKaneFisherPolchinskiPRL94}.

At the interaction-free fixed point, 
the eigenmodes are the free bosonic fields $\phi_{1/3}$ and $\phi_1$ describing the counterpropagating $\nu_c = -1/3$ and $\nu_c=1$ channels. The allowed vertex operators for the $\nu=2/3$ edge have the following general form in terms of these modes~\cite{SKaneFisherPolchinskiPRL94}:
\begin{equation}
    \Psi_{n_{1/3},n_1} \propto e^{i \big(\sqrt{1/3}\, n_{1/3} \phi_{1/3} + n_1 \phi_1 \big)},
    \label{eq:original_ops}
\end{equation}
where $n_{1/3}$ and $n_1$ are the two integers corresponding to the $\nu_c = -1/3$ and $\nu_c=1$ modes, respectively.
An alternative representation of the composite $\nu=2/3$ edge involves the charge (subscript ``c'') and neutral (subscript ``n'') fields:
\begin{equation}
    \phi_c = \sqrt{\frac{3}{2}} \left(\phi_1 + \sqrt{\frac{1}{3}} \phi_{1/3}\right), \quad \phi_n = -\sqrt{\frac{1}{2}} \left(\phi_1 + \sqrt{3} \phi_{1/3}\right).
\end{equation}
The vertex operators, Eq.~\eqref{eq:original_ops}, can be rewritten in terms of the charge and neutral modes as follows:
\begin{equation}
    e^{i \big(\sqrt{1/3}\, n_{1/3} \phi_{1/3} + n_1 \phi_1 \big)} = e^{i (l_c \phi_c + l_n \phi_n)}.
    \label{eq:kfp_ops}
\end{equation}
At the Kane-Fisher-Polchinski fixed point, the charge and neutral modes are independent eigenmodes, and the vertex operators are characterized by the scaling dimension $(l_c^2 + l_n^2)/2$~\cite{SGoldsteinGefenPRL16}.

We emphasize that, based on Eq.~\eqref{eq:kfp_ops}, even at the Kane-Fisher-Polchinski fixed point, the allowed tunneling operators for the $\nu=2/3$ edge are characterized by the two integers, $n_{1/3}$ and $n_1$, which count the numbers of $e/3$ and $e$ quasiparticles involved in the tunneling process, respectively. Inter-channel interactions only modify the scaling dimension of the tunneling operators.

\begin{table}[h!]
\centering
\begin{tabular}{||c ||} 
 \hline
 \multicolumn{1}{|c|}{Processes} \\
 \hline\hline
 \\ 
 1 \\
 2\\
 3  \\ [1ex] 
 \hline
\end{tabular}
\begin{tabular}{||c c c c c||} 
 \hline
 \multicolumn{5}{|c|}{$\nu = 2/3$-edge operators} \\
 \hline\hline
 $n_1$ & $n_{1/3}$ & $l_c$ & $l_n$ & SD \\ 
 0 & -1 & $1/\sqrt{6}$ & $1/\sqrt{2}$ & $1/3$\\
 1 & 2 & $1/\sqrt{6}$ & $-1/\sqrt{2}$ & $1/3$ \\
 1 & 1 & $\sqrt{2/3}$ & 0 & $1/3$\\ [1ex] 
 \hline
\end{tabular}
\begin{tabular}{||c c||} 
 \hline
 \multicolumn{2}{|c|}{$\nu = 1/3$-edge operators} \\
 \hline\hline
 $n_{1/3}'$ & SD\\
 1 & $1/6$ \\ 
 1 & $1/6$ \\
 2 & $2/3$ \\ [1ex] 
 \hline
\end{tabular}
\begin{tabular}{||c ||} 
 \hline
 \multicolumn{1}{|c|}{Charges} \\
 \hline\hline
  \\
 $e/3$ \\ 
 $e/3$ \\
 $2e/3$ \\ [1ex] 
 \hline
\end{tabular}
\begin{tabular}{||c ||} 
 \hline
 \multicolumn{1}{|c|}{Total SD} \\
 \hline\hline
 \\
 $1/2$ \\ 
 $1/2$ \\
 $1$ \\ [1ex] 
 \hline
\end{tabular}
\caption{Leading operators (marked from 1 to 3) for tunneling of fractional quasiparticles between $\nu=2/3$ and $1/3$ edges at the Kane-Fisher-Polchinsky fixed point of the $\nu=2/3$ edge. 
In this table, we provide the scaling dimensions (SD) both for the complete tunneling operators (last column) and separately for vertex operators of $\nu=2/3$ and $\nu = 1/3$ edges.  Process 1 corresponds to direct tunneling of a charge-$e/3$ anyon between the $\phi_{1/3}$ and $\phi_{1/3}'$ modes. Processes 2 and 3 are Andreev-like tunneling processes, where a charge-$e$ fermion, from the $\nu_c = 1$ channel of the $\nu=2/3$ edge, splits into two fractional-charge quasiparticles, $2e/3$ and $e/3$. The charges of the tunneling quasiparticles for each of the processes are indicated in the fourth column. In a single tunneling event, $\nu = 1/3$ and $2/3$ edges exchange with charge $e/3$ for processes 1 and 2 and with charge $2 e/3$ in process 3. Ignoring the presence of neutral mode ($l_n=0$ at the Kane-Fisher-Polchinsky point), process 3 can be considered as a realization of tunneling in the setup shown in Fig.~\ref{fig:different_fillings},
with $\nu=1/3$, $m_\text{A}=1$, $m\text{B}=2$, and $m=2$. In this process, charge-$2e/3$ quasiparticles tunnel between the charge mode of the $\nu=2/3$ edge and $\nu=1/3$ edge.}
\label{tab:operators}
\end{table}

Now, we combine the allowed vertex operators of the $\nu=2/3$ edge with those of the $\nu=1/3$ edge, $$\Psi_{n_{1/3}'}\propto \exp (i\sqrt{\nu} n_{1/3}' \phi_{1/3}'),$$ 
to form a tunneling operator for the QPC bridging $\nu=2/3$ and $nu=1/3$ edges. Here, $n_{1/3}'$ is another integer and $\phi_{1/3}'$ is the bosonic field for the $\nu = 1/3$ edge. Relevant and marginal operators are listed in Table~\ref{tab:operators}, 
where $\nu=2/3$ operators for tunneling of fractional-charge quasiparticles are borrowed from Ref.~\cite{SGoldsteinGefenPRL16}. Tunneling processes involving integer charges, which were also discussed in Ref.~\cite{SGoldsteinGefenPRL16}, have scaling dimensions larger than 1, and are thus neglected.

Based on Table~\ref{tab:operators},
at the quantum point contact bridging $\nu=2/3$ and $\nu=1/3$ edges, both charge-$e/3$ and charge-$2e/3$ quasiparticles are allowed to tunnel, meaning that in realistic setups, tunneling between these two edges is normally more complicated than that considered in the main text, where only one type of quasiparticles (cf. Fig.~4 of the main text) is assumed to be transferred between the channels.
Fortunately, these idealized situations are, in principle, realizable in carefully designed experiments.

According to Table~\ref{tab:operators}, tunneling of charge $e/3$ dominates over the tunneling of charge $2e/3$ at sufficiently low temperatures.
Indeed, processes 1 and 2, both transferring charge $e/3$, have a total scaling dimension $1/2$. They are thus relevant in the renormalization-group sense, and become important at low energies.
The charge-$2e/3$ tunneling, listed as process 3 in Table~\ref{tab:operators}, has a scaling dimension $1$, and is thus marginal.
Consequently, charge $e/3$ tunneling processes of types 1 and 2 are more relevant than charge-$2e/3$ tunneling processes 3. However, tunneling of charge $2e/3$ can, in principle, be comparable to that of charge $e/3$ in experiments at not extremely low temperatures, given a sufficiently strong bare amplitude for charge-$2e/3$ tunneling. Thus, given the possibility of the coexistence of tunneling processes with different transferred charges, the effective potential for setups involving composite edges may depend on the microscopic details. The study of non-equilibrium distributions in such anyonic structures is relegated to future work.

\end{widetext}


\begin{thebibliography}{96}
\bibitem [{Note0()}]{f0}%
  \BibitemOpen
  \bibinfo {note} {Remarkably, the value of the chemical potential that follows this definition coincides with the value obtained from two other physically motivated definitions, cf. SI Sec.~S5.}\BibitemShut {Stop}%
\end{thebibliography}
\begin{thebibliography}{22}
\bibitem [{Note1()}]{f1}%
  \BibitemOpen
  \bibinfo {note} {Details of the derivation will be presented in a forthcoming work by the authors; See also Ref.~\cite{SOneHalfPRB24}}\BibitemShut {NoStop}%
\end{thebibliography}
\end{document}